\input harvmac
\input epsf

%
\let\includefigures=\iftrue
%
%
%
\newfam\black
\input rotate
\input epsf
\noblackbox
%
%
\includefigures
\message{If you do not have epsf.tex (to include figures),}
\message{change the option at the top of the tex file.}
\def\figin{\epsfcheck\figin}\def\figins{\epsfcheck\figins}
\def\epsfcheck{\ifx\epsfbox\UnDeFiNeD
\message{(NO epsf.tex, FIGURES WILL BE IGNORED)}
\gdef\figin##1{\vskip2in}\gdef\figins##1{\hskip.5in}
\else\message{(FIGURES WILL BE INCLUDED)}%
\gdef\figin##1{##1}\gdef\figins##1{##1}\fi}
\def\DefWarn#1{}
\def\N{{\cal N}}
\def\figinsert{\goodbreak\midinsert}
\def\ifig#1#2#3{\DefWarn#1\xdef#1{fig.~\the\figno}
\writedef{#1\leftbracket fig.\noexpand~\the\figno}%
\figinsert\figin{\centerline{#3}}\medskip\centerline{\vbox{\baselineskip12pt
\advance\hsize by -1truein\noindent\footnotefont{\bf
Fig.~\the\figno:} #2}}
\bigskip\endinsert\global\advance\figno by1}
\else
\def\ifig#1#2#3{\xdef#1{fig.~\the\figno}
\writedef{#1\leftbracket fig.\noexpand~\the\figno}%
\global\advance\figno by1} \fi

\def\tilde{\widetilde}

\def\subsubsec#1{\bigskip\noindent{\it #1}}
\def\yboxit#1#2{\vbox{\hrule height #1 \hbox{\vrule width #1
\vbox{#2}\vrule width #1 }\hrule height #1 }}
\def\fillbox#1{\hbox to #1{\vbox to #1{\vfil}\hfil}}
\def\ybox{{\lower 1.3pt \yboxit{0.4pt}{\fillbox{8pt}}\hskip-0.2pt}}

\def\rightarrowbox#1#2{
  \setbox1=\hbox{\kern#1{${ #2}$}\kern#1}
  \,\vbox{\offinterlineskip\hbox to\wd1{\hfil\copy1\hfil}
    \kern 3pt\hbox to\wd1{\rightarrowfill}}}

\def\QZ{\Bbb{Z}}

\def\ket#1{|#1\rangle}

\def\vev#1{\langle{#1}\rangle}

\def\CA{{\cal A}}

\def\tilde{\widetilde}

\def\II{\relax{I\kern-.10em I}}

\def\bar{\overline}

\def\IZ{\relax\ifmmode\mathchoice
{\hbox{\cmss Z\kern-.4em Z}}{\hbox{\cmss Z\kern-.4em Z}}
{\lower.9pt\hbox{\cmsss Z\kern-.4em Z}} {\lower1.2pt\hbox{\cmsss
Z\kern-.4em Z}}\else{\cmss Z\kern-.4em Z}\fi}
\def\IB{\relax{\rm I\kern-.18em B}}
\def\IC{{\relax\hbox{$\inbar\kern-.3em{\rm C}$}}}
\def\ID{\relax{\rm I\kern-.18em D}}
\def\IE{\relax{\rm I\kern-.18em E}}
\def\IF{\relax{\rm I\kern-.18em F}}
\def\IG{\relax\hbox{$\inbar\kern-.3em{\rm G}$}}
\def\IGa{\relax\hbox{${\rm I}\kern-.18em\Gamma$}}
\def\IH{\relax{\rm I\kern-.18em H}}
\def\II{\relax{\rm I\kern-.18em I}}
\def\IK{\relax{\rm I\kern-.18em K}}
\def\IN{\relax{\rm I\kern-.18em N}}
\def\IP{\relax{\rm I\kern-.18em P}}

%
\def\inbar{\,\vrule height1.5ex width.4pt depth0pt}

\font\cmss=cmss10 \font\cmsss=cmss10 at 7pt
\def\IR{\relax{\rm I\kern-.18em R}}

\def\lp10{l_P^{10}}
\def\lp11{l_P^{11}}
\def\R11{R_{11}}

\def\a{\alpha}
\def\b{\beta}

\def\lt{\tilde\lambda}

\def\gb#1{ {\langle #1 ] } }
\def\tgb #1{ \left[ #1 \right\rangle}

\def\D{\dot}
\def\la{\lambda}
\def\a{\alpha}
\def\b{\beta}
\def\braket{\vev}
\def\W{\widetilde}


\def\ra{\rangle}

\def\spa#1.#2{\langle#1\,#2\rangle}
\def\spb#1.#2{[#1\,#2]}

\def\spab#1.#2.#3{\langle\mskip-1mu{#1}
                  | #2 | {#3}]}

\def\spba#1.#2.#3{[\mskip-1mu{#1}
                  | #2 | {#3}\rangle}

\def\spbb#1.#2.#3.#4{[\mskip-1mu{#1}
                     | {#2} \ {#3} | {#4}]}

\def\spaa#1.#2.#3.#4{\langle\mskip-1mu{#1}
                     | {#2} \ {#3} | {#4}\rangle}

\def\dea{\langle \ell \ d \ell \rangle}
\def\deb{[\ell \ d \ell]}

\def\dedea{\langle d \ell \ \partial_\ell \rangle}
\def\dedeb{[d \ell \ \partial_\ell]}

\newbox\tmpbox\setbox\tmpbox\hbox{\abstractfont
}
\Title{\vbox{\baselineskip12pt
\hbox{hep-ph/0602178}
\hbox{Imperial/TP/}
\hbox{ITFA-2006-07}
\hbox{UCLA/06/TEP/06}
\hbox{ZU-TH 07/06}
}}
{\vbox{\centerline{The Cut-Constructible Part of QCD Amplitudes}
}}

\centerline{Ruth Britto${}^1$, Bo
Feng${}^{2,3}$, Pierpaolo Mastrolia${}^{4,5}$}

\smallskip
\smallskip
\bigskip

\centerline{\it ${}^1$Institute for Theoretical Physics, University of Amsterdam}
\centerline{\it Valckenierstraat 65, 1018 XE Amsterdam, The Netherlands}

\smallskip
\smallskip

\centerline{\it ${}^2$Blackett Laboratory, Imperial College, London, SW7 2AZ, UK}

\smallskip
\smallskip

\centerline{\it ${}^3$The Institute for Mathematical Sciences, Imperial College
  London}
\centerline{\it 48 Princes Gardens, London SW7 2AZ, UK}

\smallskip
\smallskip

\centerline{\it ${}^4$ Department of Physics and Astronomy, UCLA}
\centerline{\it Los Angeles, CA 90090-1547, USA}

\smallskip
\smallskip

\centerline{\it ${}^5$ Institut f\"ur Theoretische Physik,
                       Universit\"at Z\"urich}
\centerline{\it CH-8057 Z\"urich, Switzerland}

\bigskip

\bigskip
\smallskip
\noindent

\input amssym.tex

Unitarity cuts are widely used in analytic computation of loop amplitudes in gauge theories such as QCD.
We expand upon
the technique introduced in hep-ph/0503132 to carry out any finite unitarity cut integral.
This technique naturally separates the contributions of bubble, triangle and box integrals in one-loop amplitudes and is not constrained to any particular helicity configurations.
Loop momentum integration is reduced to a sequence of algebraic operations.
We discuss the extraction of the residues at higher-order poles.  Additionally, we offer concise algebraic formulas for expressing coefficients of three-mass triangle integrals.
As an application, we compute all remaining coefficients of bubble and triangle integrals for nonsupersymmetric six-gluon amplitudes.

\Date{February 2006}
%

\lref\cuts{ L.D. Landau, Nucl.\ Phys. {\bf 13}, 181 (1959);
 S. Mandelstam, Phys. Rev. {\bf 112}, 1344 (1958), {\bf 115}, 1741 (1959);
 R.E. Cutskosky, J. Math. Phys. {\bf 1}, 429 (1960).
}

\lref\DennerQQ{ A.~Denner, U.~Nierste and R.~Scharf, ``A Compact
expression for the scalar one loop four point function,'' Nucl.\
Phys.\ B {\bf 367}, 637 (1991).
}

\lref\BernZX{
  Z.~Bern, L.~J.~Dixon, D.~C.~Dunbar and D.~A.~Kosower,
  ``One loop n point gauge theory amplitudes, unitarity and collinear limits,''
  Nucl.\ Phys.\ B {\bf 425}, 217 (1994)
  [arXiv:hep-ph/9403226].
}

\lref\BernCG{
  Z.~Bern, L.~J.~Dixon, D.~C.~Dunbar and D.~A.~Kosower,
  ``Fusing gauge theory tree amplitudes into loop amplitudes,''
  Nucl.\ Phys.\ B {\bf 435}, 59 (1995)
  [arXiv:hep-ph/9409265].
}

\lref\WittenNN{ E.~Witten, ``Perturbative gauge theory as a string
theory in twistor space,'' Commun.\ Math.\ Phys.\  {\bf 252}, 189
(2004) [arXiv:hep-th/0312171].
}

\lref\CachazoKJ{ F.~Cachazo, P.~Svr\v{c}ek and E.~Witten, ``MHV
vertices and tree amplitudes in gauge theory,'' JHEP {\bf 0409}, 006
(2004) [arXiv:hep-th/0403047].
}

\lref\BerkovitsJJ{ N.~Berkovits and E.~Witten, ``Conformal
supergravity in twistor-string theory,'' JHEP {\bf 0408}, 009 (2004)
[arXiv:hep-th/0406051].
}

\lref\penrose{R. Penrose, ``Twistor Algebra,'' J. Math. Phys. {\bf
8} (1967) 345.}

\lref\berends{F. A. Berends, W. T. Giele and H. Kuijf, ``On
Relations Between Multi-Gluon And Multi-Graviton Scattering," Phys.
Lett. {\bf B211} (1988) 91.}

\lref\berendsgluon{F. A. Berends, W. T. Giele and H. Kuijf, ``Exact
and Approximate Expressions for Multigluon Scattering," Nucl. Phys.
{\bf B333} (1990) 120.}

\lref\bernplusa{Z. Bern, L. Dixon and D. A. Kosower, ``New QCD
Results From String Theory,'' in {\it Strings '93}, ed. M. B.
Halpern et. al. (World-Scientific, 1995), hep-th/9311026.}

\lref\bernplusb{Z. Bern, G. Chalmers, L. J. Dixon and D. A. Kosower,
``One Loop $N$ Gluon Amplitudes with Maximal Helicity Violation via
Collinear Limits," Phys. Rev. Lett. {\bf 72} (1994) 2134.}

\lref\bernfive{Z. Bern, L. J. Dixon and D. A. Kosower, ``One Loop
Corrections to Five Gluon Amplitudes," Phys. Rev. Lett. {\bf 70}
(1993) 2677.}

\lref\bernfourqcd{Z.Bern and  D. A. Kosower, "The Computation of
Loop Amplitudes in Gauge Theories," Nucl. Phys.  {\bf B379,} (1992)
451.}

\lref\cremmerlag{E. Cremmer and B. Julia, ``The $N=8$ Supergravity
Theory. I. The Lagrangian," Phys. Lett.  {\bf B80} (1980) 48.}

\lref\cremmerso{E. Cremmer and B. Julia, ``The $SO(8)$
Supergravity," Nucl. Phys.  {\bf B159} (1979) 141.}

\lref\dewitt{B. DeWitt, "Quantum Theory of Gravity, III:
Applications of Covariant Theory," Phys. Rev. {\bf 162} (1967)
1239.}

\lref\dunbarn{D. C. Dunbar and P. S. Norridge, "Calculation of
Graviton Scattering Amplitudes Using String Based Methods," Nucl.
Phys. B {\bf 433,} 181 (1995), hep-th/9408014.}

\lref\ellissexton{R. K. Ellis and J. C. Sexton, "QCD Radiative
corrections to parton-parton scattering," Nucl. Phys.  {\bf B269}
(1986) 445.}

\lref\gravityloops{Z. Bern, L. Dixon, M. Perelstein, and J. S.
Rozowsky, ``Multi-Leg One-Loop Gravity Amplitudes from Gauge
Theory,"  hep-th/9811140.}

\lref\kunsztqcd{Z. Kunszt, A. Signer and Z. Tr\'{o}cs\'{a}nyi,
``One-loop Helicity Amplitudes For All $2\rightarrow2$ Processes in
QCD and ${\cal N}=1$ Supersymmetric Yang-Mills Theory,'' Nucl. Phys.
{\bf B411} (1994) 397, hep-th/9305239.}

\lref\mahlona{G. Mahlon, ``One Loop Multi-photon Helicity
Amplitudes,'' Phys. Rev.  {\bf D49} (1994) 2197, hep-th/9311213.}

\lref\mahlonb{G. Mahlon, ``Multi-gluon Helicity Amplitudes Involving
a Quark Loop,''  Phys. Rev.  {\bf D49} (1994) 4438, hep-th/9312276.}

\lref\klt{H. Kawai, D. C. Lewellen and S.-H. H. Tye, ``A Relation
Between Tree Amplitudes of Closed and Open Strings," Nucl. Phys.
{B269} (1986) 1.}

\lref\pppmgr{Z. Bern, D. C. Dunbar and T. Shimada, ``String Based
Methods In Perturbative Gravity," Phys. Lett.  {\bf B312} (1993)
277, hep-th/9307001.}

\lref\GiombiIX{ S.~Giombi, R.~Ricci, D.~Robles-Llana and
D.~Trancanelli, ``A Note on Twistor Gravity Amplitudes,''
hep-th/0405086.
}

\lref\sbook{ R. J. Eden, P. V. Landshoff, D. I. Olive and J. C.
Polkinghorne, {\it The Analytic S-Matrix}, Cambridge University
Press, 1966. }

\lref\WuFB{
  J.~B.~Wu and C.~J.~Zhu,
  ``MHV vertices and scattering amplitudes in gauge theory,''
  JHEP {\bf 0407}, 032 (2004)
  [arXiv:hep-th/0406085].
}

\lref\Feynman{R.P. Feynman, Acta Phys. Pol. 24 (1963) 697, and in
{\it Magic Without Magic}, ed. J. R. Klauder (Freeman, New York,
1972), p. 355.}

\lref\Peskin{M.~E. Peskin and D.~V. Schroeder, {\it An Introduction
to Quantum Field Theory} (Addison-Wesley Pub. Co., 1995).}

\lref\parke{S. Parke and T. Taylor, ``An Amplitude For $N$ Gluon
Scattering,'' Phys. Rev. Lett. {\bf 56} (1986) 2459; F. A. Berends
and W. T. Giele, ``Recursive Calculations For Processes With $N$
Gluons,'' Nucl. Phys. {\bf B306} (1988) 759. }

\lref\loopmhv{A.~Brandhuber, B.~Spence and G.~Travaglini, ``One-Loop
Gauge Theory Amplitudes In N = 4 Super Yang-Mills From MHV
Vertices,'' Nucl.\ Phys.\ B {\bf 706}, 150 (2005)
  [arXiv:hep-th/0407214];
C.~Quigley and M.~Rozali,
  ``One-loop MHV amplitudes in supersymmetric gauge theories,''
  JHEP {\bf 0501}, 053 (2005)
  [arXiv:hep-th/0410278];
J.~Bedford, A.~Brandhuber, B.~Spence and G.~Travaglini,
  ``A twistor approach to one-loop amplitudes in N = 1 supersymmetric
  Yang-Mills theory,''
  Nucl.\ Phys.\ B {\bf 706}, 100 (2005)
  [arXiv:hep-th/0410280];
Y.~t.~Huang,
``N = 4 SYM NMHV loop amplitude in superspace,''
Phys.\ Lett.\ B {\bf 631}, 177 (2005)
[arXiv:hep-th/0507117].
A.~Brandhuber, B.~Spence and G.~Travaglini,
``From trees to loops and back,''
JHEP {\bf 0601}, 142 (2006)
[arXiv:hep-th/0510253].
}

\lref\CachazoZB{ F.~Cachazo, P.~Svr\v{c}ek and E.~Witten, ``Twistor
space structure of one-loop amplitudes in gauge theory,'' JHEP {\bf
0410}, 074 (2004) [arXiv:hep-th/0406177].
}

\lref\passarino{ L.~M. Brown and R.~P. Feynman, ``Radiative
Corrections To Compton Scattering,'' Phys. Rev. 85:231 (1952);
G.~Passarino and M.~Veltman, ``One Loop Corrections For E+ E-
Annihilation Into Mu+ Mu- In The Weinberg Model,'' Nucl. Phys.
B160:151 (1979); G.~'t Hooft and M.~Veltman, ``Scalar One Loop
Integrals,'' Nucl. Phys. B153:365 (1979); R.~G.~ Stuart, ``Algebraic
Reduction Of One Loop Feynman Diagrams To Scalar Integrals,'' Comp.
Phys. Comm. 48:367 (1988); R.~G.~Stuart and A.~Gongora, ``Algebraic
Reduction Of One Loop Feynman Diagrams To Scalar Integrals. 2,''
Comp. Phys. Comm. 56:337 (1990).}

\lref\neerven{ W. van Neerven and J. A. M. Vermaseren, ``Large Loop
Integrals,'' Phys. Lett. 137B:241 (1984)}

\lref\melrose{ D.~B.~Melrose, ``Reduction Of Feynman Diagrams,'' Il
Nuovo Cimento 40A:181 (1965); G.~J.~van Oldenborgh and
J.~A.~M.~Vermaseren, ``New Algorithms For One Loop Integrals,'' Z.
Phys. C46:425 (1990); G.J. van Oldenborgh,  PhD Thesis, University
of Amsterdam (1990); A. Aeppli, PhD thesis, University of Zurich
(1992).}

\lref\bernTasi{Z.~Bern, hep-ph/9304249, in {\it Proceedings of
Theoretical Advanced Study Institute in High Energy Physics (TASI
92)}, eds. J. Harvey and J. Polchinski (World Scientific, 1993). }

\lref\morgan{ Z.~Bern and A.~G.~Morgan, ``Supersymmetry relations
between contributions to one loop gauge boson amplitudes,'' Phys.\
Rev.\ D {\bf 49}, 6155 (1994), hep-ph/9312218.
}

\lref\RoiSpV{R.~Roiban, M.~Spradlin and A.~Volovich, ``A Googly
Amplitude From The B-Model In Twistor Space,'' JHEP {\bf 0404}, 012
(2004) hep-th/0402016; R.~Roiban and A.~Volovich, ``All Googly
Amplitudes From The $B$-Model In Twistor Space,''Phys.\ Rev.\ Lett.\
{\bf 93}, 131602 (2004) [arXiv:hep-th/0402121]; R.~Roiban,
M.~Spradlin and A.~Volovich, ``On The Tree-Level S-Matrix Of
Yang-Mills Theory,'' Phys.\ Rev.\ D {\bf 70}, 026009 (2004)
hep-th/0403190, S.~Gukov, L.~Motl and A.~Neitzke, ``Equivalence of
twistor prescriptions for super Yang-Mills,'' arXiv:hep-th/0404085,
I.~Bena, Z.~Bern and D.~A.~Kosower, ``Twistor-space recursive
formulation of gauge theory amplitudes,'' arXiv:hep-th/0406133. }

\lref\CachazoBY{
  F.~Cachazo, P.~Svr\v{c}ek and E.~Witten,
  ``Gauge theory amplitudes in twistor space and holomorphic anomaly,''
  JHEP {\bf 0410}, 077 (2004)
  [arXiv:hep-th/0409245].
}

\lref\DixonWI{ L.~J.~Dixon, ``Calculating Scattering Amplitudes
Efficiently,'' hep-ph/9601359.
}

\lref\BernMQ{ Z.~Bern, L.~J.~Dixon and D.~A.~Kosower, ``One Loop
Corrections To Five Gluon Amplitudes,'' Phys.\ Rev.\ Lett.\  {\bf
70}, 2677 (1993), [hep-ph/9302280].
%
Z.~Kunszt, A.~Signer and Z.~Trocsanyi,
``One loop radiative corrections to the helicity amplitudes of QCD processes
involving four quarks and one gluon,''
Phys.\ Lett.\ B {\bf 336} (1994) 529
[hep-ph/9405386];
%
Z.~Bern, L.J.~Dixon and D.A.~Kosower,
``One loop corrections to two quark three gluon amplitudes,''
Nucl.\ Phys.\ B {\bf 437} (1995) 259
[hep-ph/9409393];
}

\lref\berends{F.~A.~Berends, R.~Kleiss, P.~De Causmaecker,
R.~Gastmans and T.~T.~Wu, ``Single Bremsstrahlung Processes In Gauge
Theories,'' Phys. Lett. {\bf B103} (1981) 124; P.~De Causmaeker,
R.~Gastmans, W.~Troost and T.~T.~Wu, ``Multiple Bremsstrahlung In
Gauge Theories At High-Energies. 1. General Formalism For Quantum
Electrodynamics,'' Nucl. Phys. {\bf B206} (1982) 53; R.~Kleiss and
W.~J.~Stirling, ``Spinor Techniques For Calculating P Anti-P $\to$
W+- / Z0 + Jets,'' Nucl. Phys. {\bf B262} (1985) 235; R.~Gastmans
and T.~T. Wu, {\it The Ubiquitous Photon: Helicity Method For QED
And QCD} Clarendon Press, 1990.}

\lref\xu{Z. Xu, D.-H. Zhang and L. Chang, ``Helicity Amplitudes For
Multiple Bremsstrahlung In Massless Nonabelian Theories,''
 Nucl. Phys. {\bf B291}
(1987) 392.}

\lref\gunion{J.~F. Gunion and Z. Kunszt, ``Improved Analytic
Techniques For Tree Graph Calculations And The G G Q Anti-Q Lepton
Anti-Lepton Subprocess,'' Phys. Lett. {\bf 161B} (1985) 333.}

\lref\GeorgiouBY{ G.~Georgiou, E.~W.~N.~Glover and V.~V.~Khoze,
``Non-MHV Tree Amplitudes In Gauge Theory,'' JHEP {\bf 0407}, 048
(2004) [arXiv:hep-th/0407027].
}

\lref\WuJX{
  J.~B.~Wu and C.~J.~Zhu,
  ``MHV vertices and fermionic scattering amplitudes in gauge theory with
  quarks and gluinos,''
  JHEP {\bf 0409}, 063 (2004)
  [arXiv:hep-th/0406146].
}

\lref\WuFB{ J.~B.~Wu and C.~J.~Zhu, ``MHV Vertices And Scattering
Amplitudes In Gauge Theory,'' JHEP {\bf 0407}, 032 (2004)
[arXiv:hep-th/0406085].
}

\lref\GeorgiouWU{ G.~Georgiou and V.~V.~Khoze, ``Tree Amplitudes In
Gauge Theory As Scalar MHV Diagrams,'' JHEP {\bf 0405}, 070 (2004)
[arXiv:hep-th/0404072].
}

\lref\Nair{V. Nair, ``A Current Algebra For Some Gauge Theory
Amplitudes," Phys. Lett. {\bf B78} (1978) 464. }

\lref\BernAD{ Z.~Bern, ``String Based Perturbative Methods For Gauge
Theories,'' hep-ph/9304249.
}

\lref\BernKR{ Z.~Bern, L.~J.~Dixon and D.~A.~Kosower,
``Dimensionally Regulated Pentagon Integrals,'' Nucl.\ Phys.\ B {\bf
412}, 751 (1994), hep-ph/9306240.
}

\lref\CachazoDR{ F.~Cachazo, ``Holomorphic Anomaly Of Unitarity Cuts
And One-Loop Gauge Theory Amplitudes,'' hep-th/0410077.
}

\lref\giel{W. T. Giele and E. W. N. Glover, ``Higher order
corrections to jet cross-sections in e+ e- annihilation,'' Phys.
Rev. {\bf D46} (1992) 1980; W. T. Giele, E. W. N. Glover and D. A.
Kosower, ``Higher order corrections to jet cross-sections in hadron
colliders,'' Nucl. Phys. {\bf B403} (1993) 633. }

\lref\kuni{Z. Kunszt and D. Soper, ``Calculation of jet
cross-sections in hadron collisions at order alpha-s**3,''Phys. Rev.
{\bf D46} (1992) 192; Z. Kunszt, A. Signer and Z. Tr\' ocs\' anyi,
``Singular terms of helicity amplitudes at one loop in QCD and the
soft limit of the cross-sections of multiparton processes,'' Nucl.
Phys. {\bf B420} (1994) 550. }

\lref\seventree{F.~A. Berends, W.~T. Giele and H. Kuijf, ``Exact And
Approximate Expressions For Multi - Gluon Scattering,'' Nucl. Phys.
{\bf B333} (1990) 120.}

\lref\mangpxu{M. Mangano, S.~J. Parke and Z. Xu, ``Duality And Multi
- Gluon Scattering,'' Nucl. Phys. {\bf B298} (1988) 653.}

\lref\colorord{ Z.~Bern and D.~A.~Kosower,
  ``Color Decomposition Of One Loop Amplitudes In Gauge Theories,''
  Nucl.\ Phys.\ B {\bf 362}, 389 (1991); F.~A.~Berends and W.~Giele,
  ``The Six Gluon Process As An Example Of Weyl-Van Der Waerden Spinor
  Calculus,''
  Nucl.\ Phys.\ B {\bf 294}, 700 (1987); M. Mangano, S.~J. Parke and Z. Xu, ``Duality And Multi - Gluon Scattering,'' Nucl. Phys. {\bf B298}
(1988) 653; M.~L.~Mangano,
  ``The Color Structure Of Gluon Emission,''
  Nucl.\ Phys.\ B {\bf 309}, 461 (1988);}

\lref\mangparke{M. Mangano and S.~J. Parke, ``Multiparton Amplitudes
In Gauge Theories,'' Phys. Rep. {\bf 200} (1991) 301.}

\lref\grisaru{M. T. Grisaru, H. N. Pendleton and P. van
Nieuwenhuizen, ``Supergravity And The S Matrix,'' Phys. Rev.  {\bf
D15} (1977) 996; M. T. Grisaru and H. N. Pendleton, ``Some
Properties Of Scattering Amplitudes In Supersymmetric Theories,''
Nucl. Phys. {\bf B124} (1977) 81.}

\lref\BenaXU{
I.~Bena, Z.~Bern, D.~A.~Kosower and R.~Roiban,
``Loops in twistor space,''
Phys.\ Rev.\ D {\bf 71}, 106010 (2005)
[arXiv:hep-th/0410054].
}

\lref\BernKY{
Z.~Bern, V.~Del Duca, L.~J.~Dixon and D.~A.~Kosower,
``All non-maximally-helicity-violating one-loop seven-gluon amplitudes in N = 4
super-Yang-Mills theory,''
Phys.\ Rev.\ D {\bf 71}, 045006 (2005)
[arXiv:hep-th/0410224].
}

\lref\BrittoNJ{
  R.~Britto, F.~Cachazo and B.~Feng,
  ``Computing one-loop amplitudes from the holomorphic anomaly of unitarity
  cuts,''
  Phys.\ Rev.\ D {\bf 71}, 025012 (2005)
  [arXiv:hep-th/0410179].
}

\lref\BidderVX{
S.~J.~Bidder, N.~E.~J.~Bjerrum-Bohr, D.~C.~Dunbar and W.~B.~Perkins,
``Twistor space structure of the box coefficients of N = 1 one-loop
amplitudes,''
Phys.\ Lett.\ B {\bf 608}, 151 (2005)
[arXiv:hep-th/0412023].
}

\lref\BidderTX{
S.~J.~Bidder, N.~E.~J.~Bjerrum-Bohr, L.~J.~Dixon and D.~C.~Dunbar,
``N = 1 supersymmetric one-loop amplitudes and the holomorphic anomaly of
unitarity cuts,''
Phys.\ Lett.\ B {\bf 606}, 189 (2005)
[arXiv:hep-th/0410296].
}

\lref\BrittoNC{
R.~Britto, F.~Cachazo and B.~Feng,
``Generalized unitarity and one-loop amplitudes in N = 4 super-Yang-Mills,''
Nucl.\ Phys.\ B {\bf 725}, 275 (2005)
[arXiv:hep-th/0412103].
}

\lref\BernIX{ Z.~Bern and G.~Chalmers, ``Factorization in one loop
gauge theory,'' Nucl.\ Phys.\ B {\bf 447}, 465 (1995)
[arXiv:hep-ph/9503236].
}

\lref\LuoMY{
M.~x.~Luo and C.~k.~Wen,
``Compact formulas for all tree amplitudes of six partons,''
Phys.\ Rev.\ D {\bf 71}, 091501 (2005)
[arXiv:hep-th/0502009].
}

\lref\LuoRX{
M.~x.~Luo and C.~k.~Wen,
``Recursion relations for tree amplitudes in super gauge theories,''
JHEP {\bf 0503}, 004 (2005)
[arXiv:hep-th/0501121].
}

\lref\BrittoAP{
R.~Britto, F.~Cachazo and B.~Feng,
``New recursion relations for tree amplitudes of gluons,''
Nucl.\ Phys.\ B {\bf 715}, 499 (2005)
[arXiv:hep-th/0412308].
}

\lref\BidderRI{
S.~J.~Bidder, N.~E.~J.~Bjerrum-Bohr, D.~C.~Dunbar and W.~B.~Perkins,
``One-loop gluon scattering amplitudes in theories with N < 4
supersymmetries,''
Phys.\ Lett.\ B {\bf 612}, 75 (2005)
[arXiv:hep-th/0502028].
}

\lref\BernJE{
  Z.~Bern, L.~J.~Dixon and D.~A.~Kosower,
  ``Progress in one-loop QCD computations,''
  Ann.\ Rev.\ Nucl.\ Part.\ Sci.\  {\bf 46}, 109 (1996)
  [arXiv:hep-ph/9602280].
}

\lref\BedfordNH{
J.~Bedford, A.~Brandhuber, B.~J.~Spence and G.~Travaglini,
``Non-supersymmetric loop amplitudes and MHV vertices,''
Nucl.\ Phys.\ B {\bf 712}, 59 (2005)
[arXiv:hep-th/0412108].
}

\lref\MahlonSI{
  G.~Mahlon,
  ``Multi - gluon helicity amplitudes involving a quark loop,''
  Phys.\ Rev.\ D {\bf 49}, 4438 (1994)
  [arXiv:hep-ph/9312276].
}

\lref\BrittoFQ{
R.~Britto, F.~Cachazo, B.~Feng and E.~Witten,
``Direct proof of tree-level recursion relation in Yang-Mills theory,''
Phys.\ Rev.\ Lett.\  {\bf 94}, 181602 (2005)
[arXiv:hep-th/0501052].
}


\lref\BernSC{
  Z.~Bern, L.~J.~Dixon and D.~A.~Kosower,
  ``One-loop amplitudes for e+ e- to four partons,''
  Nucl.\ Phys.\ B {\bf 513}, 3 (1998)
  [arXiv:hep-ph/9708239].
}

\lref\BernDN{
  Z.~Bern, L.~J.~Dixon and D.~A.~Kosower,
  ``A two-loop four-gluon helicity amplitude in QCD,''
  JHEP {\bf 0001}, 027 (2000)
  [arXiv:hep-ph/0001001].
%
Z.~Bern, A.~De Freitas and L.~J.~Dixon, ``Two-loop amplitudes for
gluon fusion into two photons,'' JHEP {\bf 0109}, 037 (2001)
[hep-ph/0109078];
%
``Two-loop helicity amplitudes for gluon gluon scattering in QCD and
supersymmetric Yang-Mills theory,'' JHEP {\bf 0203}, 018 (2002)
[hep-ph/0201161].
}

\lref\BernDB{
  Z.~Bern and A.~G.~Morgan,
  ``Massive Loop Amplitudes from Unitarity,''
  Nucl.\ Phys.\ B {\bf 467}, 479 (1996)
  [arXiv:hep-ph/9511336].
}

\lref\BrittoHA{ R.~Britto, E.~Buchbinder, F.~Cachazo and B.~Feng,
``One-loop amplitudes of gluons in SQCD,''
Phys.\ Rev.\ D {\bf 72}, 065012 (2005) [arXiv:hep-ph/0503132].
}

\lref\BernJI{ Z.~Bern, L.~J.~Dixon and D.~A.~Kosower,
``The last of the finite loop amplitudes in QCD,''
Phys.\ Rev.\ D {\bf 72}, 125003 (2005) [arXiv:hep-ph/0505055].
}

\lref\BernCQ{ Z.~Bern, L.~J.~Dixon and D.~A.~Kosower,
``Bootstrapping multi-parton loop amplitudes in QCD,''
arXiv:hep-ph/0507005.
}

\lref\BernHH{ Z.~Bern, N.~E.~J.~Bjerrum-Bohr, D.~C.~Dunbar and
H.~Ita,
``Recursive calculation of one-loop QCD integral coefficients,''
JHEP {\bf 0511}, 027 (2005) [arXiv:hep-ph/0507019].
}

\lref\BernBT{ Z.~Bern, L.~J.~Dixon and D.~A.~Kosower,
``All next-to-maximally helicity-violating one-loop gluon amplitudes in N  = 4
super-Yang-Mills theory,''
Phys.\ Rev.\ D {\bf 72}, 045014 (2005) [arXiv:hep-th/0412210].
}


\lref\TwoLoopSplitting{ Z.~Bern, L.~J.~Dixon and D.~A.~Kosower,
``Two-loop g $\to$ g g splitting amplitudes in QCD,'' JHEP {\bf
0408}, 012 (2004) [hep-ph/0404293].
}

\lref\bry{
 Z. Bern, J.S. Rozowsky and B. Yan, Phys,
 ``Two-loop four-gluon amplitudes in N = 4 super-Yang-Mills,''
Phys.\ Lett.\ B {\bf 401}, 273 (1997) [hep-ph/9702424];
Z.~Bern, L.~J.~Dixon, D.~C.~Dunbar, M.~Perelstein and
J.~S.~Rozowsky, ``On the relationship between Yang-Mills theory and
gravity and its implication for ultraviolet divergences,'' Nucl.\
Phys.\ B {\bf 530}, 401 (1998) [hep-th/9802162];
%
C.~Anastasiou, Z.~Bern, L.~J.~Dixon and D.~A.~Kosower, ``Planar
amplitudes in maximally supersymmetric Yang-Mills theory,'' Phys.\
Rev.\ Lett.\  {\bf 91}, 251602 (2003) [hep-th/0309040];
%
Z.~Bern, L.~J.~Dixon and V.~A.~Smirnov, ``Iteration of planar
amplitudes in maximally supersymmetric Yang-Mills theory at three
loops and beyond,'' Phys.\ Rev.\ D {\bf 72}, 085001 (2005)
[arXiv:hep-th/0505205].
}

\lref\VIIIcuts{ E.~I.~Buchbinder and F.~Cachazo, ``Two-loop
amplitudes of gluons and octa-cuts in N = 4 super Yang-Mills,''
JHEP {\bf 0511}, 036 (2005)
[arXiv:hep-th/0506126].
}

\lref\DimShift{ Z.~Bern, L.~J.~Dixon, D.~C.~Dunbar and
D.~A.~Kosower, ``One-loop self-dual and N = 4 super-Yang-Mills,''
Phys.\ Lett.\ B {\bf 394}, 105 (1997) [hep-th/9611127];
%
Z.~Bern, L.~J.~Dixon, M.~Perelstein and J.~S.~Rozowsky, ``Multi-leg
one-loop gravity amplitudes from gauge theory,'' Nucl.\ Phys.\ B
{\bf 546}, 423 (1999) [hep-th/9811140].
}

\lref\BidderXX{ S.~J.~Bidder, D.~C.~Dunbar and W.~B.~Perkins,
``Supersymmetric Ward identities and NMHV amplitudes involving
gluinos,'' JHEP {\bf 0508}, 055 (2005)
[arXiv:hep-th/0505249].
}

\lref\BDKrecrel{ Z.~Bern, L.~J.~Dixon and D.~A.~Kosower, ``On-shell
recurrence relations for one-loop QCD amplitudes,'' Phys.\ Rev.\ D
{\bf 71}, 105013 (2005) [hep-th/0501240].
}

\lref\FordeHH{
  D.~Forde and D.~A.~Kosower,
``All-multiplicity one-loop corrections to MHV amplitudes in QCD,''
  arXiv:hep-ph/0509358.
}


\lref\SalamDU{
  G.~P.~Salam,
  ``Developments in perturbative QCD,''
  arXiv:hep-ph/0510090.
}

\lref\GianottiFM{
  F.~Gianotti and M.~L.~Mangano,
  ``LHC physics: The first one-two year(s),''
  arXiv:hep-ph/0504221.
}

\lref\DixonCF{
  L.~J.~Dixon,
  ``Twistor string theory and QCD,''
  arXiv:hep-ph/0512111.
}

\lref\CachazoGA{
  F.~Cachazo and P.~Svrcek,
  ``Lectures on twistor strings and perturbative Yang-Mills theory,''
  PoS {\bf RTN2005}, 004 (2005)
  [arXiv:hep-th/0504194].
}
\lref\DennerES{
  A.~Denner, S.~Dittmaier, M.~Roth and L.~H.~Wieders,
  ``Complete electroweak O(alpha) corrections to charged-current e+ e- $\to$
  Phys.\ Lett.\ B {\bf 612}, 223 (2005)
  [arXiv:hep-ph/0502063].
%
  A.~Denner, S.~Dittmaier, M.~Roth and L.~H.~Wieders,
  ``Electroweak corrections to charged-current e+ e- $\to$ 4 fermion
  processes: Technical details and further results,''
  arXiv:hep-ph/0505042.
%
  F.~Boudjema, {\it et al.}",
"Electroweak corrections for the study of the Higgs potential at the LC",
  arXiv:hep-ph/0510184

}

\lref\LesHouches{
A. Denner, and S. Dittmaier,
``Reduction schemes for one-loop tensor integrals",
Nucl.\ Phys.\ B {\bf 734} 62 (2006)
[arXiv:hep-ph/0509141].
%
S. Dittmaier,
``Separation of soft and collinear singularities from one-
loop N-point  integrals",
Nucl.\ Phys.\ B {\bf 675} 447 (2003)
[arXiv:hep-ph/0308246].
%
Giele, W. T. and Glover, E. W. N.,
``A calculational formalism for one-loop integrals",
JHEP \ {\bf 04} (2004) 029
[arXiv:hep-ph/0402152].
%
Ellis, R. K. and Giele, W. T. and Zanderighi, G.,
``Virtual QCD corrections to Higgs boson plus four parton processes",
Phys. \ Rev. \ D {\bf 72} 054018 (2005)
[arXiv:hep-ph/0506196].
%
--
``Semi-numerical evaluation of one-loop corrections",
arXiv:hep-ph/0508308.
%
Binoth, T. and Guillet, J. P. and Heinrich, G.,
``Reduction formalism for dimensionally regulated one-loop 
N-point  integrals",
Nucl.\ Phys.\ B{\bf 572} 361 (2000)
[arXiv:hep-ph/9911342].
%
Binoth, T. and Guillet, J. Ph. and Heinrich, G. and Pilon, E. and Schubert, C.,
``An algebraic / numerical formalism for one-loop multi-leg amplitudes",
JHEP {\bf 10} 015 (2005)
[arXiv:hep-ph/0504267].
%
Boudjema, F. and others,
``Multi-leg calculations with the GRACE/1-LOOP system: Toward
radiative  corrections to e+ e- --> mu- anti-nu u anti-d",
Nucl. Phys. Proc. Suppl. {\bf 135} 323 (2004)
[arXiv:hep-ph/0407079].
%
Kurihara, Y.,
``Dimensionally regularized one-loop tensor-integrals with
massless internal particles",
arXiv:hep-ph/0504251".
%
del Aguila, F. and Pittau, R.,
``Recursive numerical calculus of one-loop tensor integrals",
JHEP {\bf 07} 017 (2004)
[arXiv:hep-ph/0404120].
%
van Hameren, A. and Vollinga, J. and Weinzierl, S.,
``Automated computation of one-loop integrals in massless theories",
Eur. Phys. J. C {\bf 41} 361 (2005)
[arXiv:hep-ph/0502165].
%
Ferroglia, A., Passera, M., Passarino, G.,and Uccirati, Sandro,
``All-purpose numerical evaluation of one-loop multi-leg Feynman diagrams",
Nucl. Phys. B {\bf 650} 162 (2003)
[arXiv:hep-ph/0209219].
Binoth, T. and Heinrich, G. and Kauer, N.,
``A numerical evaluation of the scalar hexagon integral in the physical region",
Nucl. Phys. B {\bf 654} 277 (2003)
[arXiv:hep-ph/0210023].
%
Nagy, Z. and Soper, D. E.,
``General subtraction method for numerical calculation of one-loop QCD  matrix
elements", 
JHEP {\bf 09} 055 (2003)
[arXiv:hep-ph/0308127].
%
--
Acta Phys. Polon. B {\bf 35} 2557 (2004).
%
Kurihara, Y. and Kaneko, T.,
``Numerical contour integration for loop integrals",
arXiv:hep-ph/0503003.
%
Anastasiou, C. and Daleo, A.,
``Numerical evaluation of loop integrals",
arXiv:hep-ph/0511176".
}

\lref\BadgerZH{
S.~D.~Badger, E.~W.~N.~Glover, V.~V.~Khoze and P.~Svrcek,
``Recursion relations for gauge theory amplitudes with massive particles,''
JHEP {\bf 0507}, 025 (2005)
[arXiv:hep-th/0504159].
}

\lref\QuigleyCU{
C.~Quigley and M.~Rozali,
``Recursion relations, helicity amplitudes and dimensional regularization,''
arXiv:hep-ph/0510148.
}

\lref\BadgerJV{
S.~D.~Badger, E.~W.~N.~Glover and V.~V.~Khoze,
``Recursion relations for gauge theory amplitudes with massive vector bosons
and fermions,''
JHEP {\bf 0601}, 066 (2006)
[arXiv:hep-th/0507161].
}

\lref\RisagerKE{
K.~Risager, S.~J.~Bidder and W.~B.~Perkins,
``One-loop NMHV amplitudes involving gluinos and scalars in N = 4 gauge
theory,''
JHEP {\bf 0510}, 003 (2005)
[arXiv:hep-th/0507170].
}

\lref\FordeUE{
D.~Forde and D.~A.~Kosower,
``All-multiplicity amplitudes with massive scalars,''
arXiv:hep-th/0507292.
}

\lref\RodrigoEU{
G.~Rodrigo,
``Multigluonic scattering amplitudes of heavy quarks,''
JHEP {\bf 0509}, 079 (2005)
[arXiv:hep-ph/0508138].
}

\lref\FerrarioNP{
P.~Ferrario, G.~Rodrigo and P.~Talavera,
``Compact multigluonic scattering amplitudes with heavy scalars and fermions,''
arXiv:hep-th/0602043.
}

\lref\BrandhuberJW{
A.~Brandhuber, S.~McNamara, B.~J.~Spence and G.~Travaglini,
``Loop amplitudes in pure Yang-Mills from generalised unitarity,''
JHEP {\bf 0510}, 011 (2005)
[arXiv:hep-th/0506068].
}

\newsec{Introduction}

Within the experimental programme of the forthcoming Large Hadron Collider,
and the exigencies of efficient ways to analyse huge
amounts of data, perturbative QCD will play the role of precision physics.
To dig out interesting signals we must be able to
distinguish them from backgrounds, largely dominated by QCD processes.
In order to have enough theoretical accuracy for comparisons
against the experimental counterpart, one-loop (and even higher) many-leg
amplitudes of several QCD processes are needed \refs{\SalamDU,\GianottiFM}.
However, the calculation of next-to-leading order (NLO)
amplitudes is extremely difficult; within the current status of the
available analytical results, `many legs' means five \BernMQ\ as
regarding QCD corrections and six \DennerES\ as for electroweak.
The most recent theoretical efforts for tackling the one-loop multi-leg
amplitudes have been using algebraic/semi-numerical approaches 
\LesHouches.

For the analysis of jet-events produced
at the high energies of the LHC,
it is mandatory to overcome the bottleneck of
the one-loop six-gluon amplitude.

Although it is very difficult to calculate QCD amplitudes, various
methods have been developed to attack this problem. One
efficient approach is
 the unitarity cut method with the spinor-helicity formalism \refs{\berends, \xu, \gunion} (a review
may be found in \refs{\DixonWI}).
It is shown in
\refs{\BernMQ,\BernZX,\BernCG} that one-loop amplitudes with all
external gluons and a gluon
 circulating around the loop can be decomposed into following three
 pieces (the so-called supersymmetry decomposition),
\eqn\themiste{\CA^{\rm QCD}=\CA^{\N=4}-4\CA^{\N=1}+\CA^{\rm scalar},
}
where $\CA^{\rm QCD}$ denotes an amplitude with only a gluon
circulating in the loop, $\CA^{\N=4}$ has the full $\N=4$ multiplet
circulating in the loop,  $\CA^{\N=1}$ has an $\N=1$ chiral
supermultiplet in the loop, and $\CA^{\rm scalar}$ has only a
complex scalar in the loop.  This last term is sometimes referred to as $\CA^{\N=0}$.

The main advantage of this decomposition is that supersymmetric
amplitudes, $\CA^{\N=4}$ and $\CA^{\N=1}$, are four-dimensional
cut-constructible \refs{\BernZX,\BernCG}, meaning that they do not
suffer any ambiguity due to the presence of rational terms; they
are completely determined by their finite unitarity cuts.

However, the term $\CA^{\rm scalar}$ cannot be completely reconstructed from
its absorptive part, and the presence of rational functions of the kinematic invariants make its calculation quite involved, though still
 simpler than the full $\CA^{\rm QCD}$. In fact, it is our aim in this
paper to complete the program introduced in \BrittoHA\ to give a systematic way to evaluate the cut-constructible piece of
$\CA^{\rm scalar}$.
 The determination of the rational terms could perhaps be
later achieved by implementing the recursive technique introduced
in \refs{\BernCQ,\BernJI,\FordeHH}.  An alternative is to apply the unitarity method in $(4-2\epsilon)$ dimensions, for there the entire amplitude is cut-constructible \BernJE.

Our systematic method to deal with the 
$\CA^{\rm scalar}$ part is related methods and ideas from the twistor
string theory initiated in \refs{\WittenNN} and further developed
in \refs{\CachazoKJ,\CachazoZB,\CachazoBY,\CachazoDR}.
In particular, we make heavy use of
the new way to write phase space integrals and perform the integration
given in \refs{\CachazoKJ,\CachazoDR}.\foot{For a review, see
\refs{\CachazoGA} and its citations. See also \refs{\DixonCF}.}
The algorithm initiated in \BrittoHA\ and developed here {\sl
reduces phase space integration to algebraic manipulations.}

With the complete algorithm developed in this paper, we
calculate the (heretofore missing) cut-constructible part of the NLO six-gluon
amplitudes in QCD.

\subsec{The Current Status of Amplitudes}

Under the separation given by \themiste, the unitarity method
established itself as an effective means of computation
\refs{\BernZX,\BernCG,\BernSC,\TwoLoopSplitting,\BernKY,\BrittoNC,\BernBT}.
This
method has been applied successfully in several contexts,
 supersymmetric
\refs{\BernZX,\BernCG,\bry,\VIIIcuts,\BernBT,\BrittoHA}
 as well as nonsupersymmetric
 \refs{\BernDB,\BernSC,\BernDN,\TwoLoopSplitting}.
A recent innovation \refs{\BrittoNC,\VIIIcuts},
making use of leading singularities, allows a simple
determination of coefficients of integrals associated to box
topologies, without any explicit integration.\foot{With
the unitarity method, tree-level amplitudes are the bricks of the
cut integral.
Off-shell recursion relations \refs{\parke,\mangparke} are a
well-known method to construct trees.
As valid alternatives, there are two new
efficient techniques, exploiting the on-shellness of the amplitudes:
the MHV diagrammatic rules of \CachazoKJ\ and
the on-shell recursion relations of  \refs{\BrittoAP,
\BrittoFQ}.}
Some related methods are
the  use of MHV diagrams in
 the cut calculations \loopmhv\ and the
use  \refs{\CachazoDR,\BrittoNJ,\BidderTX} of the holomorphic
anomaly \CachazoBY\ to determine certain cut integrals
\refs{\BenaXU,\CachazoDR}.

Although supersymmetric multiplets
contain more particles,
the reading of the QCD amplitudes in this
supersymmetric fashion introduces
a degree of simplicity in terms of computation.

The simplest term is the contribution
of an $\N=4$ super Yang-Mills multiplet.  The
$\N=4$ amplitudes can be expressed as a combination of scalar box integral
functions with rational coefficients \BernZX. These
coefficients have been evaluated in a closed form for the case with
maximal helicity violation (MHV), namely helicity configurations where
two gluons are of negative helicity and the rest are of positive
helicity \refs{\BernZX,\loopmhv}, and for the case of next-to-MHV (NMHV)
amplitudes \refs{\BernCG,\BrittoNJ,\BernKY,\BernBT,\BrittoNC}. The ingredients here are the  Parke-Taylor tree-level MHV
amplitudes \refs{\parke,\mangparke}.

In the case of an $\N=1$ chiral
multiplet, the all-$n$ one-loop MHV
and one-loop six-gluon amplitudes are known
\refs{\BernCG,\loopmhv,\BidderTX,\BidderVX,\BidderRI,\BrittoHA}.

As we have said, the most difficult part in the decomposition \themiste\ is
 $\CA^{\rm scalar}$ part, which is known only in special cases.
All amplitudes with at most one negative-helicity gluon were computed in
\refs{\bernplusa,\bernplusb,\MahlonSI}. The cut parts of the MHV
amplitudes are known \refs{\BernCG,\BernZX,\BedfordNH} for an
arbitrary number of legs. Very recently an explicit form of the
rational functions has been presented  for the all-multiplicity MHV
amplitude in which the negative-helicity gluons are adjacent
\refs{\BernCQ,\FordeHH}.
All box coefficients of the six-gluon amplitude are given in \BidderRI.
All cut-constructible coefficients of
one-loop
amplitudes where the gluons are
ordered in two adjacent bunches of opposite helicity (a `split helicity' configuration), for ${\cal
N}=0$ and ${\cal N}=1$, have been computed in \BernHH.

To achieve the complete calculation of $\CA^{\rm scalar}$ for the six-gluon
amplitude, there are two possible paths to follow. The first
is to apply the unitarity method in $(4-2\epsilon)$ dimensions instead
of four \BernJE.
Cases with four external particles, and up to six,
in special helicity configurations, have been worked out along this
way
\refs{\BernDB,\DimShift,\BernDN,\TwoLoopSplitting,\BrandhuberJW}.
Recent progress in deriving the tree-amplitude ingredients has appeared in \refs{\BadgerZH, \BadgerJV, \FordeUE, \RodrigoEU, \QuigleyCU, \FerrarioNP}.
Alternatively, one can split $\CA^{\rm scalar}$ into a
 cut-constructible piece and a remaining rational function, and tackle these
 two pieces separately. The reason is as follows.
 Due to a better understanding of the recursive structure of QCD
amplitudes, at tree level \refs{\BrittoAP,\BrittoFQ} and at one-loop \BDKrecrel, and exploiting the knowledge of their collinear and
soft-behaviour,
it has recently been shown \refs{\BernCQ,\BernJI,\FordeHH} that the
rational term of one-loop QCD amplitudes does have, in itself, a
recursive character: given the knowledge of the coefficients of the
logarithmic and polylogarithmic terms of an $n$--point amplitude, the
leftover rational coefficient has been reconstructed by feeding into
the recursion the rational coefficients of the $(n-1)$--point
amplitudes which represent the all-channel collinear limits of the
$n$--point one. Thus, if we calculate the cut-constructible part,
we can try to obtain the corresponding rational part by recursion
relations. Combining the results,
one might obtain the complete answer for $\CA^{\rm scalar}$.

\subsec{The Plan of the Paper}

The plan of this paper is as follows. In Section 2, we give the general
setting for our paper.
We exploit the divergent behavior of the amplitude to reduce the size of the
basis of known integrals.
Then we analyze the structure of our phase space integration.
We show that we can
neatly divide  contributions into rational and logarithmic
parts. For the logarithmic parts, we show where the three-mass triangle
and
four-mass box functions show up. Furthermore, we give our general
strategy to reduce integration to algebraic manipulation.
Specifically, we show how to read out residues of higher-order poles, which
is one of the most important steps in this procedure.

Starting from Section 3, we describe
the calculation of the cut-constructible part of
$\CA^{\rm scalar}$ of six-gluon
scattering amplitudes. Section 3 is dedicated to the helicity
configuration $(1^-,2^-,3^-,4^+,5^+, 6^+)$; Section 4, to the
configuration $(1^-,2^+,3^-,4^+,5^-, 6^+)$; and Section 5,
to configuration $(1^-,2^-,3^+,4^-,5^+, 6^+)$.

The results of the rational coefficients of bubble and triangle functions are
expressed, in compact form, in terms of sums over spinor products. 

Concluding remarks are given in Section 6.

We supply the paper with three appendices which contain the main technical
details of the calculation.
Appendix A gives the NMHV tree-level
amplitudes which enter the cuts.
Appendix B contains a detailed analysis of logarithmic contributions in
cut-integration.
In Appendix C, we show how to read out the coefficients
of three-mass triangles, which require the computation of
one-Feynman-parameter integrals,
and we define some functions we will use throughout the
manuscript.

%
%
%
We leave to future work the automatic
implementation of the whole algorithm for providing the numerical 
counterpart of our results. 
That would allow numerical checks of our results against those already 
existing in the literature.  The non-uniqueness of the expressions in the framework
of the spinor formalism, 
due to hidden identities among spinor products, 
(e.g. Schouten identities and momentum
conservation), can make direct analytic comparison quite difficult.
Currently,
spinor algebra manipulation is frequently carried out in  
{\bf Mathematica}, while the final integration, 
employing the Cauchy residue theorem and, when needed,  
one-Feynman-parameter representation, has not been automatized. 



\newsec{General Setting: Preliminaries}

In this paper, we freely use the word `amplitude' to refer to the
cut-constructible part of $\CA^{\rm scalar}$ for the leading-color
partial amplitude of gluons.

Our purpose is to expand and develop the procedure introduced in
\BrittoHA, so we shall not repeat all the background information
here.  Rather we shall mainly just point out the new features showing up
 in this application.  We also make use of the same
notation and conventions as in \BrittoHA, which follows the spinor-helicity formalism \refs{\berends,\xu,\gunion} and conventions of \WittenNN.  In particular,  in the
following calculations we generally omit an overall factor of  $-i
{(4\pi)^{2-\epsilon} \over (2\pi)^{4-2\epsilon}}$.

By reduction techniques, the cut-constructible portion of the
amplitude may be expanded in a basis of scalar integral functions
known as boxes ($I_4$), triangles ($I_3$), and bubbles ($I_2$)
\refs{\BernKR, \BernCG}.
\eqn\ampl{\eqalign{ A_n = {r_\Gamma (\mu^2)^{\epsilon} \over
 (4\pi)^{2-\epsilon}}\sum & \left(c_2 I_2 + c_3^{1m} I_3^{1m}+ c_3^{2m}
I_3^{2m}+ c_3^{3m} I_3^{3m} \right. \cr &~~~ \left. + c_4^{1m}
I_4^{1m} + c_4^{2m~e} I_4^{2m~e}+ c_4^{2m~h} I_4^{2m~h} + c_4^{3m}
I_4^{3m} + c_4^{4m} I_4^{4m} \right).}}
This defines what we mean by the cut-constructible portion of the
amplitude. Here $\epsilon \equiv (4-D)/2$ is the dimensional
regularization parameter, $\mu$ is the renormalization scale, and
$r_\Gamma$ is defined by \eqn\rgamma{r_\Gamma =
{\Gamma(1+\epsilon)\Gamma^2(1-\epsilon) \over \Gamma(1-2\epsilon)}.
} The sum runs over all the cyclic permutations within each type of
integral.

For a gluon amplitude with a complex scalar in the loop, the
infrared and ultraviolet singular behavior is
\refs{\kuni,\giel,\BernIX} 
\eqn\nzerosing{
 A_{n}^{\rm scalar}|_{\rm singular} = {r_\Gamma \over
3 \epsilon (4\pi)^{2-\epsilon}} A_{n}^{\rm tree}. } 
Because the
divergence has exactly the same $\epsilon$-dependence as $
A_{n}^{\N=1}|_{\rm singular}$, we may follow the same argument as in
\BrittoHA\ to express the amplitude in terms of a smaller, modified
basis of scalar integrals with {\sl no one-mass or two-mass triangle
functions}:
\eqn\kole{\eqalign{
 \CA_{n}^{\rm scalar} = & {r_\Gamma (\mu^2)^{\epsilon} \over
 (4\pi)^{2-\epsilon}} \sum\left( c_2 I_2 + c_3^{3m} I_3^{3m}
+ c_4^{1m} I_{4F}^{1m} + c_4^{2m~e} I_{4F}^{2m~e} 
\right. \cr &~~~~~~~~~~~~~~~~~~~~~~~ \left.
 + c_4^{2m~h}
I_{4F}^{2m~h} + c_4^{3m} I_{4F}^{3m} + c_4^{4m} I_{4}^{4m} \right).}}
This basis differs from the one in \ampl\ in that each integral
function (except the bubble functions) has had its divergences
stripped away. See Figures 1 and 2.  Precise definitions appear in Appendix A of
\BrittoHA.

\ifig\ponhug{Scalar bubble and triangle integrals. (a) One-mass triangle $I^{1m}_{3;i}$. (b) Two-mass triangle $I^{2m}_{3:r;i}$.  (c) Three-mass triangle $I^{3m}_{3:r:r';i}$.  (d) Bubble $I_{2:r;i}$.  Note that the modified basis in \kole\ involves only three-mass triangles and bubbles.}
{\epsfxsize=0.70\hsize\epsfbox{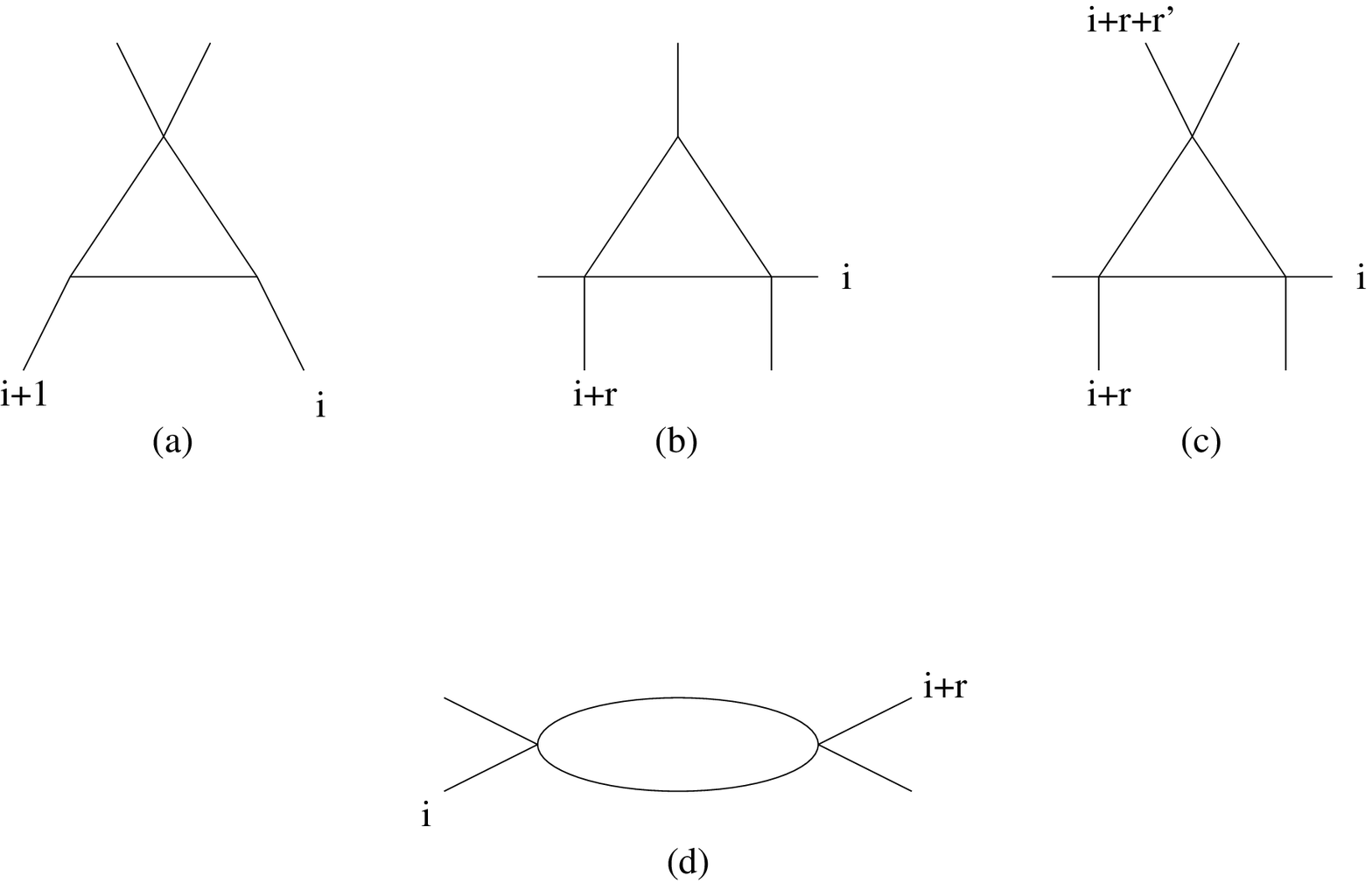}}

\ifig\kkmbp{Scalar box integrals.  (a) The outgoing external
momenta at each of the vertices are $K_1,K_2,K_3,K_4$, defined to
correspond to sums of the momenta of gluons in the exact
orientation shown. (b) One-mass $I^{1m}_{4;i}$. (c) Two-mass
``easy" $I^{2m~e}_{4:r;i}$. (d) Two-mass ``hard"
$I^{2m~h}_{4:r;i}$. (e) Three-mass $I^{3m}_{4:r:r';i}$. (f)
Four-mass $I^{4m}_{4:r:r':r";i}$. }
{\epsfxsize=0.70\hsize\epsfbox{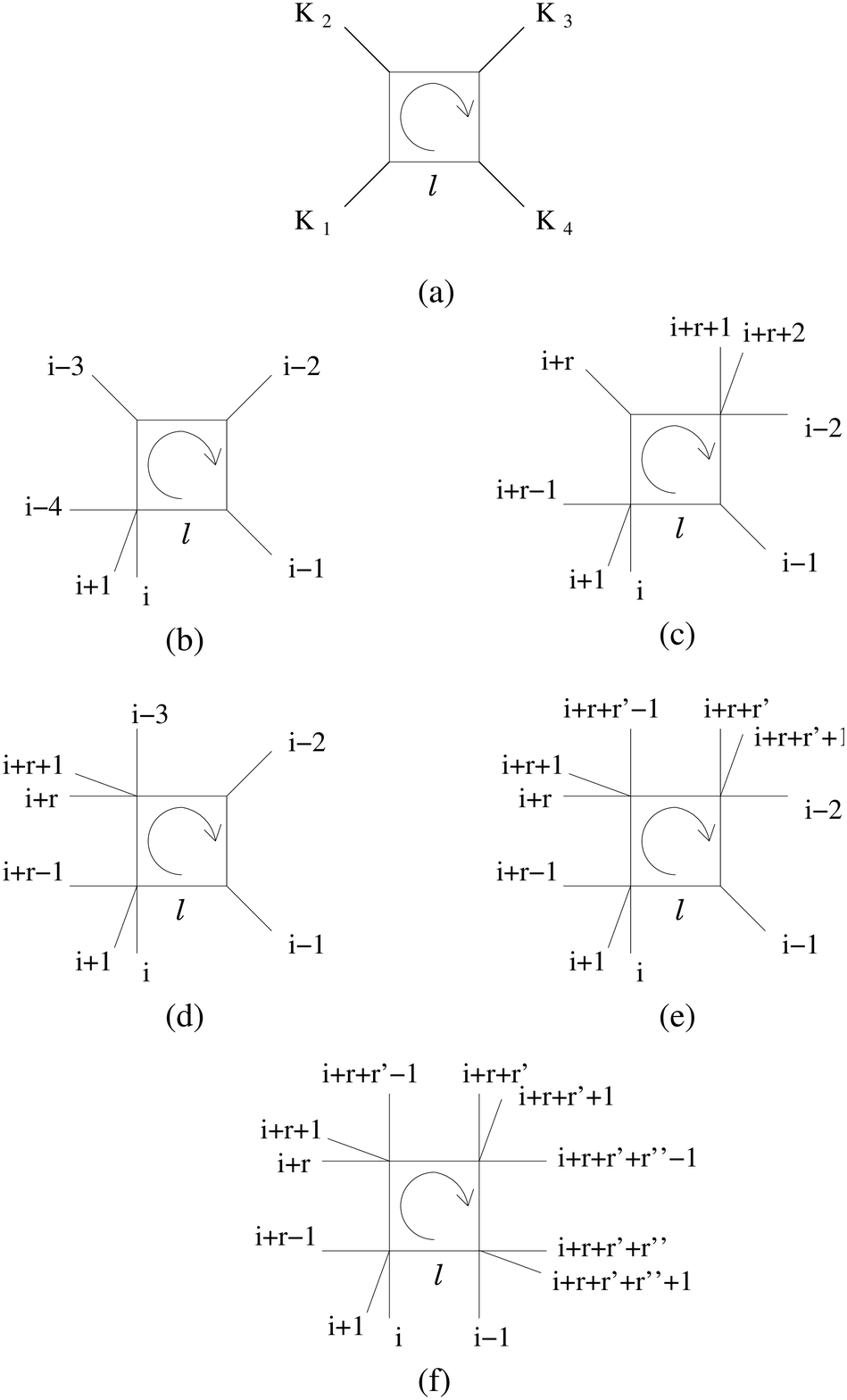}}

The justification for eliminating one-mass and two-mass triangles is based on 
the observation that according to \nzerosing, $\CA_{n}^{\rm scalar}$ diverges as $1/\epsilon.$  Since one-mass and two-mass triangle integrals
are the only ones with larger, $1/\epsilon^2$, divergences, these leading terms must conspire to cancel.  After observing further that these $1/\epsilon^2$ divergences arise only in the particular combination $(-s)^{-\epsilon}/\epsilon^2$, where $s$ is a momentum invariant, it follows that the contributions of one-mass and two-mass triangles may be neglected altogether as per the modified basis in \kole. 

To compute the amplitude, it is sufficient to compute each of these
coefficients separately.  The principle of the unitarity-based
method \refs{\BernZX, \BernCG,\BernDB} is to exploit the unitarity
cuts of the scalar integrals to extract the coefficients.

\subsec{Coefficients from Unitary Cuts}

Our goal is to compute the coefficients in \kole\ by applying
unitarity cuts \cuts\foot{
This technique is thoroughly discussed in \sbook.
While these early works were not
 intended to apply to massless theories,
we find there is no obstacle in adapting them to our purposes.
  The modern interpretation is found in
\refs{\BernZX,\BernCG}.} directly in four dimensions.

\ifig\convi{Representation of the cut integral. Left and right
tree-level amplitudes are on-shell. Internal lines represent the
legs coming from the cut propagators.}
{\epsfxsize=0.50\hsize\epsfbox{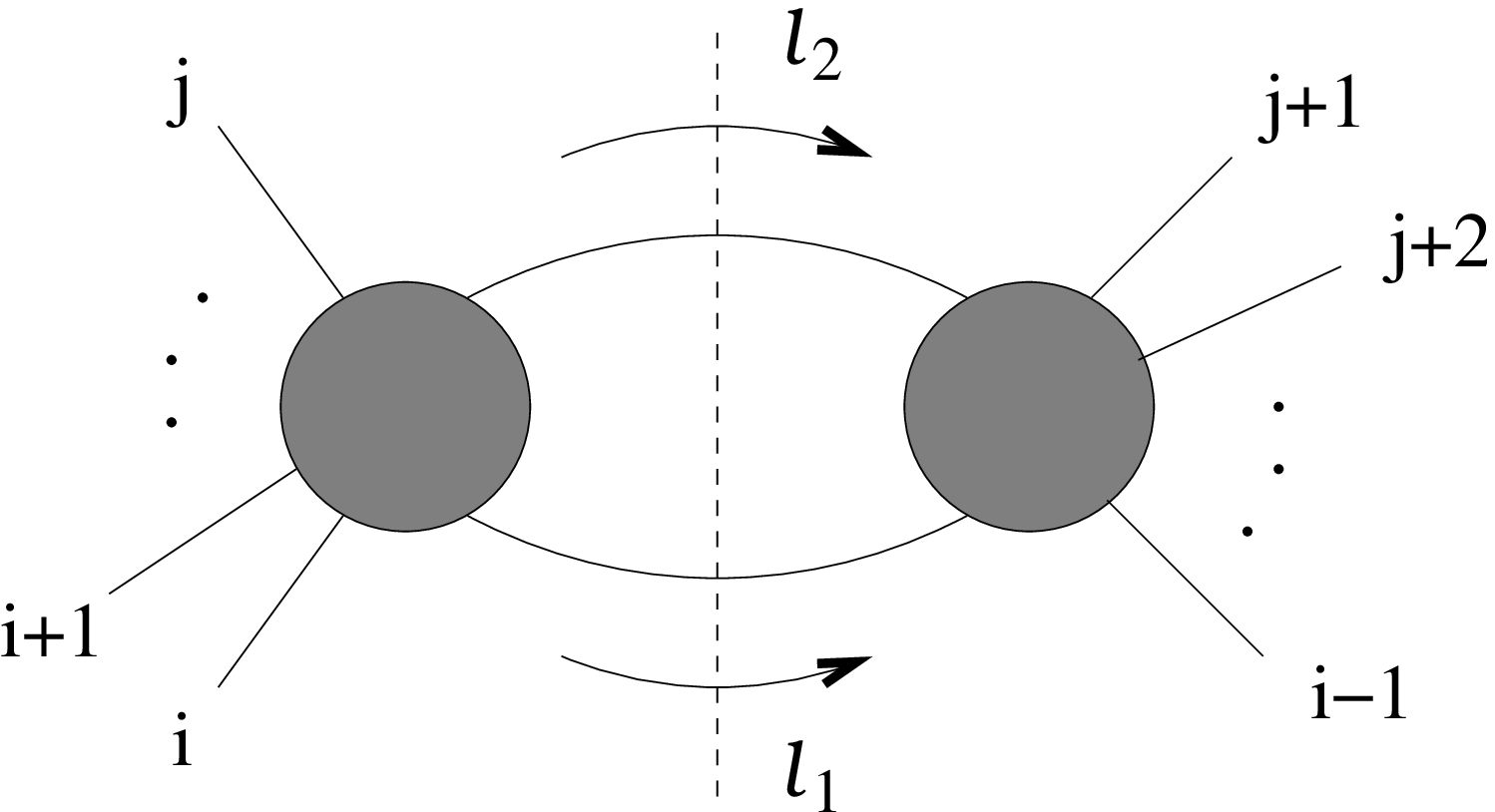}}

On the right-hand side, the scalar integrals are known functions, so
their discontinuities are also known explicitly.  We can read out
the coefficients of box integrals from quadruple cuts, while
coefficients of bubbles and three-mass triangle integrals from
double cuts.  The procedure of \BrittoHA\ can be extended to
box integrals as well, but since it is more efficient to use
quadruple cuts, we do not explore that possibility any further here.

To be able to read out these coefficients of bubbles and three-mass
triangles we need to know their discontinuities in double cuts. They
are given by \eqn\bubbledelta{ \Delta I_2(P_{cut})= -1,}
 and
\eqn\threemdelta{
 \Delta
I_3^{3m}(K_1) =\int_0^1 dz {1\over (z Q+ (1-z) K_1)^2},~~~~Q=K_3
+{K_3^2\over K_1^2} K_1 }
 where $K_1$ is the
cut momentum (see \BrittoHA\ for detailed derivations). In fact we
could carry out the integration for $\Delta I_3^{3m}(K_1)$ as in
\BrittoHA.  But
 to read off the coefficients of three-mass triangles, we need
only compare the expressions on both sides without really doing the
$z$-integration, so the form \threemdelta\ is more useful.\foot{The
discontinuities of box functions may be found in \CachazoDR.}

On the left-hand side of \kole, the discontinuity of the amplitude
in the $P_{i,j}$ momentum channel is computed by the integral
\eqn\cutIn{ \eqalign{ & C_{i,i+1,\ldots ,j-1,j} = \cr &   \int d\mu
A^{\rm tree}(\ell_1,i,i+1,\ldots ,j-1,j,\ell_2)A^{\rm
tree}((-\ell_2),j+1,j+2,\ldots ,i-2,i-1,(-\ell_1)),}}
where $d\mu= d^4\ell_1 d^4\ell_2
\delta^{(+)}(\ell_1^2)\delta^{(+)}(\ell_2^2)\delta^{(4)}
(\ell_1+\ell_2 - P_{ij})$ is the Lorentz invariant phase space
measure of two light-like vectors $(\ell_1, \ell_2)$ constrained by
momentum conservation.  See Figure 3.

We need to bring the integral \cutIn\ into a form convenient to work
with. In a nutshell, we begin by expressing the two tree-level
amplitudes in terms of spinor products. We then use the
four-dimensional delta function to integrate one of the propagator
momenta. Then, we use the technique of \CachazoKJ\ to rewrite the
measure in terms of spinors.
\eqn\meas{ \int d^4\ell \delta^{(+)}(\ell^2) ~ (\bullet ) =
\int_0^{\infty}dt~t\int\vev{\lambda,
d\lambda}[\tilde\lambda,d\tilde\lambda] ( \bullet ),}
where the bullets represent generic arguments, and the integration
contour for the spinors is the diagonal $\Bbb{CP}^1$ defined by
$\tilde\lambda =\bar\lambda$.

Next, we use the remaining delta function to perform the
$t$-integral. The fact that $\lambda$ and $\lt$ are independent
homogeneous coordinates on two copies of $\Bbb{CP}^1$ means that the
result must be homogeneous in $\lambda$ and $\lt$.  In particular,
it may be written as a sum of terms of the form
\eqn\genform{
 {1\over \gb{\ell|P_{cut}|\ell}^n} {\prod
[a_i ~\ell] \prod \vev{b_j~\ell}\prod \gb{\ell|Q_k|\ell} \over \prod
[c_i ~\ell] \prod \vev{d_j~\ell}\prod \gb{\ell|
O_k|\ell}},~~~~O_k\neq P_{cut},~~O_k^2\neq 0, }
 where $P_{cut}$ is the momentum in the channel
of the corresponding double cut and the $O_k$ are massive.

There are two key features of the form \genform. The first key
feature is that the degrees of both $\la$ and $\W \la$ are $-2$,
which is consistent with the scaling of the integration measure $\int
\vev{\ell~d\ell}[\ell~d\ell]$.

The second key feature is that  among all
 kinds of factors in denominators, namely
 $\gb{\ell|P_{cut}|\ell}$, $[c~\ell]$, $\vev{d~\ell}$ and $\gb{\ell|O|\ell}$,
only the factor $\gb{\ell|P_{cut}|\ell}$ can appear with power
greater than one. The reason is clear: the factor
$\gb{\ell|P_{cut}|\ell}$ comes from $t$-integration, so in principle
we can have an arbitrary power,\foot{The cases $n=1$ and $n=0$ are
degenerate. For example, the $n=0$ case shows up in Appendix A of
\BrittoNC.} while the other three factors $[c~\ell]$,
$\vev{d~\ell}$ and $\gb{\ell|O|\ell}$, with $O^2\neq 0$, come from
tree-level amplitudes, which have only {\sl single} poles.

\subsec{Canonical Decomposition}

 Now we discuss the canonical decomposition procedure given in \BrittoHA.
 When we do the decomposition we need to choose which variables, $\la$ or $\W\la$, to be reduced.
 For the general discussion in this section and in Appendix B we reduce
 $\W\la$ variables, but in the explicit examples, we may reduce $\la$,
 depending on the situation.

First, we should split the various single poles by partial fractioning,
using the following identity:
 \eqn\splitone{\eqalign{
 { [\ell~c]\over [\ell~a][\ell~b]} & =   {
[a~b][\ell~c]\over [a~b][\ell~a][\ell~b]}= {
[\ell~a][c~b]+[\ell~b][a~c]\over [a~b][\ell~a][\ell~b]} \cr & =  {
[c~b]\over [a~b]}{1\over [\ell~b]}+{ [a~c]\over [a~b]}{1\over
[\ell~a]}, }}
and its generalization,
where the degree of
$\W\lambda$ in both numerator and denominator decreases by one. It
is worth noting that in the process of splitting, we may have the
following factors in the denominator: $\gb{\ell|O_k|c_i}$ or
$\vev{\ell|O_k O_j|\ell}$. But the important point is that
these factors appear only once, i.e., {\sl they are all single
poles.}

After splitting all single poles we end up with factors\foot{As we
have remarked, for degenerate cases we may end up with factors like
${(\gb{\ell|O|\ell} \gb{\ell|Q|\ell})}^{-1}$ or
${([\ell~a][\ell~b])}^{-1}$, but the discussion applies to these
cases as well.}
$({\gb{\ell|P_{cut}|\ell}^n [\ell~a]})^{-1}$ or
${(\gb{\ell|P_{cut}|\ell}^n \gb{\ell|Q|\ell})}^{-1}$. Then we need
to perform the same splitting of $\gb{\ell|P_{cut}|\ell}$ and
$[\ell~a]$ (or $\gb{\ell|P_{cut}|\ell}$ and $\gb{\ell|Q|\ell}$).
After finishing this step we have the following types of terms:
\eqn\fourtypes{\eqalign{
 (1)& ~~
{G(\la,\W\la)\over \gb{\ell|P_{cut}|a}^{m+1}
\gb{\ell|P_{cut}|\ell}^{n-m}},~~~(2)~~ {G(\la,\W\la)\over
\vev{\ell|P_{cut}Q|\ell}^{m+1} \gb{\ell|P_{cut}|\ell}^{n-m}}, \cr
(3)& ~~  {F(\la)\over \gb{\ell|P_{cut}|a}^{n-1}}{1\over
\gb{\ell|P_{cut}|\ell}[\ell~a]},~~~(4)~~ {F(\la)\over
\vev{\ell|P_{cut}Q|\ell}^{n-1}}{1\over
\gb{\ell|P_{cut}|\ell}\gb{\ell|Q|\ell}}, }}
where $G(\la,\W\la)$ is a function of both $\la$ and $\W\la$ while
$F(\la)$ is a function of $\la$ only. One important thing for both
functions $G(\la,\W\la)$ and $F(\la)$ is that they have only {\sl
single} poles.

The results in \fourtypes\ are our final results for the canonical
decomposition. There are several points to be explained. First, we
have multiple poles like $\gb{\ell|P_{cut}|a}$ or
$\vev{\ell|P_{cut}Q|\ell}$, so we need to discuss how to read out
residues  of these multiple poles. Second, as we will analyze
carefully in Appendix B, terms of types (1) and (2) will contribute
to rational functions while terms of types (3) and (4) will
contribute to {\sl pure} logarithmic functions. Thirdly, type (3)
will only contribute to  one-mass, two-mass and three-mass box
functions, while type (4) will contribute in addition to three-mass
triangle and four-mass box functions. The reason is because $Q^2\neq
0$ in type (4).
Since box coefficients are easily obtained from quadruple cuts, we will pay the most attention to type (4).

\subsec{Rewriting as Total Derivative}

Before proceeding to extract residues of  multiple poles, let us
recall the strategy of integration. The key idea is to write $\int
\vev{\ell~d\ell}[\ell~d\ell] G(\la, \W \la)$ in the form $\int
\vev{\ell~d\ell}[d\ell~\partial_\ell] \W G(\la, \W \la)$. Then the
integration is reduced to algebraic manipulation by reading off
residues at  poles in $\W G(\la, \W \la)$. One useful formula is
given by \BrittoHA\
\eqn\geninte{\eqalign{ &  { [\ell~d\ell] \prod_{i=1}^{j+1}
[\eta_i~\ell][\eta_{j+2}~\ell]^{n-j-1}\over\gb{\ell|P|\ell}^{n+2}}=
[d\ell~\partial_\ell] \left[ {\prod_{i=1}^{j+1} \gb{\ell|P|\eta_i}
\over  \gb{\ell|P|\ell}^{n+1}} \right. \cr & \times \left.\left(
\sum_{k=0}^{j+1} { (-1)^{j+1-k} (j+1-k)!\over (n+1-(j+1)) (n+1-j)
\cdots (n+1-k)} g_k[x_s] {[\eta_{j+2}~\ell]^{n+1-k} \over
\gb{\ell|P|\eta_{j+2}}^{(j+1)+1-k}}\right) \right], }}
 where $\eta$
is an arbitrary but fixed auxiliary spinor and
\eqn\hubidr{ g_k[x_i]=\sum_{i_1<i_2<\cdots<i_k} x_{i_1} x_{i_2}
\cdots x_{i_k},~~~{\rm with}~~ x_i= { [\eta_i~\ell] \over
\gb{\ell|P|\eta_i}}. }
One special case is when all the $\eta_i$ are the same and we choose the
auxiliary spinor to be same $\eta$ as well.  Then we have
\eqn\speziell{ [\ell~d\ell] \left( { [\eta~\ell]^n \over
\gb{\ell|P|\ell}^{n+2}}\right)= [d\ell~\partial_{\ell}]\left(
{1\over (n+1)} {1\over \gb{\ell|P|\eta}} {[\eta~\ell]^{n+1}\over
\gb{\ell|P|\ell}^{n+1}}\right), }
where no multiple poles show up. We will use \speziell\ often in
later calculations.

For terms of types (1) and (2) we can use the formula \geninte\
directly, so the  results are just  pure rational functions from
residues of poles. But for type (3) and (4) we need to use Feynman
parametrization first to write it in our standard form, as in
\geninte, and then read out residues.  After that we need to integrate
the Feynman parameter. Notice that the order of  {\sl reading
out residues} versus {\sl Feynman parameter integration} is important.
In Appendix B we discuss with care
 the properties of this
integration.

\subsec{The Residues of Multiple Poles}

As we have seen, in general there are multiple poles in $\W G(\la,
\W \la)$ after we rewrite the integral in the form $\int
\vev{\ell~d\ell}[d\ell~\partial_\ell] \W G(\la, \W \la)$. We need to
know how to read off residues at these multiple poles.\foot{In
\BrittoHA, no multiple poles were encountered. 
We do not know whether it is a general
feature of supersymmetric theories
that no multiple poles show
up.}

The main idea is the following. Recall the underlying complex
analysis. To obtain the residue of $\oint dz~z^{-n}f(z)$, we need to
take the $(n-1)$-th derivative of the function $f(z)$ and set $z=0$.
One important consequence of the above result is that  if the degree
of polynomial function of $f(z)$ is less than $(n-1)$, we get zero
contribution.

For our problem, notice that the degree of $\la$ is $-2$ in $\W
G(\la, \W \la)$, so if we split  $\W G(\la, \W \la)$ using the
identity \eqn\sep{
 {\braket{\ell~a}\over \braket{\ell~\eta}\braket{\ell~b}}
={\braket{\eta~a}\over \braket{\eta~b}} {1\over
\braket{\ell~\eta}}+{\braket{b~a}\over \braket{b~\eta}}{1\over
\braket{\ell~b}}, } then  at the end of splitting process we will
have terms like ${1\over \braket{\ell~\eta}^2}$,
${\braket{\ell~a}\over \braket{\ell~\eta}^3}$,
${\braket{\ell~a}\braket{\ell~b}\over \braket{\ell~\eta}^4}$ etc.\
(or in general, taking the form ${P_{n-2}(\la)\over
\braket{\ell~\eta}^n}$) for the multiple poles. However, since the
degree of $\la$ in numerator is two less than the degree of $\la$ in
denominator, by similar reasoning the residues of all these pieces
will be zero.

Now we demonstrate our strategy in several examples.

\subsubsec{The Double Pole Contribution}

Let us start from a term with a double pole, \eqn\mitig{ I_2
={1\over \braket{\ell~\eta}^2}{\prod_{j=1}^N \braket{\ell~a_j}\over
\prod_{k=1}^N \braket{\ell~b_k}}. } Using \sep\ once, we get
$$
I_2  = {\braket{\eta~a_1}\over \braket{\eta~b_1}} {1\over
\braket{\ell~\eta}^2}{\prod_{j=2}^N \braket{\ell~a_j}\over
\prod_{k=2}^N \braket{\ell~b_k}}+ {1\over
\braket{\ell~\eta}}{\braket{a_1~b_1}\over
\braket{\eta~b_1}\braket{\ell~a_1}}{\prod_{j=1}^N
\braket{\ell~a_j}\over \prod_{k=1}^N \braket{\ell~b_k}}.
$$
The second term already has just a single pole, namely $\braket{\ell~\eta}$, 
while the first term still has a double pole. We use \sep\ for the first
term again and get \eqn\gneiss{\eqalign{
 I_2 & =  {1\over
\braket{\ell~\eta}}{\braket{a_1~b_1}\over
\braket{\eta~b_1}\braket{\ell~a_1}}{\prod_{j=1}^N
\braket{\ell~a_j}\over \prod_{k=1}^N \braket{\ell~b_k}}
+{\braket{\eta~a_1}\over \braket{\eta~b_1}}{1\over
\braket{\ell~\eta}}{\braket{a_2~b_2}\over
\braket{\eta~b_2}\braket{\ell~a_2}}{\prod_{j=2}^N
\braket{\ell~a_j}\over \prod_{k=2}^N \braket{\ell~b_k}} \cr &  +
{\braket{\eta~a_1}\over \braket{\eta~b_1}}{\braket{\eta~a_2}\over
\braket{\eta~b_2}} {1\over \braket{\ell~\eta}^2}{\prod_{j=3}^N
\braket{\ell~a_j}\over \prod_{k=3}^N \braket{\ell~b_k}}. 
}} 
Iterating
this step, we eventually reach the final result:
\eqn\rectwo{ I_2 =
{1\over \braket{\ell~\eta}}\sum_{m=0}^{N-1} {\prod_i^m
\braket{\eta~a_i}\over \prod_i^m \braket{\eta~b_i}}
{\braket{a_{m+1}~b_{m+1}}\over
\braket{\eta~b_{m+1}}\braket{\ell~a_{m+1}}}{\prod_{j=m+1}^N
\braket{\ell~a_j}\over \prod_{k=m+1}^N \braket{\ell~b_k}}+{1\over
\braket{\ell~\eta}^2} {\prod_i^N \braket{\eta~a_i}\over \prod_i^N
\braket{\eta~b_i}}. }
 The second term does not
contribute, as we have argued, while the first term gives the
following residue at the pole $\ket{\ell}=\ket{\eta}$: \eqn\twopole{
P_{2}[\ket{\eta}, L_a,L_b] = {\prod_i^N \braket{\eta~a_i}\over
\prod_i^N \braket{\eta~b_i}} \sum_{i=1}^N{\braket{a_i~b_i}\over
\braket{\eta~b_i}\braket{\eta~a_i}}, } where the subscript ``2''
indicates a  double pole at the spinor $\ket{\eta}$. For ease of
presentation we have also defined two lists, $L_a=\{ a_1, a_2,...,
a_N\}$ and $L_b=\{ b_1,b_2,...,b_N\}$.

%


Let us do one brief example to illustrate the result \twopole.
Consider the integral
$$\int d\mu {
\braket{\ell~a}[b~\ell]\over \gb{\ell|P|\ell}^3}.$$ Using \geninte\
we can write it  as
$$
\int \braket{\ell~d\ell}[d\W \ell~\partial_{\W \ell}] \left(
{\braket{\ell~a}[b~\ell][\eta~\ell]\over \gb{\ell|P|\ell}^2
\gb{\ell|P|\eta}} -{\braket{\ell~a}\gb{\ell|P|b} [\eta~\ell]^2 \over
2 \gb{\ell|P|\ell}^2 \gb{\ell|P|\eta}^2}\right).
$$

Now we can do the integral in two ways. The first is to choose
$|\eta]=|b]$ to eliminate the double pole. The result is given by
$-{\gb{a|P|b}\over 2(P^2)^2}$.

For the second way we just let $\eta$ remain arbitrary. The first
term gives
$$
 -{\gb{a|P|\eta} \gb{\eta|P|b} \over (P^2)^2
\gb{\eta|P|\eta}}=-{\gb{a|P|b}  \over (P^2)^2
}-{\braket{a~\eta}[\eta~b] \over (P^2) \gb{\eta|P|\eta}}
$$
 The
second term has a double pole at $\ket{\ell}=|P|\eta]$. Then we use
the formula \twopole\ to get the result.
First, we have ${[\eta~\ell]^2}={\gb{\eta|P|\eta}^2}$. Second, we
have $\ket{a_1}=\ket{a}$, $\ket{a_2}=|P|b]$,
$\ket{b_1}=\ket{b_2}=|P|\ell]=(-P^2)\ket{\eta}$ where we have used
the fact that at the pole $|\ell]\to |P|\eta\rangle$.
Putting it together,
we have
\eqn\porph{\eqalign{ & {\gb{\eta|P|\eta}^2\over 2} {\gb{a|P|\eta}
[\eta~b]\over (P^2) \gb{\eta|P|\eta}^2}\left( {\braket{a~\eta}\over
\gb{\eta|P|\eta}\gb{a|P|\eta}}+{\gb{\eta|P|b}\over \gb{\eta|P|\eta}
P^2 [\eta~b]}\right) \cr & =  {\gb{a|P|\eta} [\eta~b]\over 2(P^2) }
{(2 P^2 \braket{a~\eta}[\eta~b]+ \gb{a|P|b}\gb{\eta|P|\eta})\over
\gb{\eta|P|\eta}\gb{a|P|\eta}P^2 [\eta~b]}\cr & =   {(2 P^2
\braket{a~\eta}[\eta~b]+ \gb{a|P|b}\gb{\eta|P|\eta})\over 2 P^2
\gb{\eta|P|\eta}P^2 }\cr & =  {\braket{a~\eta}[\eta~b] \over (P^2)
\gb{\eta|P|\eta}}+{\gb{a|P|b}  \over 2(P^2)^2 } 
\cr & = -{\gb{a|P|b}\over 2(P^2)^2},
 }}
in agreement with the first method.

\subsubsec{The Triple Pole Contribution}

Now we consider the triple pole given by \eqn\pigowl{ I_3
={\braket{\ell~\xi}\over \braket{\ell~\eta}^3}{\prod_{j=1}^N
\braket{\ell~a_j}\over \prod_{k=1}^N \braket{\ell~b_k}}. } 
Using
\sep\ once, we get
$$
 I_3 = {\braket{\eta~a_1}\over \braket{\eta~b_1}}
{\braket{\ell~\xi}\over \braket{\ell~\eta}^3}{\prod_{j=2}^N
\braket{\ell~a_j}\over \prod_{k=2}^N \braket{\ell~b_k}}+ {1\over
\braket{\ell~\eta}^2}{\braket{a_1~b_1}\braket{\ell~\xi}\over
\braket{\eta~b_1}\braket{\ell~a_1}}{\prod_{j=1}^N
\braket{\ell~a_j}\over \prod_{k=1}^N \braket{\ell~b_k}}.
$$
The second term has a double pole, which can be processed using
\twopole. Using \sep\ again on the  first term we get
\eqn\giocl{\eqalign{
 I_3 = &  {1\over
\braket{\ell~\eta}^2}{\braket{a_1~b_1}\braket{\ell~\xi}\over
\braket{\eta~b_1}\braket{\ell~a_1}}{\prod_{j=1}^N
\braket{\ell~a_j}\over \prod_{k=1}^N \braket{\ell~b_k}}+ {1\over
\braket{\ell~\eta}^2}{\braket{\eta~a_1}\over
\braket{\eta~b_1}}{\braket{a_2~b_2}\braket{\ell~\xi}\over
\braket{\eta~b_2}\braket{\ell~a_2}}{\prod_{j=2}^N
\braket{\ell~a_j}\over \prod_{k=2}^N \braket{\ell~b_k}}\cr &  +
{\braket{\eta~a_1}\over \braket{\eta~b_1}}{\braket{\eta~a_2}\over
\braket{\eta~b_2}} {\braket{\ell~\xi}\over
\braket{\ell~\eta}^3}{\prod_{j=3}^N \braket{\ell~a_j}\over
\prod_{k=3}^N \braket{\ell~b_k}}. }} 
Continuing in this way we get
\eqn\recthree{ I_3 = {1\over \braket{\ell~\eta}^2}\sum_{m=0}^{N-1}
{\prod_i^m \braket{\eta~a_i}\over \prod_i^m \braket{\eta~b_i}}
{\braket{a_{m+1}~b_{m+1}}\braket{\ell~\xi}\over
\braket{\eta~b_{m+1}}\braket{\ell~a_{m+1}}}{\prod_{j=m+1}^N
\braket{\ell~a_j}\over \prod_{k=m+1}^N
\braket{\ell~b_k}}+{\braket{\ell~\xi}\over \braket{\ell~\eta}^3}
{\prod_i^N \braket{\eta~a_i}\over \prod_i^N \braket{\eta~b_i}}. 
}

Now we can use the result \twopole\ to read out the contribution at
the triple pole. It is given by \eqn\threepole{\eqalign{ &
P_3[\ket{\eta},L_a,L_b]
 = \cr
&  {\braket{\eta~\xi} \prod_{i=1}^N \braket{\eta~a_i}\over
\prod_{i=1}^N \braket{\eta~b_i}}\left( \sum_{1\leq i\leq j\leq N}
{\braket{a_i~b_i}\over
\braket{\eta~b_i}\braket{\eta~a_i}}{\braket{a_j~b_j}\over
\braket{\eta~b_j}\braket{\eta~a_j}}+\sum_{i=1}^N
{\braket{a_i~b_i}\over
\braket{\eta~b_i}\braket{\eta~a_i}}{\braket{\xi~a_{i}}\over
\braket{\eta~\xi}\braket{\eta~a_i}}\right) }}
%
%
where the two lists given as arguments are $L_a=\{ a_1, a_2,...,
a_N,\xi\}$ and $L_b=\{ b_1,b_2,...,b_N\}$.

\subsubsec{Higher Multiplicity Poles}

We will not give the residue of a general multiple pole explicitly,
but one can now see how the procedure continues step by step.
 Defining
\eqn\genIn{ I_n  = {\prod_{j=1}^{n-2}\braket{\ell~\xi_j}\over
\braket{\ell~\eta}^n}{\prod_{k=1}^N \braket{\ell~a_k}\over
\prod_{k=1}^N \braket{\ell~b_k}}, } we have the following relation:
\eqn\recgen{
  I_n  =
{\prod_{j=2}^{n-2}\braket{\ell~\xi_j}\over
\braket{\ell~\eta}^{n-1}}\sum_{m=0}^{N-1} {\prod_i^m
\braket{\eta~a_i}\over \prod_i^m \braket{\eta~b_i}}
{\braket{a_{m+1}~b_{m+1}}\braket{\ell~\xi_1}\over
\braket{\eta~b_{m+1}}\braket{\ell~a_{m+1}}}{\prod_{j=m+1}^N
\braket{\ell~a_j}\over \prod_{k=m+1}^N
\braket{\ell~b_k}}+{\prod_{j=1}^{n-2}\braket{\ell~\xi_j}\over
\braket{\ell~\eta}^n} {\prod_i^N \braket{\eta~a_i}\over \prod_i^N
\braket{\eta~b_i}}, } where the last term does not contribute to the
residue, while the first term will give the residue of poles of one
lower multiplicity according to the formula for
$P_{n-1}[\ket{\eta},L_a,L_b]$.

\newsec{$A(1^-,2^-,3^-,4^+,5^+,6^+)$ }

In this section we demonstrate the principles outlined above, in the case of
$A(1^-,2^-,3^-,4^+,5^+,6^+)$, the simplest of the three NMHV
helicity configurations of six gluons.
The contribution to the cut constructible part of $\CA^{\rm scalar}$
in this 'split helicity' configuration  has 
already appeared in the literature \BernHH\ and was derived there by 
a recursive technique.

%

For this configuration of external gluons,
there are no contributions from box or triangle
integrals, as explained in \BrittoHA.
The amplitude may be expressed in
terms of bubble integrals alone. So the cut integrals will turn out
to be rational functions.

The nonvanishing cuts are $C_{34}$,
$C_{61}$, $C_{234}$ and $C_{345}$.
This amplitude is invariant under a $\QZ_2$ symmetry generated by $\b: 1\leftrightarrow 3, 4\leftrightarrow
6$.  Under this symmetry, the cuts are related through the relations $\b(C_{234})=C_{612}$ and
$\b(C_{61})=C_{34}$. Thus there are only two independent integrals, say
 $C_{34}$ and $C_{612}$.

These two cuts are given as follows. For the cut
$C_{34}$ we have
\eqn\beryl{\eqalign{
 C_{34} & =  \int d\mu~
A^{tree}(\ell_1^{\pm},5^+,6^+,1^-,2^-,
\ell_2^{\mp})A^{tree}((-\ell_2)^{\pm},3^-,4^+,(-\ell_1)^{\mp}) \cr
& =  \int d\mu \left[ {\gb{\ell_2|1+2|6}^3 \over
[6~1][1~2]\braket{\ell_2~\ell_1}\braket{\ell_1~5} P_{612}^2
\gb{5|6+1|2}} \left( -{ \gb{\ell_1|1+2|6} \over
\gb{\ell_2|1+2|6}}\right)^2 {\braket{3~\ell_1}^2 \braket{3~\ell_2}^2
\over
\braket{\ell_2~3}\braket{3~4}\braket{4~\ell_1}\braket{\ell_1~\ell_2}}\right. \cr
&  \left.+ {\gb{1|5+6|\ell_1}^3 \over
[2~\ell_2][\ell_2~\ell_1]\braket{5~6}\braket{6~1} P_{561}^2
\gb{5|6+1|2}} \left( {\gb{1|5+6|\ell_2}\over
\gb{1|5+6|\ell_1}}\right)^2 {[4~\ell_1]^2[4~\ell_2]^2 \over [\ell_2~3][3~4][4~\ell_1][\ell_1~\ell_2]}\right] \cr
&  + \left[ { \gb{\ell_1|1+2|6}^4 \over
[6~1][1~2]\braket{\ell_2~\ell_1}\braket{\ell_1~5}
P_{612}^2\gb{5|6+1|2} \gb{\ell_2|1+2|6}} \left(
{\gb{\ell_2|1+2|6}\over \gb{\ell_1|1+2|6}}\right)^2
{\braket{3~\ell_1}^2 \braket{3~\ell_2}^2 \over
\braket{\ell_2~3}\braket{3~4}\braket{4~\ell_1}\braket{\ell_1~\ell_2}}\right. \cr
&  \left. + { \gb{1|5+6|\ell_2}^4 \over
[2~\ell_2][\ell_2~\ell_1]\braket{5~6}\braket{6~1}
P_{561}^2\gb{5|6+1|2}\gb{1|5+6|\ell_1}}\left(
-{\gb{1|5+6|\ell_1}\over
\gb{1|5+6|\ell_2}}\right)^2{[4~\ell_1]^2[4~\ell_2]^2 \over
[\ell_2~3][3~4][4~\ell_1][\ell_1~\ell_2]}\right],
}}
 where the
first square bracket uses the upper choice of helicities of the cut propagators and the second, the lower.
The expression may be simplified to get
\eqn\exaonecthreefour{\eqalign{
 C_{34}  =  2\int d\mu & \left[ -{\gb{\ell_2|1+2|6}
\gb{\ell_1|1+2|6}^2\over
[6~1][1~2]\braket{\ell_2~\ell_1}\braket{\ell_1~5} P_{612}^2
\gb{5|6+1|2}} {\braket{3~\ell_1}^2 \braket{3~\ell_2} \over
\braket{3~4}\braket{4~\ell_1}\braket{\ell_1~\ell_2}}\right.  \cr
& ~~ \left. + {\gb{1|5+6|\ell_1}\gb{1|5+6|\ell_2}^2 \over
[2~\ell_2][\ell_2~\ell_1]\braket{5~6}\braket{6~1} P_{561}^2
\gb{5|6+1|2}}  {[4~\ell_1][4~\ell_2]^2 \over
[\ell_2~3][3~4][\ell_1~\ell_2]}\right].
}}

For the cut $C_{612}$ we have
\eqn\exaonecsixonetwo{\eqalign{
C_{612} & =  \int d\mu~
A^{tree}(\ell_1^{\pm},6^+,1^-,2^-,\ell_2^\mp)A^{tree}((-\ell_2)^\pm,3^-,4^+,5^+,(-\ell_1)^\mp)
 \cr
& =  \int d\mu { [\ell_1~6]^4 \over
[\ell_1~6][6~1][1~2][2~\ell_2][\ell_2~\ell_1]} \left(
{[\ell_2~6]\over [\ell_1~6]}\right)^2 { \braket{3~\ell_1}^4 \over
\braket{\ell_2~3}\braket{3~4}\braket{4~5}\braket{5~\ell_1}\braket{\ell_1~\ell_2}}
\left( {\braket{3~\ell_2}\over \braket{3~\ell_1}}\right)^2  \cr
&  + { [6~\ell_2]^4 \over
[\ell_1~6][6~1][1~2][2~\ell_2][\ell_2~\ell_1]} \left(
{[\ell_1~6]\over [\ell_2~6]}\right)^2 { \braket{3~\ell_2}^4 \over
\braket{\ell_2~3}\braket{3~4}\braket{4~5}\braket{5~\ell_1}\braket{\ell_1~\ell_2}}
\left( {\braket{3~\ell_1}\over \braket{3~\ell_2}}\right)^2 \cr
& =  2 \int d\mu { [6~\ell_2]^2 [6~\ell_1]^2 \over
[\ell_1~6][6~1][1~2][2~\ell_2][\ell_2~\ell_1]}  {
\braket{3~\ell_2}^2 \braket{3~\ell_1}^2 \over
\braket{\ell_2~3}\braket{3~4}\braket{4~5}\braket{5~\ell_1}\braket{\ell_1~\ell_2}}.
}}

\subsec{The Integration of Cut $C_{34}$}

There are two terms in \exaonecthreefour. Let us start with the
first term. We wish to
eliminate $\ell_2$ using the identities\foot{In our notation we have $P_{12}^2=\vev{1~2}[1~2]$
and $2p_a\cdot p_b=-\gb{a|b|a}$.}
$${\gb{\ell_2|1+2|6}
\over \braket{\ell_2~\ell_1}}
 = -{[\ell_1 | P_{34} P_{12} |6] \over P_{34}^2 }
~~~~~{\rm and}~~~~~
 {\braket{3~\ell_2}\over \braket{\ell_1~\ell_2}}={[4~\ell_1]\over
[4~3]}.$$
When this is done, we have
$$
C_{34}^{(1)}   =  -{2\over (P_{34}^2)^2 P_{612}^2
[6~1][1~2]\gb{5|P_{612}|2}} \int d\mu { [\ell_1 | P_{34} P_{12} |6]
\gb{\ell_1|P_{612}|6}^2\braket{3~\ell_1}^2 [4~\ell_1]\over
\braket{\ell_1~5}\braket{4~\ell_1}}.
$$
In the formula above, the measure is given by \CachazoKJ\
\eqn\measure{
 d\mu=\int_0^{+\infty}
t dt \braket{\ell_1~d{\ell_1}}[{\ell_1}~d{\ell_1}]\delta(
(P_{34}-\ell_1)^2).
}
For simplicity, we write $\ell$ instead of $\ell_1$ from now on. We also represent
 $\ket{\la}$ by $\ket{\ell}$ and $|\W\la]$ by $|\ell]$.
The cut integral is then
\eqn\newpattern{\eqalign{
C_{34}^{(1)}  = -{2\over (P_{34}^2)^2 P_{612}^2
[6~1][1~2]\gb{5|P_{612}|2}} \int_0^{+\infty} &  tdt
\braket{{\ell}~d{\ell}}[{\ell}~d{\ell}]
\delta\left( P_{34}^2+t \gb{{\ell}|P_{34}|{\ell}}\right) \cr
& \times { t^2 [\ell| P_{34} P_{12} |6]
\gb{\ell|P_{612}|6}^2\braket{3~\ell}^2 [4~\ell]\over
\braket{\ell~5}\braket{4~\ell}}
}}
 where it is essential to note that an {\sl
extra factor} of $t^2$ shows up in the second line. The reason is
that we have pulled out an overall $t$ factor when we write measure
\measure.  That is to say, we have written \refs{\CachazoKJ,
\CachazoDR}
\eqn\reason{ P_{\a \D \a}=\la_{\a}^{old} \W \la_{\D \a}^{old} \to
(\sqrt{t}\la_{\a}^{new})(\sqrt{t} \W \la_{\D \a}^{new}) }
in the measure formula given by \measure. Because of this scaling,
the extra two pairs of $\la^{old}$ and $\W \la^{old}$ in the
numerator will give an extra $t^2$ after changing to the variables
$\la^{new}, \W \la^{new}$.

Since we are working in the
dynamical region $P_{34}^2>0$, to find a nonzero result from
the delta function we must have $\gb{\ell|P_{34}|\ell}<0$.
This is important when we integrate the delta function, because
$$
\delta(a x)={1\over |a|}\delta(x).
$$

Now we can perform the $t$-integration to get
\eqn\ctfofo{
C_{34}^{(1)}  =
-{2 P_{34}^2\over  P_{612}^2 [6~1][1~2]\gb{5|P_{612}|2}}\int
\braket{{\ell}~d{\ell}}[{\ell}~d{\ell}]{  [\ell| P_{34} P_{12} |6]
\gb{\ell|P_{612}|6}^2\braket{3~\ell}^2 [4~\ell]\over
\gb{\ell|P_{34}|\ell}^4
\braket{\ell~5}\braket{4~\ell}}.
}
Notice
that \ctfofo\ is in the form of type (1) in
\fourtypes, so in principle we can already apply \geninte\ to do
the integration. But by observing the identity
$$
\braket{\ell~3}[\ell|P_{34} P_{612}|6] =
- \gb{\ell|P_{34}|\ell}\gb{3|P_{612}|6} + \gb{\ell|P_{612}|6}
\gb{3|P_{34}|\ell},
$$
we can split it into two  terms
that are much easier to integrate:
$$
C_{34}^{(1)}   =  C\int
\braket{{\ell}~d{\ell}}[{\ell}~d{\ell}] \left(
-{\gb{\ell|P_{612}|6}^2\braket{\ell~3} [4~\ell]\gb{3|P_{612}|6}\over
\gb{\ell|P_{34}|\ell}^3 \braket{\ell~5}\braket{4~\ell}}
+{\gb{\ell|P_{612}|6}^3\braket{\ell~3} [4~\ell]^2\braket{3~4}\over
\gb{\ell|P_{34}|\ell}^4 \braket{\ell~5}\braket{4~\ell}}\right),
$$
where $$C=-{2P_{34}^2\over P_{612}^2 [6~1][1~2]\gb{5|P_{612}|2}}.$$
The reason for doing so is simple: we find that by  a judicious choice
of the auxiliary spinor $\eta$ in \geninte\ we can reduce the
integration to the special case \speziell\ in which  multiple
poles have been canceled.

Using \speziell\ for the first term of $C_{34}^{(1)}$, we get
$$ C_{34}^{(1;1)}  =  -{C \over 2[3~4]}\int
\braket{{\ell}~d{\ell}}[d\ell~\partial_{\ell}]\left[
{\gb{\ell|P_{612}|6}^2 \gb{3|P_{612}|6}[4~\ell]^2\over
 \braket{\ell~5}\braket{4~\ell}}{1\over \gb{\ell|P_{34}|\ell}^2}\right].
$$
Now we can read off the  contributions at the poles. Naively, there are two
poles, $\ket{\ell} =  \ket{5}$
 and $\ket{\ell} =  \ket{4}$. However, due to the factor of
 $[4~\ell]^2$ in the numerator, the residue at the second pole is zero.
 In the end we arrive at\foot{
Here we remark on
 signs  in these calculations. There are several places
to pay attention to signs.  First, we need to write the pole in
 the right form ($\braket{\ell~a}$ or $[\ell~a]$) in order to apply the formulas of Section 2.
 Second, the contribution to the integral (as for example in the formulas \twopole\ and \threepole\ ) is the  {\sl negative} of the residue obtained by substituting the value of $\ell$ at the pole.
 However,
 to read out  the coefficient of the corresponding bubble function we need
 to put another minus sign in front of the rational part of the cut
 integration, because of the minus sign in \bubbledelta.
}
$$
 C_{34}^{(1;1)}  =  -{\braket{3~4} [4~5]^2 \gb{3|P_{612}|6} \gb{5|P_{612}|6}^2 \over
  P_{612}^2 [6~1][1~2]\braket{4~5} \gb{5|P_{612}|2} \gb{5|P_{34}|5}^2}.
$$

Similarly we can find the second term of $C_{34}^{(1)}$
and the second term of $C_{34}$. Putting it all together, we find that the
coefficient of the bubble integral $I_{2:2;3}$ is
\eqn\ithreefour{\eqalign{
 c_{2:2;3} & =
-{2\braket{2~3}^3 [3~4]^2 \gb{1|P_{561}|2}^3\over 3 [2~3]
\gb{2|P_{34}|2}^3\braket{5~6}\braket{6~1} P_{561}^2 \gb{5|P_{561}|2}
}+{\braket{2~3}^2 [3~4] \gb{1|P_{561}|2}^2\gb{1|P_{561}|4}\over
[2~3] \gb{2|P_{34}|2}^2\braket{5~6}\braket{6~1} P_{561}^2
\gb{5|P_{561}|2} }  \cr &  +{\braket{3~4}
[4~5]^2 \gb{3|P_{612}|6} \gb{5|P_{612}|6}^2 \over
  P_{612}^2 [6~1][1~2]\braket{4~5} \gb{5|P_{612}|2} \gb{5|P_{34}|5}^2}- {2\braket{3~4}^2
[4~5]^3 \gb{5|P_{612}|6}^3 \over 3 P_{612}^2
[6~1][1~2]\braket{4~5}\gb{5|P_{612}|2}\gb{5|P_{34}|5}^3}.
}}

\subsec{The Cut $C_{612}$}

For the cut $C_{612}$ given by \exaonecsixonetwo\ we perform similar
manipulations to reach
\eqn\mqpkchl{\eqalign{
C_{612}   =  { 2 P_{345}^2\over
[6~1][1~2]\braket{3~4}\braket{4~5}}\int
\braket{\ell~d\ell}[\ell~d\ell] & \left(  {\gb{3|P_{345}|6}
\gb{\ell|P_{345}|6} \braket{3~\ell}^2\over \gb{\ell|P_{345}|2}
\braket{\ell~5}} {[6~\ell]\over \gb{\ell|P_{345}|\ell}^3}\right.
\cr & ~~~~ \left. - {P_{345}^2
\gb{\ell|P_{345}|6} \braket{3~\ell}^3\over \gb{\ell|P_{345}|2}
\braket{\ell~5}}{[6~\ell]^2\over
\gb{\ell|P_{345}|\ell}^4}\right).
}}
After doing the integration we
find that  the coefficient of the bubble integral $I_{2:3;6}$ is
\eqn\isot{\eqalign{
 c_{2:3;6}& =  \left[{P_{345}^2\gb{3|P_{345}|6} \over
[6~1][1~2]\braket{3~4}\braket{4~5} \gb{5|P_{345}|2}} \left( {
\braket{3~5}^2 [5~6]^2 \over \gb{5|P_{345}|5}^2}-{
[6~1]^2\braket{1~2}^2 \gb{3|P_{345}|2}^2\over (P_{345}^2)^2
\gb{2|P_{345}|2}^2}\right) \right. \cr &   \left.+{
2(P_{345}^2)^2 \over 3 [6~1][1~2]\braket{3~4}\braket{4~5}
\gb{5|P_{345}|2}} \left( { \braket{3~5}^3 [5~6]^3 \over
\gb{5|P_{345}|5}^3}-{ [6~1]^3\braket{1~2}^3 \gb{3|P_{345}|2}^3\over
(P_{345}^2)^3 \gb{2|P_{345}|2}^3}\right)\right].
}}

We have verified that the above results satisfy the singular
behavior given by \nzerosing.

\newsec{$A(1^-,2^+,3^-,4^+,5^-,6^+)$}

In this configuration, there are one-mass and two-mass-hard box functions,
three-mass triangle functions, and bubble functions. This amplitude is invariant under a $\QZ_6$ symmetry generated by $\a: i\to i+1$ accompanied by
conjugation. Because of this,  we need to calculate just one
coefficient for each type of function and act on it by $\a$ to
obtain all the others. Representative box coefficients are given by \BidderRI\
\eqn\secconfibox{\eqalign{
c^{2m~h}_{4:2;2} &=  {2[1~2]\vev{5~6}\gb{5|P_{123}|1}^2
\gb{6|P_{123}|2}^2 P_{123}^2 \over [2~3]\vev{4~5} \gb{4|P_{123}|1}
\gb{6|P_{123}|3} \gb{6|P_{123}|1}^4};
 \cr
c^{1m}_{4;4}  &=  {2[1~2][2~3]\gb{5|P_{123}|1}^2 \gb{5|P_{123}|3}^2
\over [1~3]^4 \vev{4~5} \vev{5~6} \gb{4|P_{123}|1}
\gb{6|P_{123}|3}P_{123}^2}.
}}

We need to choose just two representative integrals, one in a three-particle channel and one in a two-particle channel.  We choose the following cuts.

$$
C_{123}   =  {2\over
[1~2][2~3]\braket{4~5}\braket{5~6}(P_{123}^2)^2}\int d\mu {
[2~\ell]^2 \gb{\ell|P_{123}|2}^2 \over [\ell~1] \gb{\ell|P_{123}|3}
}{\braket{5~\ell}^2 \gb{5|P_{123}|\ell}^2\over \gb{4|P_{123}|\ell}
\braket{6~\ell}},
$$

\eqn\iifivesix{\eqalign{
 C_{56}  & =  2\int d\mu { [\ell_1~2]^2 \braket{3~\ell_2}^2
\gb{3|P_{\ell_1 12}|2}^2 \over
[\ell_1~1][1~2]\braket{3~4}\braket{4~\ell_2} P_{\ell_1 12}^2
\gb{4|P_{\ell_1 12}|\ell_1} \gb{\ell_2|P_{\ell_1 12}|2}} {
\braket{5~\ell_1}^2 \braket{5~\ell_2}^2\over
\braket{\ell_2~5}\braket{5~6}
\braket{6~\ell_1}\braket{\ell_1~\ell_2}} \cr
&  + { \braket{\ell_1~1}^2 \braket{\ell_2~1}^2[2~4]^4 \over
\braket{\ell_2~\ell_1}\braket{\ell_1~1}[2~3][3~4]\gb{1|P_{234}|4}\gb{\ell_2|P_{234}|2}
P_{234}^2} { \braket{5~\ell_1}^2
\braket{5~\ell_2}^2\over \braket{\ell_2~5}\braket{5~6}\braket{6~\ell_1}\braket{\ell_1~\ell_2}} \cr
&  + { [4~\ell_1]^2[4~\ell_2]^2 \braket{3~1}^4 \over
\braket{1~2}\braket{2~3}[4~\ell_2][\ell_2~\ell_1] P_{123}^2
\gb{3|P_{123}|\ell_1}\gb{1|P_{123}|4}}{ \braket{5~\ell_1}^2
\braket{5~\ell_2}^2\over
\braket{\ell_2~5}\braket{5~6}\braket{6~\ell_1}\braket{\ell_1~\ell_2}}.
}}

\subsec{Cut $C_{123}$}

We start with the cut $C_{123}$ because it does not contain
three-mass triangle functions and should thus be easier to deal with.
Since we know the box coefficients, we need only to extract the rational functions giving the bubble coefficients.

After the $t$-integration we get
$$
C_{123}  =  {2
(P_{123}^2)\over [1~2][2~3]\braket{4~5}\braket{5~6}}\int
\vev{\ell~d\ell}[\ell~d\ell] {1\over \gb{\ell|P_{123}|\ell}^4}{
[2~\ell]^2 \gb{\ell|P_{123}|2}^2 \over [\ell~1] \gb{\ell|P_{123}|3}
}{\braket{5~\ell}^2 \gb{5|P_{123}|\ell}^2\over \gb{4|P_{123}|\ell}
\braket{6~\ell}}.
$$
We look for singularities in $|\ell]$ in the denominator; we find
two single poles $[\ell~1]$ and $\gb{4|P_{123}|\ell}$ and one
quadruple pole $\gb{\ell|P_{123}|\ell}$.

To demonstrate our general strategy, we give some details of the
calculation. Since there are only two single poles, we can separate
them by application of \splitone\ to find
$$
C_{123}  =  C\int \vev{\ell~d\ell}[\ell~d\ell]
{1\over \gb{\ell|P_{123}|\ell}^4}{ [2~\ell]^2 \gb{\ell|P_{123}|2}^2
\over  \gb{\ell|P_{123}|3} }{\braket{5~\ell}^2
\gb{5|P_{123}|\ell}\over  \braket{6~\ell}}  \left( {P_{123}^2
\braket{5~4}\over \gb{4|P_{123}|\ell}}+{ \gb{5|P_{123}|1} \over
[\ell~1]}\right),
$$
where
$$C={2 (P_{123}^2)\over
[1~2][2~3]\braket{4~5}\braket{5~6} \gb{4|P_{123}|1}}.$$ Next we
split the simple pole from the quadruple pole by repeating \splitone.
Finally we get
 \eqn\iiottsplit{
 C_{123} =
C_{123}^{rational}+ C_{123}^{log},}
where
\eqn\mambo{\eqalign{
C_{123}^{rational} & = -{2 (P_{123}^2)^2\over
[1~2][2~3]\braket{4~5}\braket{5~6} }\int
\vev{\ell~d\ell}[\ell~d\ell] {[2~\ell]^2\over
\gb{\ell|P_{123}|\ell}^4}{ \braket{5~\ell}^4\gb{\ell|P_{123}|2}^2
\over
\braket{\ell~4}\gb{\ell|P_{123}|1}\gb{\ell|P_{123}|3}\braket{6~\ell}
}\cr
&  + {2 (P_{123}^2)\braket{4~5}^2\over
[1~2][2~3]\braket{4~5}\braket{5~6} \gb{4|P_{123}|1}}\int
\vev{\ell~d\ell}[\ell~d\ell] {
 \gb{\ell|P_{123}|2}^3 \braket{5~\ell}^2\over
\braket{\ell~4}^2\gb{\ell|P_{123}|3}\braket{6~\ell} }{[2~\ell]\over
\gb{\ell|P_{123}|\ell}^3}\cr
&  -{2 \braket{4~5}^2\over [1~2][2~3]\braket{4~5}\braket{5~6}
\gb{4|P_{123}|1}}\int \vev{\ell~d\ell}[\ell~d\ell]{
 \gb{\ell|P_{123}|2}^3
\braket{5~\ell}^2\gb{4|P_{123}|2}\over
\braket{\ell~4}^3\gb{\ell|P_{123}|3}\braket{6~\ell} }{1\over
\gb{\ell|P_{123}|\ell}^2}\cr &   -{2
(P_{123}^2)\gb{5|P_{123}|1}^2\over
[1~2][2~3]\braket{4~5}\braket{5~6} \gb{4|P_{123}|1}}\int
\vev{\ell~d\ell}[\ell~d\ell]{
 \gb{\ell|P_{123}|2}^3 \braket{5~\ell}^2\over
\gb{\ell|P_{123}|1}^2\gb{\ell|P_{123}|3}\braket{6~\ell}
}{[2~\ell]\over \gb{\ell|P_{123}|\ell}^3}\cr
&  -{2 (P_{123}^2)\gb{5|P_{123}|1}^2[2~1]\over
[1~2][2~3]\braket{4~5}\braket{5~6} \gb{4|P_{123}|1}}\int
\vev{\ell~d\ell}[\ell~d\ell]{
 \gb{\ell|P_{123}|2}^3 \braket{5~\ell}^2\over \gb{\ell|P_{123}|1}^3\gb{\ell|P_{123}|3}\braket{6~\ell}
}{1\over \gb{\ell|P_{123}|\ell}^2}
}}
and
\eqn\asscher{\eqalign{
& C_{123}^{log}  =  \cr
& {2
\braket{4~5}^2\over [1~2][2~3]\braket{4~5}\braket{5~6}
\gb{4|P_{123}|1}}\int \vev{\ell~d\ell}[\ell~d\ell] {
 \gb{\ell|P_{123}|2}^2
\braket{5~\ell}^2\gb{4|P_{123}|2}^2\over
\braket{\ell~4}^3\gb{\ell|P_{123}|3}\braket{6~\ell} }{1\over
\gb{\ell|P_{123}|\ell}\gb{4|P_{123}|\ell}}\cr
&  +{2 (P_{123}^2)\gb{5|P_{123}|1}^2[2~1]^2\over
[1~2][2~3]\braket{4~5}\braket{5~6} \gb{4|P_{123}|1}}\int
\vev{\ell~d\ell}[\ell~d\ell]{
 \gb{\ell|P_{123}|2}^2 \braket{5~\ell}^2\over \gb{\ell|P_{123}|1}^3\gb{\ell|P_{123}|3}\braket{6~\ell}}{1\over
\gb{\ell|P_{123}|\ell} [\ell~1]}.
}}

The key feature of the above expansion is that each term in
\iiottsplit\ is in the standard form given in
\fourtypes, so we know how to deal with
each one. Furthermore, as indicated by our notation,
$C_{123}^{rational}$ gives only a rational contribution while
$C_{123}^{log}$ contributes only to pure logarithmic terms (see
Appendix B for a discussion of the latter), from which we can read the
coefficients of corresponding one-mass and two-mass hard box
functions. Since we have calculated these box functions by quadruple
cuts, it can serve as an independent check of our method. However, we
will not do it here.

Now we focus on the rational part, $C_{123}^{rational}$. If we use \speziell\
for the first term we find that we get only a single pole, easily dealt with.
 The second and fourth terms have double poles which are
new to us, so we will use them to demonstrate our general strategy
as laid out in Section 2. However, before we proceed to a detailed
treatment of double poles, we want to remark upon the third and
fifth terms. Naively we have triple poles, but recall that in the
relation
$$
[\ell~d\ell]{1\over
\gb{\ell|P_{123}|\ell}^2}=[d\ell~\partial_\ell]{ [\eta~\ell]\over
\gb{\ell|P_{123}|\ell}\gb{\ell|P_{123}|\eta}}
$$
we have some
freedom in the choice of $|\eta]$. If we choose $|\eta]=|4]$ for the
third term, the numerator factor $[4~\ell]$ will make the
contribution from triple pole ${1\over \vev{\ell~4}^3}$ to be zero.
A similar manipulation can be done in fifth term by choosing
$|\eta]=|P_{123}|1\rangle$.

\subsubsec{Double Pole: The Second Term of $C_{123}^{rational}$}

We can write the second term of $C_{123}^{rational}$ as
$$
C_{123}^{rational;2}= C\int
\vev{\ell~d\ell}[d\ell~\partial_\ell]\left( {
 \gb{\ell|P_{123}|2}^3 \braket{5~\ell}^2\over
\braket{\ell~4}^2\gb{\ell|P_{123}|3}\braket{6~\ell}
}{[2~\ell]^2\over
\gb{\ell|P_{123}|\ell}^2\gb{\ell|P_{123}|2}}\right)
$$
 with
$$C={(P_{123}^2)\braket{4~5}\over
[1~2][2~3]\braket{5~6} \gb{4|P_{123}|1}}.$$
There are
several single poles but we will be concerned only with the double pole
$\braket{\ell~4}$. To read out the residue at this double pole, we write
the integrand as
$$
 -C[2~4]^2 {1\over \vev{\ell~4}^2}
{\gb{\ell|P_{123}|2}^3 \braket{\ell~5}^2\over
\gb{\ell|P_{123}|3}\braket{\ell~6}\gb{\ell|P_{123}|4}^2\gb{\ell|P_{123}|2}},$$
where we have made the replacement $|\ell]=|4]$. Now it is in our standard
form for a double pole, so we can use \rectwo\ and
\twopole\ to write down the answer as
$$
 -C[2~4]^2
P_2[\ket{4}, L_1^{(II;C_{123})}, L_2^{(II;C_{123})}],
$$
 where we will make use of the following lists:
\eqn\listiiott{\eqalign{
L_1^{(II;C_{123})} & =  \{ \ket{5},\ket{5}, |P_{123}|2\rangle,
|P_{123}|2\rangle\},
\cr L_2^{(II;C_{123})} & =  \{\ket{6}, |P_{123}|3\rangle, |P_{123}|4\rangle, |P_{123}|4\rangle  \}
\cr L_3^{(II;C_{123})} & =  \{\ket{6}, |P_{123}|3\rangle,
\ket{1},\ket{1}  \}.
}}

\subsubsec{The Coefficient of Bubble $I_{2:3;1}$}

Combining all these results, we find that
the coefficient of the bubble integral $I_{2:3;1}$ is given by
\eqn\iiottcoef{\eqalign{ & c_{2:3;1}  = \cr
& {1\over
[1~2][2~3]\braket{4~5}\braket{5~6}}\left( -{2 (P_{123}^2)^2\over 3
}\sum_{i=1}^4\lim_{\ket{\ell}\to \ket{\ell_i}}
\vev{\ell~\ell_i}\left[ {[2~\ell]^3\over \gb{\ell|P_{123}|\ell}^3 }{
\braket{5~\ell}^4\gb{\ell|P_{123}|2} \over
\braket{\ell~4}\gb{\ell|P_{123}|1}\gb{\ell|P_{123}|3}\braket{6~\ell}
}\right] \right. \cr &  +{2 \braket{4~5}^2\over
\gb{4|P_{123}|1}}\sum_{i=1,2,6}\lim_{\ket{\ell}\to \ket{\ell_i}}
\vev{\ell~\ell_i}\left[-{
 \gb{\ell|P_{123}|2}^3
\braket{5~\ell}^2\gb{4|P_{123}|2}\over
\braket{\ell~4}^3\gb{\ell|P_{123}|3}\braket{6~\ell} }{[4~\ell]\over
\gb{\ell|P_{123}|\ell}\gb{\ell|P_{123}|4}}\right] \cr
&   -{2 (P_{123}^2)\gb{5|P_{123}|1}^2[2~1]\over
\gb{4|P_{123}|1}}\sum_{i=1,2,5}\lim_{\ket{\ell}\to \ket{\ell_i}}
\vev{\ell~\ell_i}\left[{
 \gb{\ell|P_{123}|2}^3 \braket{5~\ell}^2\over \gb{\ell|P_{123}|1}^3\gb{\ell|P_{123}|3}\braket{6~\ell}
}{\gb{1|P_{123}|\ell}\over \gb{\ell|P_{123}|\ell} P_{123}^2
\braket{\ell~1}}\right] \cr
& + \sum_{i=1,2} \lim_{\ket{\ell}\to \ket{\ell_i}}
\vev{\ell~\ell_i}\left[ { (P_{123}^2)\braket{4~5}^2\over
 \gb{4|P_{123}|1}} {
 \gb{\ell|P_{123}|2}^2 \braket{5~\ell}^2\over
\braket{\ell~4}^2\gb{\ell|P_{123}|3}\braket{6~\ell}
}{[2~\ell]^2\over \gb{\ell|P_{123}|\ell}^2 }\right. \cr &
\left.-{ (P_{123}^2)\gb{5|P_{123}|1}^2\over \gb{4|P_{123}|1}}{
 \gb{\ell|P_{123}|2}^2 \braket{5~\ell}^2\over
\gb{\ell|P_{123}|1}^2\gb{\ell|P_{123}|3}\braket{6~\ell}
}{[2~\ell]^2\over \gb{\ell|P_{123}|\ell}^2}\right]\cr &
+{ (P_{123}^2)\braket{4~5}^2[2~4]^2\over
 \gb{4|P_{123}|1}}
P_2[\ket{4},L_1^{(II;C_{123})},L_2^{(II;C_{123})}] \cr &
\left. -{ \gb{5|P_{123}|1}^2\vev{1~3}^2[2~3]^2\over
 \gb{4|P_{123}|1}(P_{123}^2)}
P_2[|P_{123}|1],L_1^{(II;C_{123})},
L_3^{(II;C_{123})}]\right),
}}
 where the list
$L_i^{(II;C_{123})}$ has been given in \listiiott, while the
various poles are given by
$$
\ket{\ell_1}=|P_{123}|3],~~\ket{\ell_2}=\ket{6},~~\ket{\ell_3}=\ket{4},~~\ket{\ell_4}=|P_{123}|1],
~~\ket{\ell_5}=\ket{1},~~\ket{\ell_6}=|P_{123}|4].
$$

Let us give a brief explanation for the result \iiottcoef. The first
three lines give  single pole contributions of the first, the third
and the fifth terms in \mambo. The fourth and fifth lines give
 single pole contributions of the second  and the fourth terms in \mambo.
 The last two lines are double pole contributions of the second  and the fourth terms in \mambo.
\subsec{Cut $C_{56}$}

Now we consider to the two-particle channel where the cut integral is given by \iifivesix. Of the three terms,
only the first one will contribute to the three-mass
triangle, so we leave it for last.

\subsubsec{The second term}

For the second term in \iifivesix\ we achieve the following separation by
our standard splitting process:
\eqn\rarify{\eqalign{
C_{56}^{(2r)}  & =  {2[2~4]^4
P_{56}^2\braket{5~6}\over [2~3][3~4]\gb{1|P_{234}|4}P_{234}^2}\int
\braket{\ell~d\ell}[\ell~d\ell]
{\braket{\ell~5}\braket{\ell~1}^3[6~\ell]^2 \over
\gb{\ell|P_{234}|2}\braket{\ell~6}\gb{\ell|P_{56}|\ell}^4}\cr
&  - {2[2~4]^4 P_{56}^2\braket{6~1}\over
[2~3][3~4]\gb{1|P_{234}|4}P_{234}^2}\int
\braket{\ell~d\ell}[\ell~d\ell]
{\braket{\ell~5}^2\braket{\ell~1}^2[6~\ell] \over
\gb{\ell|P_{234}|2}\braket{\ell~6}^2\gb{\ell|P_{56}|\ell}^3}\cr
&  - {2[2~4]^4 P_{56}^2\braket{6~1}\over
[2~3][3~4]\gb{1|P_{234}|4}P_{234}^2}\int
\braket{\ell~d\ell}[\ell~d\ell] {\braket{\ell~5}^2\braket{\ell~1}^2
\over
\gb{\ell|P_{234}|2}\braket{\ell~6}^3\gb{\ell|P_{56}|\ell}^2},
}}
and
$$
C_{56}^{(2l)}  =  {2[2~4]^4 P_{56}^2\braket{6~1}\over
[2~3][3~4]\gb{1|P_{234}|4}P_{234}^2}\int
\braket{\ell~d\ell}[\ell~d\ell] {\braket{\ell~5}\braket{\ell~1}^2
\over
\gb{\ell|P_{234}|2}\braket{\ell~6}^3[5~\ell]\gb{\ell|P_{56}|\ell}},
$$
 where $r,l$ indicate rational and logarithmic parts, respectively.

It can be shown that in this case, with a judicious choice of the
auxiliary spinor $\eta$, we can get rid of all multiple poles and
are left with only single pole contributions. The contribution to
the coefficient of the bubble $I_{2:2;5}$ is given by
\eqn\iifstwor{\eqalign{ c_{2:2;5}^{(2)}& = {2[2~4]^4
\braket{5~6}\braket{6~1}\over [2~3][3~4]\gb{1|P_{234}|4}P_{234}^2}
{\gb{5|P_{234}|2}\gb{1|P_{234}|2}^2 \gb{2|P_{234}|6} \over
\gb{6|P_{234}|2}^3 \gb{2|P_{234}P_{56}P_{234}|2}}   \cr &  +{[2~4]^4
\braket{5~6}\braket{6~1}\over [2~3][3~4]\gb{1|P_{234}|4}P_{234}^2} {
 \gb{1|P_{234}|2}^2 \gb{2|P_{234}|6}^2 \gb{5|P_{234}|2} \over
\gb{6|P_{234}|2}^2 \gb{2|P_{234}P_{56}P_{234}|2}^2} \cr
&   -{2[2~4]^4 \braket{5~6}^2\gb{1|P_{234}|2}^3
\gb{2|P_{234}|6}^3 \over
3[2~3][3~4]\gb{1|P_{234}|4}P_{234}^2\gb{6|P_{234}|2}
\gb{2|P_{234}P_{56}P_{234}|2}^3}.
}}

\subsubsec{The third term}

Now we split the third term of \iifivesix\ into
\eqn\lapiz{\eqalign{
C_{56}^{(3r)}
& =  {2\braket{3~1}^4 P_{56}^2[5~6]\over
\braket{1~2}\braket{2~3}P_{123}^2\gb{1|P_{123}|4}}\int
\braket{\ell~d\ell}[\ell~d\ell]{ [4~\ell]^3 [6~\ell]
\braket{\ell~5}^2\over \gb{3|P_{123}|\ell}
[5~\ell]\gb{\ell|P_{56}|\ell}^4} \cr &  -{2\braket{3~1}^4
P_{56}^2[4~5]\over
\braket{1~2}\braket{2~3}P_{123}^2\gb{1|P_{123}|4}}\int
\braket{\ell~d\ell}[\ell~d\ell]{ [4~\ell]^2 [6~\ell]^2
\braket{\ell~5}\over \gb{3|P_{123}|\ell}
[5~\ell]^2\gb{\ell|P_{56}|\ell}^3} \cr &  -{2\braket{3~1}^4
P_{56}^2[4~5]\over
\braket{1~2}\braket{2~3}P_{123}^2\gb{1|P_{123}|4}}\int
\braket{\ell~d\ell}[\ell~d\ell]{ [4~\ell]^2 [6~\ell]^2 \over
\gb{3|P_{123}|\ell} [5~\ell]^3\gb{\ell|P_{56}|\ell}^2}
}}
and
$$ C_{56}^{(3l)}  =  {2\braket{3~1}^4 P_{56}^2[4~5]\over
\braket{1~2}\braket{2~3}P_{123}^2\gb{1|P_{123}|4}}\int
\braket{\ell~d\ell}[\ell~d\ell]{ [4~\ell]^2 [6~\ell] \over
\gb{3|P_{123}|\ell}
[5~\ell]^3\braket{\ell~6}\gb{\ell|P_{56}|\ell}}.
$$

From here we read out the bubble coefficient part as
\eqn\iifsthreer{\eqalign{ c_{2:2;5}^{(3)}& =  -{2\braket{1~3}^4
[5~6]^2 \gb{3|P_{123}|4}^3 \gb{5|P_{123}|3}^3 \over
3\braket{1~2}\braket{2~3}P_{123}^2 \gb{1|P_{123}|4}\gb{3|P_{123}|5}
\gb{3|P_{123}P_{56}P_{123}|3}^3} \cr &  +{\braket{1~3}^4
[4~5][5~6]\gb{3|P_{123}|4}^2 \gb{3|P_{123}|6} \gb{5|P_{123}|3}^2
\over \braket{1~2}\braket{2~3}P_{123}^2\gb{1|P_{123}|4}
\gb{3|P_{123}|5}^2 \gb{3|P_{123}P_{56}P_{123}|3}^2}
 \cr
&  +{2\braket{1~3}^4 [4~5][5~6]\gb{3|P_{123}|4}^2 \gb{3|P_{123}|6}
\gb{5|P_{123}|3} \over
\braket{1~2}\braket{2~3}P_{123}^2\gb{1|P_{123}|4} \gb{3|P_{123}|5}^3
\gb{3|P_{123}P_{56}P_{123}|3}}.
}}

\subsubsec{The first term}

Now we move to the first term in \iifivesix. After doing the
$t$-integration and setting $\ell=\ell_1$, we get
\eqn\baby{\eqalign{
C_{56}^{(1)}
& =  -{2 \over
[1~2]\braket{3~4} P_{56}^2} \int
\braket{\ell~d\ell}[\ell~d\ell]{ (P_{56}^2)^2\over
\gb{\ell|P_{56}|\ell}^3}{ [\ell~2]^2 \gb{3|P_{56}|\ell}^2
\braket{5~\ell}^2 [6~\ell]\over [\ell~1] \gb{4|P_{ 12}|\ell}
\gb{4|P_{56}|\ell}[2|P_{5612}P_{56}|\ell]\braket{6~\ell}}
\cr
& \times { \gb{3|P_{12}|2}^2 -2 {P_{56}^2\over
\gb{\ell|P_{56}|\ell}}\gb{3|P_{12}|2}\braket{3~\ell}[\ell~2]+{(P_{56}^2)^2\over
\gb{\ell|P_{56}|\ell}^2}\braket{3~\ell}^2[\ell~2]^2\over
P_{12}^2 +{P_{56}^2\gb{\ell|P_{12}|\ell}\over
\gb{\ell|P_{56}|\ell}}}.
}}

Defining
$$ Q={P_{12}+
{P_{12}^2\over P_{56}^2} P_{56}},
$$
 we get
\eqn\aphid{\eqalign{
C_{56}^{(1)} & =  -{2 \over [1~2]\braket{3~4} } \int
\braket{\ell~d\ell}[\ell~d\ell]{1 \over
\gb{\ell|P_{56}|\ell}^3}{ [\ell~2]^2 \gb{3|P_{56}|\ell}^2
\braket{5~\ell}^2 [6~\ell]\over [\ell~1] \gb{4|P_{ 12}|\ell}
\gb{4|P_{56}|\ell}[2|P_{5612}P_{56}|\ell]\braket{6~\ell}}
\cr &  \left( { \gb{3|P_{12}|2}^2\gb{\ell|P_{56}|\ell}\over
\gb{\ell|Q|\ell}} - {2P_{56}^2
\gb{3|P_{12}|2}\braket{3~\ell}[\ell~2]\over
\gb{\ell|Q|\ell}}+{(P_{56}^2)^2\braket{3~\ell}^2[\ell~2]^2\over
\gb{\ell|P_{56}|\ell}\gb{\ell|Q|\ell}}\right).
}}
 To simplify the calculation further we define
\eqn\habi{
 g(\W\ell) \equiv -{[\ell~6][\ell~2]^2
\tgb{\ell|P_{56}|3}^2 \over [\ell~1]
\tgb{\ell|P_{12}|4}\tgb{\ell|P_{56}|4}
[\ell|P_{56}P_{34}|2]}.
}
The tilde in $g(\W\ell)$ indicates that this function is antiholomorphic.

Now we can use our standard splitting method to split each of them
to reach the form given in \fourtypes. The
result is
$$
 C_{56}^{(1)}=
C_{56}^{(1r)}+C_{56}^{(1l)}+C_{56}^{(1;3m)},
$$
 where $1r,1l,3m$ respectively indicate
rational contributions, logarithmic contributions for box
functions and logarithmic contributions for three-mass triangle
functions. Since we do not compute box coefficients from double
cuts, we do not record them here, but give only the other two parts as:

\eqn\iifsoner{\eqalign{
 C_{56}^{(1r)} & =  -{2 (P_{56}^2)^2 \over [1~2]\braket{3~4} }
\int \braket{\ell~d\ell}[\ell~d\ell] {g(\W
\ell)[\ell~2]^2\gb{3|P_{56}|\ell}^2\over
\gb{6|P_{56}|\ell}[\ell|P_{56}Q|\ell]}{ \braket{5~\ell}^2\over
\gb{\ell|P_{56}|\ell}^4} \cr
&  + {2 (P_{56}^2)^2 \over
[1~2]\braket{3~4} } \int \braket{\ell~d\ell}[\ell~d\ell]{g(\W
\ell)[\ell~2]^2\braket{3~6}^2\gb{5|P_{56}|\ell}\over \gb{6|Q|\ell}
\gb{6|P_{56}|\ell}^2}{ \braket{5~\ell}\over \gb{\ell|P_{56}|\ell}^3}
\cr
&
 -  {2 (P_{56}^2)^2 \over [1~2]\braket{3~4} } \int
\braket{\ell~d\ell}[\ell~d\ell] {g(\W
\ell)[\ell~2]^2\gb{3|Q|\ell}^2\gb{5|P_{56}|\ell}\over \gb{6|Q|\ell}
[\ell|P_{56}Q|\ell]^2}{\braket{5~\ell}\over
\gb{\ell|P_{56}|\ell}^3}\cr
&  + {4 P_{56}^2
\braket{1~3}\over \braket{3~4} } \int
\braket{\ell~d\ell}[\ell~d\ell] {g(\W
\ell)[\ell~2][6~\ell]\gb{3|P_{56}|\ell}\over
[5~\ell][\ell|P_{56}Q|\ell]}{\braket{5~\ell}\over
\gb{\ell|P_{56}|\ell}^3}\cr
& +{4 P_{56}^2
\braket{1~3}\over \braket{3~4} } \int
\braket{\ell~d\ell}[\ell~d\ell]{g(\W \ell)[\ell~2][6~\ell]^2
\braket{6~5}\gb{3|Q|\ell}\over [\ell~5][\ell|P_{56}
Q|\ell]^2}{1\over \gb{\ell|P_{56}|\ell}^2}
\cr
&  -{4
P_{56}^2 \braket{1~3}\over \braket{3~4} } \int
\braket{\ell~d\ell}[\ell~d\ell]{g(\W \ell)[\ell~2][6~\ell]
\gb{3|P_{56}|\ell}\over [\ell~5]^2[\ell|P_{56} Q|\ell]}{1\over
\gb{\ell|P_{56}|\ell}^2} \cr
&  - {2 \braket{1~3}^2[1~2]\over \braket{3~4} } \int
\braket{\ell~d\ell}[\ell~d\ell] g(\W \ell) {g(\W
\ell)\braket{5~6}[6~\ell]^2\over  [5~\ell][\ell|P_{56} Q|\ell]}
{1\over \gb{\ell|P_{56}|\ell}^2}\cr
&  + {2 (P_{56}^2)^2
\over [1~2]\braket{3~4} } \int \braket{\ell~d\ell}[\ell~d\ell] {g(\W
\ell)[\ell~2]^2\braket{3~6}^2\gb{5|P_{56}|\ell}\braket{5~6}\over
\gb{6|Q|\ell} \gb{6|P_{56}|\ell}^3}{ 1\over
\gb{\ell|P_{56}|\ell}^2}\cr
&  - {2 (P_{56}^2)^2  \over
[1~2]\braket{3~4} } \int \braket{\ell~d\ell}[\ell~d\ell] {g(\W
\ell)[\ell~2]^2\gb{3|Q|\ell}^2\gb{5|P_{56}|\ell}\gb{5|Q|\ell}\over
\gb{6|Q|\ell} [\ell|P_{56}Q|\ell]^3}{1\over
\gb{\ell|P_{56}|\ell}^2},
}}

\eqn\iifsonethreem{\eqalign{
 C_{56}^{(1;3m)} & =  -{2 \braket{1~3}^2[1~2]\over \braket{3~4}
} \int \braket{\ell~d\ell}[\ell~d\ell] g(\W
\ell){\gb{5|Q|\ell}^2\over \gb{6|Q|\ell} [\ell|P_{56} P_{12}|\ell]}
{1\over \gb{\ell|P_{56}|\ell}\gb{\ell|Q|\ell}}\cr
&  +{2
(P_{56}^2)^2  \over [1~2]\braket{3~4} } \int
\braket{\ell~d\ell}[\ell~d\ell] g(\W \ell)[\ell~2]^2
{\gb{3|Q|\ell}^2\gb{5|Q|\ell}^2\over \gb{6|Q|\ell} [\ell|P_{56}
P_{12}|\ell]^3}{1\over
\gb{\ell|P_{56}|\ell}\gb{\ell|Q|\ell}}\cr
&  + {4 P_{56}^2 \braket{1~3}\over \braket{3~4} } \int
\braket{\ell~d\ell}[\ell~d\ell] g(\W
\ell)[\ell~2]{\gb{3|Q|\ell}\gb{5|Q|\ell}^2\over
\gb{6|Q|\ell}[\ell|P_{56} P_{12}|\ell]^2}{1\over
\gb{\ell|P_{56}|\ell}\gb{\ell|Q|\ell}}.
}}

Now we will discuss them one by one. But before doing that we need
to know how to deal with poles from $[\ell|P_{56} Q|\ell]$.

\bigskip

{\bf Dealing  with $[\ell|P_{56}Q|\ell]$:}
\bigskip

Now we need to deal with the factor $[\ell|P_{56}Q|\ell]$ in the denominator.  We expand it as
$$
|\ell]= |a] + x |b]
$$
where $a,b$ are two arbitrary massless spinors.
With this substitution we find
\eqn\jolish{\eqalign{
0 & =   [a|P_{56} Q|a]+ x([a|P_{56} Q|b]+[b|P_{56}
Q|a]) + x^2[b|P_{56} Q|b] \cr
x_{\pm} & =  { -([a|P_{56} Q|b]+[b|P_{56} Q|a])\pm
[a~b]\sqrt{\Delta_{3m}}\over 2[b|P_{56} Q|b]},\cr
\Delta_{3m} & =  (P_{12}^2)^2+ (P_{34}^2)^2+ (P_{56}^2)^2- 2
P_{12}^2 P_{34}^2 -2 P_{12}^2 P_{56}^2-2 P_{34}^2 P_{56}^2
}}
 Now
we can write the factor as
\eqn\hobb{\eqalign{
 [\ell~\eta_1][\ell~\eta_2] & =
[\ell~a]^2 + (x_+ + x_-) [\ell~a][\ell~b] + x_ + x_- [\ell~b]^2 \cr
& =  { [a~b]^2 [\ell|P_{56} Q|\ell] \over [b|P_{56}Q|b]}
}}
where $|\eta_1]=|a] + x_ + |b]$ and $|\eta_2]= |a] + x_ - |b]$.
In other words, we write
\eqn\newexp{[\ell|P_{56} Q|\ell] = [\ell~\eta_1][\ell~\eta_2]
{[b|P_{56}Q|b]\over [a~b]^2},}
 where
$$
 ~~|\eta_1]=|a]+x_+
|b],~~~ |\eta_2]=|a]+ x_ - |b],~~ x_{\pm}  =  { -([a|P_{56}
Q|b]+[b|P_{56} Q|a])\pm [a~b]\sqrt{\Delta_{3m}}\over 2[b|P_{56}
Q|b]}.
$$

{\bf The Rational Part:}

\bigskip

Now we discuss the rational part, $C_{56}^{(1r)}$, given in \iifsoner.
Upon a few moments' study, we see that among these nine terms,
only the ninth one has a triple pole and only the third and the fifth
have double poles. All of the  remaining six terms have only
single poles. For example, although the second term has the pole ${1\over
\braket{6~5}^2[5~\ell]^2}$, the factor of $\braket{5~\ell}$ in
the numerator sets its contribution to zero. Also, the first term in
\iifsoner\ has a factor of ${1\over [6~\ell]}$, but $g(\W\ell)$ has
the same factor in the numerator, so it is not a pole.

To read out the rational contributions we define the following three
functions:

\eqn\aone{\eqalign{
 A_1 & =  {2 (P_{56}^2)^2 [a~b]^2\over 3 [1~2]\braket{3~4}
\vev{5~6}^2[b|P_{56} Q|b]}  {g(\W
\ell)[\ell~2]^2\gb{3|P_{56}|\ell}^2 \vev{5~\ell}^3\over
[5~\ell][\ell~\eta_1][\ell~\eta_2][6~\ell]\gb{\ell|P_{56}|\ell}^3}
\cr
&   + { [5~6]^2 \over [1~2]\braket{3~4} } {g(\W
\ell)[\ell~2]^2\braket{3~6}^2\vev{5~\ell}^2\over \gb{6|Q|\ell}
[5~\ell]^2 \gb{\ell|P_{56}|\ell}^2}\cr
&  + {2 P_{56}^2
\braket{1~3}[a~b]^2\over \braket{3~4} \vev{5~6}[b|P_{56} Q|b] }
{g(\W \ell)[\ell~2]\gb{3|P_{56}|\ell}\vev{5~\ell}^2\over
[5~\ell][\ell~\eta_1][\ell~\eta_2]\gb{\ell|P_{56}|\ell}^2}
\cr
&  -{4 P_{56}^2 \braket{1~3}[a~b]^2\over \braket{3~4}
\vev{5~6}[b|P_{56} Q|b]} {g(\W \ell)[\ell~2]
\gb{3|P_{56}|\ell}\vev{5~\ell}\over
[\ell~5]^2[\ell~\eta_1][\ell~\eta_2]\gb{\ell|P_{56}|\ell}}\cr
&  - {2 \braket{1~3}^2[1~2][a~b]^2\over \braket{3~4} [b|P_{56}
Q|b]} {g(\W \ell)[6~\ell]\vev{5~\ell}\over
[5~\ell][\ell~\eta_1][\ell~\eta_2]\gb{\ell|P_{56}|\ell}}
\cr
&  -{2 (P_{56}^2)^2 \braket{3~6}^2 \over [1~2]\braket{3~4}
\vev{5~6}^2} {g(\W \ell)[\ell~2]^2\vev{5~\ell}\over \gb{6|Q|\ell}
[5~\ell]^3\gb{\ell|P_{56}|\ell}}
}}

\eqn\atwo{\eqalign{
A_2 & =  - { (P_{56}^2)^2 [a~b]^4\over
[1~2]\braket{3~4}[b|P_{56}Q|b]^2 } {g(\W
\ell)[\ell~2]^2\gb{3|Q|\ell}^2\braket{5~\ell}^2\over \gb{6|Q|\ell}
 [\ell~\eta_1]^2[\ell~\eta_2]^2\gb{\ell|P_{56}|\ell}^2 }\cr
 &  -{4 P_{56}^2 \braket{1~3}[a~b]^4\over
\braket{3~4} [b|P_{56}Q|b]^2} {g(\W \ell)[\ell~2][6~\ell]
\gb{3|Q|\ell}\vev{5~\ell}\over [\ell~5]
[\ell~\eta_1]^2[\ell~\eta_2]^2\gb{\ell|P_{56}|\ell}}
}}

\eqn\athree{
A_3 =  - {2 (P_{56}^2)^2 [a~b]^6 \over
[1~2]\braket{3~4}[b|P_{56}Q|b]^3 } {g(\W
\ell)[\ell~2]^2\gb{3|Q|\ell}^2\gb{5|P_{56}|\ell}\gb{5|Q|\ell}\vev{\eta_1~\ell}\over
\gb{6|Q|\ell}\gb{\eta_1|P_{56}|\ell} [\ell~\eta_1]^3[\ell~\eta_2]^3
\gb{\ell|P_{56}|\ell}}
}

The function $A_1$ is the collection of the first, second, fourth,
sixth, seventh and eighth terms after writing them in the form of
$\braket{d\ell~\partial_\ell}(\bullet)$. Similarly, $A_2$ is the
collection of the third and fifth terms, while $A_3$ is the ninth
term,  after writing them in the form
$\braket{d\ell~\partial_\ell}(\bullet)$. Unlike in $A_1$, we have
double poles in $A_2$ and triple poles in $A_3$, so we need to
separate these contributions. One feature $A_3$ is that we have
chosen the auxiliary spinor carefully to cancel one of the two triple
poles (notice the factor $\vev{\eta_1~\ell}$ in numerator).

First, $A_1$ gives following contribution to the bubble coefficient:
\eqn\iifsoneaone{
c_{2:2;5}^{1r;A1}=\sum_{i=1}^7 \lim_{\ell\to \ell_i}
([\ell~\ell_i]A_1)
}
 where
$$|\ell_1] =
|1],~~|\ell_2]=|P_{12}|4\rangle,~~|\ell_3]=|P_{56}|4\rangle,~~|\ell_4]=|P_{56}P_{5612}|2],~~
|\ell_5]=|Q|6\rangle, ~~|\ell_6] =
|\eta_1],~~|\ell_7]=|\eta_2]
$$

$A_2$ gives following single pole contribution
\eqn\iifsoneatwoone{
c_{2:2;5}^{1r;A2-1}=\sum_{i=1}^5 \lim_{\ell\to \ell_i}
([\ell~\ell_i]A_2)
}
and double pole
contribution
\eqn\iifsoneatwotwo{\eqalign{
 &  c_{2:2;5}^{1r;A2-2}  = \cr
& { (P_{56}^2)^2 [a~b]^4
\vev{5~\eta_1}^2\over [1~2]\braket{3~4}[b|P_{56}Q|b]^2 }
P_2[|\eta_1],L_1,L_2] - {4 P_{56}^2
\braket{1~3}[a~b]^4\vev{5~\eta_1}\over \braket{3~4}
[b|P_{56}Q|b]^2}P_2[|\eta_1],L_3,L_4]
 + \{ |\eta_1] \leftrightarrow |\eta_2]
\}
}}
 where these lists are
\eqn\iifslistone{\eqalign{
L_1 &=
\{|6],
|2],|2], |P_{56}|3\rangle, |P_{56}|3\rangle, |2],|2], |Q|3\rangle, |Q|3\rangle \} \cr
L_2 & =  \{ |1], |P_{12}|4\rangle, |P_{56}|4\rangle,
|P_{56}P_{34}|2], |Q|6\rangle, |\eta_2],|\eta_2],
|P_{56}|\eta_1\rangle, |P_{56}|\eta_1\rangle\}\cr
 L_3 &=
\{|6], |2],|2], |P_{56}|3\rangle, |P_{56}|3\rangle,
|2],|6], |Q|3\rangle \} \cr
L_4 & =  \{ |1], |P_{12}|4\rangle, |P_{56}|4\rangle,
|P_{56}P_{34}|2], |5], |\eta_2],|\eta_2],
|P_{56}|\eta_1\rangle\}
}}
 and the function
$P_2$ is given in \twopole.

For $A_3$, the single pole contribution is
\eqn\iifsoneathreeone{
c_{2:2;5}^{1r;A3-1}=\sum_{i=1}^5 \lim_{\ell\to \ell_i}
([\ell~\ell_i]A_3)+\lim_{\ell\to |P_{56}|\eta_1\rangle}
(\tgb{\ell|P_{56}|\eta_1}A_3),
}
 and the
triple pole contribution is
\eqn\iifsoneathreetwo{
c_{2:2;5}^{1r;A3-2}=- {2
(P_{56}^2)^2 [a~b]^6 \vev{5~6} \vev{\eta_1~\eta_2}\over
[1~2]\braket{3~4}[b|P_{56}Q|b]^3 } P_3[|\eta_2],
L_5,L_6]
}
 with
\eqn\iifslisttwo{\eqalign{
 L_5 &=   \{|6],
|2],|2], |P_{56}|3\rangle, |P_{56}|3\rangle, |2],|2], |Q|3\rangle,
|Q|3\rangle, |6], |Q|5\rangle \} \cr
L_6 & =  \{ |1], |P_{12}|4\rangle, |P_{56}|4\rangle,
|P_{56}P_{34}|2], |Q|6|\rangle, |\eta_1],|\eta_1],|\eta_1],
|P_{56}|\eta_1\rangle,
|P_{56}|\eta_1\rangle\},
}}
 and the function
$P_3$ is given in \threepole.

Putting it all together, we find that the coefficient $c_{2:2;5}$ is given
by the sum of \iifstwor, \iifsthreer, \iifsoneaone, \iifsoneatwoone,
\iifsoneatwotwo, \iifsoneathreeone\ and \iifsoneathreetwo.

\bigskip

{\bf The coefficient of triangle $I_{3:2,2;1}$:}

\bigskip

To read off the coefficient of three-mass triangle function
$I_{3:2,2;1}$ from \iifsonethreem, we need to be careful about
different poles. As discussed carefully in Appendix B, poles from
the factor $[\ell|P_{56} Q|\ell]$ are special. In this example, the
poles from factors other than $[\ell|P_{56} Q|\ell]$ are all single
poles, and their contributions are given by
\eqn\iifsthreemone{\eqalign{
 c_{3:2,2;1}^{(1)} & =  \sum_{i=1}^6
\lim_{|\ell]\to |\ell_i]} [\ell~\ell_i] {g(\W\ell)\over
[\ell~\W\eta]}
R_1[\ell, \W \eta, P_{56}, Q]
\left[ -{2 \braket{1~3}^2[1~2]\over
\braket{3~4} } {\gb{5|Q|\ell}^2\over \gb{6|Q|\ell} [\ell|P_{56}
P_{12}|\ell]} \right.\cr &  \left.+{2 (P_{56}^2)^2  \over
[1~2]\braket{3~4} } {[\ell~2]^2\gb{3|Q|\ell}^2\gb{5|Q|\ell}^2\over
\gb{6|Q|\ell} [\ell|P_{56} P_{12}|\ell]^3} + {4 P_{56}^2
\braket{1~3}\over \braket{3~4} }
{[\ell~2]\gb{3|Q|\ell}\gb{5|Q|\ell}^2\over \gb{6|Q|\ell}[\ell|P_{56}
P_{12}|\ell]^2}\right]
}}
with the following
poles:
\eqn\selerie{\eqalign{
 & |\ell_1] =
|1],~~|\ell_2]=|P_{12}|4\rangle,~~|\ell_3]=|P_{56}|4\rangle,~~|\ell_4]=|P_{56}P_{5612}|2],~~
|\ell_5]=|Q|6\rangle,~~|\ell_6]=|\W\eta]
\cr & |\ell_7] =
|\eta_1],~~|\ell_8]=|\eta_2]
}}

Now consider the contributions from the poles from $[\ell|P_{56}
Q|\ell]$. The first term in \iifsonethreem, is a single pole, so

\eqn\iifsthreemtwo{
 c_{3:2,2;1}^{(2)}  = -{2 \braket{1~3}^2[1~2]\over
\braket{3~4} }{[a~b]^2\over [b|P_{56} Q|b]}\left( {g(\W\ell_7)\W
R_2[\ell_7,\W\eta, P_{56}]\gb{5|Q|\ell_7}^2\over
\gb{6|Q|\ell_7}[\ell_7~\W\eta] [\ell_7~\ell_8]}+{g(\W\ell_8)\W
R_2[\ell_8,\W\eta,P_{56}]\gb{5|Q|\ell_8}^2\over
\gb{6|Q|\ell_8}[\ell_8~\W\eta]
[\ell_8~\ell_7]}\right)
}
 where $\W R_2[\ell_7,\W\eta, P_{56}]$ is the conjugated version of $R_2$ given  in Appendix B,
 i.e., $\ket{\bullet}\leftrightarrow  |\bullet]$.

The third term in \iifsonethreem\ has a double pole, so the
contribution is
\eqn\iifsthreemthree{\eqalign{
 c_{3:2,2;1}^{(3)}  =&
 {4 P_{56}^2 \braket{1~3}\over
\braket{3~4} }{[a~b]^4\over [b|P_{56} Q|b]^2}{
g(\W\ell_7)[\ell_7~2]\gb{3|Q|\ell_7}\gb{5|Q|\ell_7}^2\over
[\ell_7~\ell_8]^2\gb{6|Q|\ell_7} [\ell_7~\W\eta]} \W
R_2[\ell_7,\W\eta,P_{56}]\sum_{i=1}^8 {[L_{7,i}~L_{8,i}]\over
[\ell_7~L_{7,i}] [\ell_7~L_{8,i}]} \cr
 &  +
 {4 P_{56}^2 \braket{1~3}\over \braket{3~4} }{[a~b]^4\over
[b|P_{56} Q|b]^2}{
g(\W\ell_7)[\ell_7~2]\gb{3|Q|\ell_7}\gb{5|Q|\ell_7}^2\over
[\ell_7~\ell_8]^2\gb{6|Q|\ell_7} [\ell_7~\W\eta][\ell_7~2]} \W R_3
[\ell_7,\W\eta,|2],P_{56}]+ \{ |\eta_1] \leftrightarrow |\eta_2]
\}
}}
with the following two lists:
\eqn\iifsthreemlistone{\eqalign{
 L_7 &=
 \{|6],
|2],|2], |P_{56}|3\rangle, |P_{56}|3\rangle,  |Q|5\rangle, |Q|5\rangle ,|Q|3\rangle,|2]\}
\cr
L_8 & =  \{ |1], |P_{12}|4\rangle, |P_{56}|4\rangle,
|P_{56}P_{34}|2], |Q|6\rangle, |\ell_8],|\ell_8], |\W\eta],
|P|\ell_7\rangle\}
}}

The second term in \iifsonethreem\ has a triple pole, so the
contribution is
\eqn\iifsthreemfour{\eqalign{
 c_{3:2,2;1}^{(4)}&=  {2 (P_{56}^2)^2 \over
[1~2]\braket{3~4} }{[a~b]^6\over [b|P_{56} Q|b]^3}  {g(\W
\ell_7)[\ell_7~2]^2\gb{3|Q|\ell_7}^2\gb{5|Q|\ell_7}^2\over
[\ell_7~\ell_8]^3\gb{6|Q|\ell_7} [\ell_7~\W\eta]}
\cr &
\left( \sum_{1\leq i\leq j\leq 9} {[L_{9,i}~L_{10,i}] \over
[\ell_7~L_{9,i}][\ell_7~L_{10,i}]}{[L_{9,j}~L_{10,j}] \over
[\ell_7~L_{9,j}][\ell_7~L_{10,j}]} \W R_2[\ell_7,\W\eta,P_{56}] \right. \cr
&  +\sum_{i=1}^9 {[L_{9,i}~L_{10,i}] \over
[\ell_7~L_{9,i}][\ell_7~L_{10,i}]}{[L_{9,11}~L_{9,i}] \over
[\ell_7~L_{9,11}][\ell_7~L_{9,i}]}\W R_2[\ell_7,\W\eta,P_{56}]
\cr &
+{[L_{9,11}~L_{9,i}] \over
[\ell_7~L_{9,11}][\ell_7~L_{9,i}][\ell_7~2]}\W
R_3[\ell_7,\W\eta,|2],P_{56}]\cr &  \left. + \sum_{1\leq
i\leq 9} {[L_{9,i}~L_{10,i}] \over
[\ell_7~L_{9,i}][\ell_7~L_{10,i}][\ell_7~2]}\W
R_3[\ell_7,\W\eta,|2],P_{56}]+ {1\over [\ell_7~2]^2} \W
R_4[\ell_7,\W\eta,|2],P_{56}]\right) \cr &  + \{ |\eta_1]
\leftrightarrow |\eta_2] \}
}}
 with the
following two lists:
\eqn\iifsthreemlisttwo{\eqalign{
 L_9 &=   \{|6], |2],|2], |P_{56}|3\rangle,
|P_{56}|3\rangle, |Q|3\rangle, |Q|3\rangle ,
|Q|5\rangle,|Q|5\rangle,|2],|2]\} \cr
L_{10} & =  \{ |1], |P_{12}|4\rangle, |P_{56}|4\rangle,
|P_{56}P_{34}|2], |Q|6\rangle, |\ell_8],|\ell_8],|8], |\W\eta],
|P|\ell_7\rangle\}.
}}

\subsubsec{The result of cut $C_{56}$}

The coefficient of bubble $I_{2:2;5}$ is the sum of
\iifstwor, \iifsthreer, \iifsoneaone,
\iifsoneatwoone, \iifsoneatwotwo,\iifsoneathreeone, and \iifsoneathreetwo:
\eqn\libelle{
c_{2:2;5} = c_{2:2;5}^{(3r)} + c_{2:2;5}^{1r;A1} + c_{2:2;5}^{1r;A2-1} +
c_{2:2;5}^{1r;A2-2} + c_{2:2;5}^{1r;A3-1} +
c_{2:2;5}^{1r;A3-2}.
}

The coefficient of the three-mass triangle function $I_{3:2,2;1}$ is
given by the sum of \iifsthreemone, \iifsthreemtwo, \iifsthreemthree, and \iifsthreemfour:
\eqn\rossignol{
c_{3:2,2;1} = c_{3:2,2;1}^{(1)} +c_{3:2,2;1}^{(2)} +c_{3:2,2;1}^{(3)} +c_{3:2,2;1}^{(4)}.
}

The result could have been written directly using the functions defined at the end of Appendix C.  Here we have given some intermediate steps for illustration.  For further details, see Appendix C.

\newsec{$A(1^-, 2^-,3^+,4^-,5^+, 6^+)$}

The last of the NMHV six-gluon helicity configurations requires the
heaviest computation.  However, there are no essentially new
features encountered.  In this section we present, with minimal discussion,
some of our intermediate steps in order to allow the reader to
confirm our final formulas for the coefficients.

This helicity configuration is invariant under a $\QZ_2$ symmetry generated by $\alpha:
i\leftrightarrow 7-i$ accompanied by conjugation. There are  box, triangle and bubble contributions. The box
coefficients are straightforward to calculate by quadruple cuts and have been given in \BidderRI.  We list them again here in the notation consistent with the rest of this paper.\foot{An apparent discrepancy is due to a typo in the numerator of the coefficient $c_5^{\N=0}$ in \BidderRI.}

\eqn\utth{\eqalign{
c^{2m~h}_{4:2;2}  &=
2 {[1~3]^2 \vev{4~6}^2 \gb{4|P_{123}|1}\gb{6|P_{123}|3}P_{123}^2 \over
[1~2][2~3]\vev{4~5}\vev{5~6}\gb{6|P_{123}|1}^4}
\cr
c^{2m~h}_{4:2;4}  &=
2 {[6~2]^2\vev{3~4}\gb{4|P_{345}|2}^2\gb{3|P_{345}|6}P_{345}^2 \over
\vev{4~5}[6~1][1~2]\gb{5|P_{345}|2}\gb{3|P_{345}|2}^4}
\cr
c^{2m~h}_{4:2;6} &=
2 {\vev{1~5}^2[3~4]\gb{5|P_{234}|3}^2 \gb{1|P_{234}|4}P_{234}^2 \over
[2~3]\vev{5~6}\vev{6~1}\gb{5|P_{234}|2}\gb{5|P_{234}|4}^4}
\cr
c^{1m}_{4;5}  &=
2 {[2~3][3~4]\gb{1|P_{234}|2}^2 \gb{1|P_{234}|4} \over
[2~4]^4 \vev{5~6}\vev{6~1} \gb{5|P_{234}|2} P_{234}^2}
\cr
c^{1m}_{4;6}  &=
2 {\vev{3~4}\vev{4~5}\gb{5|P_{345}|6}^2 \gb{3|P_{345}|6} \over
\vev{3~5}^4 [6~1][1~2]\gb{5|P_{345}|2}P_{345}^2}
}}

In this configuration, there is only one nonvanishing three-mass-triangle
coefficient,  with the distribution
$(23|45|61)$.


For the bubble part, we have the following cuts: three particle channels
$C_{123}$, $C_{612}$ and $C_{234}$; two particle channels $C_{23}$,
$C_{34}$, $C_{45}$ and $C_{61}$. Among these, the pairs
$(C_{612},C_{234})$ and $( C_{23}, C_{45})$ are related by the $Z_2$ symmetry,
while the others are invariant. So in total we have five independent
double cuts $C_{123}$, $C_{234}$, $C_{23}$, $C_{34}$ and $C_{61}$.
We address these one by one.

Throughout this section we freely omit the integral sign when its presence may be inferred from spinor differentials.

\subsec{Cut $C_{123}$}

For $C_{123}$, there is no three-mass triangle contribution and the
calculation will be relatively simple. The expression is given by

\eqn\name{\eqalign{ C_{123} &=  \int d\mu \ \Big[
                   A(\ell_1^+, 1^-, 2^-, 3^+, \ell_2^-)
                   A(\ell_2^+, 4^-, 5^+, 6^+, \ell_1^-)
\cr & \qquad \qquad +  A(\ell_1^-, 1^-, 2^-, 3^+, \ell_2^+)
                   A(\ell_2^-, 4^-, 5^+, 6^+, \ell_1^+)
                \Big]
\cr &=
 \int d\mu \ {
{2 \spa{4}.{\ell_1}^2 \spa{4}.{\ell_2}^2 \spb{3}.{\ell_1}^2
\spb{3}.{\ell_2}} \over
 {\spa{4}.{ 5} \spa{5}.{ 6} \spa{6}.{\ell_1} \spa{\ell_1}.{\ell_2} \spa{\ell_2}.{ 4} \spb{1}.{ 2} \spb{2}.{ 3}
  \spb{\ell_1}.{ 1} \spb{\ell_2}.{\ell_1}}
} }}
After the $t$-integration, the rational contribution can be read as
a sum of three contributions, \eqn\name{\eqalign{ C_{123}^{rat} &=
                      C_{123}^{(1r)}
                    + C_{123}^{(2r)}
                    + C_{123}^{(3r)},
}}
each of which will be discussed separately.

\subsubsec{The term $C_{123}^{(1r)} $}

\eqn\name{\eqalign{ C_{123}^{(1r)} &= { {2 \dea \deb \spa{4}.{
\ell}^2 \spab{4}.{ P_{123}}.{ 1} \spab{\ell}.{ P_{123}}.{ 3}^2
  \spb{3}.{ 1} (P_{123}^2)} \over {\spa{4}.{ 5} \spa{5}.{ 6} \spa{6}.{ \ell}
  \spab{\ell}.{ P_{123}}.{ 1}^3 \spab{\ell}.{ P_{123}}.{ \ell}^2 \spb{1}.{ 2} \spb{2}.{ 3}}
}. }}
This integrand may be turned into  a full derivative by choosing as
a reference spinor $|\eta] = P_{123}|1\ra$, to neutralize the
multiple pole $\spab{\ell}.{ P_{123}}.{ 1}^3$:
\eqn\name{\eqalign{ C_{123}^{(1r)} &= { {2 \dea \dedeb \spa{4}.{
\ell}^2 \spab{1}.{ P_{123}}.{ \ell} \spab{4}.{ P_{123}}.{ 1}
  \spab{\ell}.{ P_{123}}.{ 3}^2 \spb{3}.{ 1}} \over {\spa{1}.{ \ell} \spa{4}.{ 5} \spa{5}.{ 6}
  \spa{6}.{ \ell} \spab{\ell}.{ P_{123}}.{ 1}^3 \spab{\ell}.{ P_{123}}.{ \ell} \spb{1}.{ 2}
  \spb{2}.{ 3}}
}. }}
The sum of the residues can therefore be performed as follows:
\eqn\cCXXIIIresI{\eqalign{ C_{123}^{(1r:s)} &= {2 \spab{4}.{
P_{123}}.{ 1} \spb{3}.{ 1} \over \spa{4}.{ 5} \spa{5}.{ 6} \spb{1}.{
2}  \spb{2}.{ 3} } \ \sum_{i=1,2} \lim_{\ell \to \ell_i}
\spa{\ell}.{\ell_i} \ { {\spa{4}.{ \ell}^2 \spab{1}.{ P_{123}}.{
\ell}
  \spab{\ell}.{ P_{123}}.{ 3}^2 } \over {\spa{1}.{ \ell}
  \spa{6}.{ \ell} \spab{\ell}.{ P_{123}}.{ 1}^3 \spab{\ell}.{ P_{123}}.{ \ell}
} },
}}
with $|\ell_1\ra = |1 \ra$ and $|\ell_2\ra = |6 \ra$.

\subsubsec{The term $C_{123}^{(2r)} $}

\eqn\name{\eqalign{ C_{123}^{(2r)} &= { {2 \dea \deb \spa{4}.{
\ell}^2 \spab{4}.{ P_{123}}.{ 1} \spab{\ell}.{ P_{123}}.{ 3}^2
  \spb{3}.{ \ell} (P_{123}^2)} \over {\spa{4}.{ 5} \spa{5}.{ 6} \spa{6}.{ \ell}
  \spab{\ell}.{ P_{123}}.{ 1}^2 \spab{\ell}.{ P_{123}}.{ \ell}^3 \spb{1}.{ 2} \spb{2}.{ 3}}
}. }}
In this case, one can write the integrand as a full derivative by
choosing the reference spinor $\eta = 3$, so that
\eqn\name{\eqalign{ C_{123}^{(2r)} &=
 - { {{\dea \dedeb \spa{4}.{ \ell}^2 \spab{4}.{ P_{123}}.{ 1} \spab{\ell}.{ P_{123}}.{ 3} \spb{3}.{ \ell}^2
   (P_{123}^2)} \over {\spa{4}.{ 5} \spa{5}.{ 6} \spa{6}.{ \ell} \spab{\ell}.{ P_{123}}.{ 1}^2
   \spab{\ell}.{ P_{123}}.{ \ell}^2 \spb{1}.{ 2} \spb{2}.{ 3}}}
}. }} Note the presence of a double pole, $\spab{\ell}.{ P_{123}}.{
1}^2$.

The residue of the single pole $\ell = 6$ is
\eqn\cCXXIIIresII{\eqalign{ C_{123}^{(2r:s)} &=
 - { {{\spa{4}.{ 6}^2 \spab{4}.{ P_{123}}.{ 1} \spab{6}.{ P_{123}}.{ 3} \spb{3}.{ 6}^2
   (P_{123}^2)} \over {\spa{4}.{ 5} \spa{5}.{ 6} \spab{6}.{ P_{123}}.{ 1}^2
   \spab{6}.{ P_{123}}.{ 6}^2 \spb{1}.{ 2} \spb{2}.{ 3}}}
},
}}
while the residue of the double pole $|\ell \ra = P_{123}|1]$ is
\eqn\cCXXIIIresIII{\eqalign{ C_{123}^{(2r:d)} &=
 -{ \spab{4}.{ P_{123}}.{ 1} \spba{3}.{ P_{123}}.1^2
\over \spa{4}.{ 5} \spa{5}.{ 6} \spb{1}.{ 2} \spb{2}.{ 3}
(P_{123}^2) } P_2\Big[ P_{123}|1], L_1^{II:C_{123}},L_2^{II:C_{123}}
\Big]
}} with \eqn\name{\eqalign{ L_1^{II:C_{123}} &=  \{|4\ra, |4\ra,
P_{123}|3] \} \cr L_2^{II:C_{123}} &=  \{|6\ra, |1\ra, |1\ra \} }}
since, having chosen $|\ell] = P_{123}|1\ra$, we used $\spab{\ell}.{
P_{123}}.{ \ell} \to P_{123}^2 \spa{\ell}.{1}$.

\subsubsec{The term $C_{123}^{(3r)} $}

\eqn\name{\eqalign{ C_{123}^{(3r)} &= { {2 \dea \deb \spa{4}.{
\ell}^3 \spab{\ell}.{ P_{123}}.{ 3} \spb{3}.{ \ell}^2 (P_{123}^2)^2}
\over
 {\spa{4}.{ 5} \spa{5}.{ 6} \spa{6}.{ \ell} \spab{\ell}.{ P_{123}}.{ 1}
  \spab{\ell}.{ P_{123}}.{ \ell}^4 \spb{1}.{ 2} \spb{2}.{ 3}}
}. }} It is straightforward to write it as a full derivative with
$\eta = 3$:
\eqn\name{\eqalign{ C_{123}^{(3r)} &=
 - {{ 2 \dea \dedeb \spa{4}.{ \ell}^3 \spb{3}.{ \ell}^3 (P_{123}^2)^2} \over
 {3 \spa{4}.{ 5} \spa{5}.{ 6} \spa{6}.{ \ell} \spab{\ell}.{ P_{123}}.{ 1}
  \spab{\ell}.{ P_{123}}.{ \ell}^3 \spb{1}.{ 2} \spb{2}.{ 3}}
}. }}
We obtain an expression where only single poles are present, whose
sum of residues is \eqn\cCXXIIIresIV{\eqalign{ C_{123}^{(3r:s)} &= -
{ 2 (P_{123}^2)^2 \over 3 \spa{4}.{ 5} \spa{5}.{ 6} \spb{1}.{ 2}
\spb{2}.{ 3} } \sum_{i=2,3} \lim_{\ell \to \ell_i}
\spa{\ell}.{\ell_i} \
  {{ \spa{4}.{ \ell}^3 \spb{3}.{ \ell}^3 } \over
 { \spa{6}.{ \ell} \spab{\ell}.{ P_{123}}.{ 1}
  \spab{\ell}.{ P_{123}}.{ \ell}^3 }
},
}}
with $|\ell_2\ra = |6 \ra$ and $|\ell_3\ra = P_{123}|1]$

\bigskip

Finally, the coefficient of the bubble $I_{2:3;1}$ is obtained by
adding \cCXXIIIresI, \cCXXIIIresII, \cCXXIIIresIII, and
\cCXXIIIresIV:
\eqn\name{\eqalign{ c_{2:3;1} =
              C_{123}^{(1r:s)}
            + C_{123}^{(2r:s)}
            + C_{123}^{(2r:d)}
            + C_{123}^{(3r:s)}
}}


\subsec{Cut $C_{234}$}

The cut $C_{234}$ is given by
 \eqn\name{\eqalign{ C_{234} &=  \int
d\mu \ \Big[
                   A(\ell_1^+, 2^-, 3^+, 4^-, \ell_2^-)
                   A(\ell_2^+, 5^+, 6^+, 1^-, \ell_1^-)
     \cr & \qquad \qquad  +  A(\ell_1^-, 2^-, 3^+, 4^-, \ell_2^+)
                   A(\ell_2^-, 5^+, 6^+, 1^-, \ell_1^+)
                \Big]
\cr &=
 \int d\mu \ {
{2 \spa{1}.{\ell_1} \spa{1}.{\ell_2}^2 \spb{3}.{\ell_1}^2
\spb{3}.{\ell_2}^2} \over
 {\spa{5}.{ 6} \spa{6}.{ 1} \spa{\ell_1}.{\ell_2} \spa{\ell_2}.{ 5} \spb{2}.{ 3} \spb{3}.{ 4} \spb{4}.{\ell_2}
  \spb{\ell_1}.{ 2} \spb{\ell_2}.{\ell_1}}
} }}
After the $t$-integration, the rational contribution can be read as
a sum of three contributions, \eqn\name{\eqalign{ C_{234}^{rat} &=
                      C_{234}^{(1r)}
                    + C_{234}^{(2r)}
                    + C_{234}^{(3r)},
}}
each of which will be discussed separately.

\subsubsec{The term $C_{234}^{(1r)} $}

\eqn\name{\eqalign{ C_{234}^{(1r)} &=
 - {{ 2 \dea \deb \spa{1}.{ \ell}^2 \spab{1}.{ P_{234}}.{ 4} \spab{\ell}.{ P_{234}}.{ 3}^3
  (P_{234}^2)} \over {\spa{5}.{ 6} \spa{5}.{ \ell} \spa{6}.{ 1} \spab{\ell}.{ P_{234}}.{ 2}
  \spab{\ell}.{ P_{234}}.{ 4}^3 \spab{\ell}.{ P_{234}}.{ \ell}^2 \spb{2}.{ 3}}
}. }}
We write it as a full derivative
 by choosing the reference spinor $|\eta] = P_{234}|4\ra$ in order to neutralize the multiple pole
$\spab{\ell}.{ P_{23}}.{ 4}^3$:
\eqn\name{\eqalign{ C_{234}^{(1r)} &= - {{ 2 \dea \dedeb \spa{1}.{
\ell}^2 \spab{1}.{ P_{234}}.{ 4} \spab{4}.{ P_{234}}.{ \ell}
  \spab{\ell}.{ P_{234}}.{ 3}^3} \over {\spa{4}.{ \ell} \spa{5}.{ 6} \spa{5}.{ \ell} \spa{6}.{ 1}
  \spab{\ell}.{ P_{234}}.{ 2} \spab{\ell}.{ P_{234}}.{ 4}^3 \spab{\ell}.{ P_{234}}.{ \ell}
  \spb{2}.{ 3}}
}. }}
The sum of the residues can be therefore performed as follows:
\eqn\cCCXXXIVresI{\eqalign{ C_{234}^{(1r:s)} &= - { 2 \spab{1}.{
P_{234}}.{ 4} \over \spa{5}.{ 6}\spa{6}.{ 1} \spb{2}.{ 3} } \
\sum_{i=1,2,3} \lim_{\ell \to \ell_i} \spa{\ell}.{\ell_i} \ {{
\spa{1}.{ \ell}^2  \spab{4}.{ P_{234}}.{ \ell}
  \spab{\ell}.{ P_{234}}.{ 3}^3} \over {\spa{4}.{ \ell}  \spa{5}.{ \ell}
  \spab{\ell}.{ P_{234}}.{ 2} \spab{\ell}.{ P_{234}}.{ 4}^3 \spab{\ell}.{ P_{234}}.{ \ell}
  }
},
}}
with $|\ell_1\ra = |4 \ra, |\ell_2\ra = |5 \ra$, and $|\ell_3\ra =
P_{234}|2]$.

\subsubsec{The term $C_{234}^{(2r)} $}

\eqn\name{\eqalign{ C_{234}^{(2r)} &=
 - {{ 2 \dea \deb \spa{1}.{ \ell}^2 \spab{1}.{ P_{234}}.{ 4} \spab{\ell}.{ P_{234}}.{ 3}^3
  \spb{3}.{ \ell} (P_{234}^2)} \over {\spa{5}.{ 6} \spa{5}.{ \ell} \spa{6}.{ 1}
  \spab{\ell}.{ P_{234}}.{ 2} \spab{\ell}.{ P_{234}}.{ 4}^2 \spab{\ell}.{ P_{234}}.{ \ell}^3
  \spb{2}.{ 3} \spb{3}.{ 4}}
}. }} We write it as a full derivative by choosing the reference
spinor $\eta = 3$: \eqn\name{\eqalign{ C_{234}^{(2r)} &= { {\dea
\dedeb \spa{1}.{ \ell}^2 \spab{1}.{ P_{234}}.{ 4} \spab{\ell}.{
P_{234}}.{ 3}^2 \spb{3}.{ \ell}^2
  (P_{234}^2)} \over {\spa{5}.{ 6} \spa{5}.{ \ell} \spa{6}.{ 1} \spab{\ell}.{ P_{234}}.{ 2}
  \spab{\ell}.{ P_{234}}.{ 4}^2 \spab{\ell}.{ P_{234}}.{ \ell}^2 \spb{2}.{ 3} \spb{3}.{ 4}}
}. }}

The sum of the residues of the single pole is
\eqn\cCCXXXIVresII{\eqalign{ C_{234}^{(2r:s)} &= { \spab{1}.{
P_{234}}.{ 4} (P_{234}^2) \over \spa{5}.{ 6} \spa{6}.{ 1} \spb{2}.{
3} \spb{3}.{ 4} } \ \sum_{i=2,3} \lim_{\ell \to \ell_i}
\spa{\ell}.{\ell_i} \ { {\spa{1}.{ \ell}^2  \spab{\ell}.{ P_{234}}.{
3}^2 \spb{3}.{ \ell}^2
  } \over { \spa{5}.{ \ell}  \spab{\ell}.{ P_{234}}.{ 2}
  \spab{\ell}.{ P_{234}}.{ 4}^2 \spab{\ell}.{ P_{234}}.{ \ell}^2 }
},
}}
with $|\ell_2\ra = |5 \ra$, and $|\ell_3\ra = P_{234}|2]$. The
residue of double pole $|\ell\ra = P_{234}|4]$ can be written as
\eqn\cCCXXXIVresIII{\eqalign{ C_{234}^{(2r:d)} &= { \spab{1}.{
P_{234}}.{ 4} \spab{4}.{ P_{234}}.{3}^2  \over (P_{234}^2) \spa{5}.{ 6}
\spa{6}.{ 1} \spb{2}.{ 3} \spb{3}.{ 4}} P_2\Big[ P_{234}|4],
L_1^{II:C_{234}},L_2^{II:C_{234}} \Big],
}}

with \eqn\name{\eqalign{ L_1^{II:C_{234}} &=  \{|1\ra, |1\ra,
P_{234}|3], P_{234}|3] \} \cr L_2^{II:C_{234}} &=  \{|5\ra,
P_{234}|2] , |4\ra, |4\ra \} }} since, having chosen $|\ell] =
P_{234}|4\ra$, we used $\spab{\ell}.{ P_{234}}.{ \ell} \to P_{234}^2
\spa{\ell}.{4}$.

\subsubsec{The term $C_{234}^{(3r)} $}

\eqn\name{\eqalign{ C_{234}^{(3r)} &=
 - {{ 2 \dea \deb \spa{1}.{ \ell}^3 \spab{\ell}.{ P_{234}}.{ 3}^2 \spb{3}.{ \ell}^2
  (P_{234}^2)^2} \over {\spa{5}.{ 6} \spa{5}.{ \ell} \spa{6}.{ 1} \spab{\ell}.{ P_{234}}.{ 2}
  \spab{\ell}.{ P_{234}}.{ 4} \spab{\ell}.{ P_{234}}.{ \ell}^4 \spb{2}.{ 3} \spb{3}.{ 4}}
}. }}

In this case only simple poles are present. We can write it as a
full derivative with $\eta = 3$:
\eqn\name{\eqalign{ C_{234}^{(3r)} &= { {2 \dea \dedeb \spa{1}.{
\ell}^3 \spab{\ell}.{ P_{234}}.{ 3} \spb{3}.{ \ell}^3 (P_{234}^2)^2}
\over
 {3 \spa{5}.{ 6} \spa{5}.{ \ell} \spa{6}.{ 1} \spab{\ell}.{ P_{234}}.{ 2}
  \spab{\ell}.{ P_{234}}.{ 4} \spab{\ell}.{ P_{234}}.{ \ell}^3 \spb{2}.{ 3} \spb{3}.{ 4}}
}. }}
We obtain an expression where only single poles are present, whose
sum of residues is \eqn\cCCXXXIVresIV{\eqalign{ C_{123}^{(3r:s)} &=
{2 (P_{234}^2)^2 \over 3 \spa{5}.{ 6} \spa{6}.{ 1} \spb{2}.{ 3}
\spb{3}.{ 4} } \ \sum_{i=2,3,4} \lim_{\ell \to \ell_i}
\spa{\ell}.{\ell_i} \ { { \spa{1}.{ \ell}^3 \spab{\ell}.{ P_{234}}.{
3} \spb{3}.{ \ell}^3 } \over
 { \spa{5}.{ \ell}  \spab{\ell}.{ P_{234}}.{ 2}
  \spab{\ell}.{ P_{234}}.{ 4} \spab{\ell}.{ P_{234}}.{ \ell}^3 }
},
}}
with $|\ell_2\ra = |5 \ra, |\ell_3\ra = P_{234}|2]$ and $|\ell_4\ra
= P_{234}|4]$.

\bigskip

Finally the coefficient of the bubble $I_{2:3;2}$ is obtained by
adding \cCCXXXIVresI, \cCCXXXIVresII, \cCCXXXIVresIII, and
\cCCXXXIVresIV: \eqn\name{\eqalign{ c_{2:3;2} =   C_{234}^{(1r:s)}
            + C_{234}^{(2r:s)}
            + C_{234}^{(3r:d)}
            + C_{234}^{(3r:s)}
}}


\subsec{Cut $C_{34}$}

For this double cut, there is no triangle contribution and the
result is simpler. The cut is given by

\eqn\name{\eqalign{ C_{34} &=  \int d\mu \ \Big[
                   A(\ell_1^+, 3^+, 4^-, \ell_2^-)
                   A(\ell_2^+, 5^+, 6^+, 1^-, 2^-, \ell_1^-)
     \cr & \qquad \qquad           +  A(\ell_1^-, 3^+, 4^-, \ell_2^+)
                   A(\ell_2^-, 5^+, 6^+, 1^-, 2^-, \ell_1^+)
                \Big]
\cr &=
 \int d\mu \ \bigg\{
{ {2 \spa{4}.{\ell_1}^2 \spa{4}.{\ell_2} \spab{1}.{
P_{56}}.{\ell_1}^2 \spab{1}.{ P_{56}}.{\ell_2}} \over
  {\spa{3}.{ 4} \spa{5}.{ 6} \spa{6}.{ 1} \spa{\ell_1}.{ 3} \spa{\ell_2}.{\ell_1} \spab{5}.{ P_{61}}.{ 2}
   \spb{2}.{\ell_1} \spb{\ell_1}.{\ell_2} (P_{561}^2)}  } \cr
&  \qquad \qquad + {
 {2 \spa{4}.{\ell_1}^2 \spa{4}.{\ell_2} \spab{\ell_1}.{ P_{12}}.{ 6} \spab{\ell_2}.{ P_{12}}.{ 6}^2} \over
  {\spa{3}.{ 4} \spa{\ell_1}.{ 3} \spa{\ell_1}.{\ell_2} \spa{\ell_2}.{ 5} \spa{\ell_2}.{\ell_1} \spab{5}.{ P_{61}}.{ 2}
   \spb{1}.{ 2} \spb{6}.{ 1} (P_{345}^2)}
} \bigg\} \cr &= C_{34}^{(1)} + C_{34}^{(2)} }}
Due to the possibility of writing $A(\ell_1^+, 3^+, 4^-, \ell_2^-)$
in terms of either holomorphic or antiholomorphic spinor products,
as follows, \eqn\name{\eqalign{ A(\ell_1^+, 3^+, 4^-, \ell_2^-) = {
{\spa{4}.{\ell_1}^2 \spa{4}.{\ell_2}^2} \over {\spa{\ell_1}.{ 3}
\spa{3}.{ 4} \spa{4}.{\ell_2}  \spa{\ell_2}.{\ell_1}} } = {
{\spb{3}.{\ell_1}^2 \spb{3}.{\ell_2}^2} \over {\spb{\ell_1}.{ 3}
\spb{3}.{ 4} \spb{4}.{\ell_2}  \spb{\ell_2}.{\ell_1}} }, }}
one can rewrite $C_{34}^{(2)}$ as \eqn\name{\eqalign{ C_{34}^{(2)}
&= {
 {2 \spa{4}.{\ell_1}^2 \spa{4}.{\ell_2} \spab{\ell_1}.{ P_{12}}.{ 6} \spab{\ell_2}.{ P_{12}}.{ 6}^2} \over
  {\spa{3}.{ 4} \spa{\ell_1}.{ 3} \spa{\ell_1}.{\ell_2} \spa{\ell_2}.{ 5} \spa{\ell_2}.{\ell_1} \spab{5}.{ P_{61}}.{ 2}
   \spb{1}.{ 2} \spb{6}.{ 1} (P_{345}^2)}
} \cr &= {
 {2 \spb{3}.{\ell_2}^2 \spb{3}.{\ell_1} \spab{\ell_1}.{ P_{12}}.{ 6} \spab{\ell_2}.{ P_{12}}.{ 6}^2} \over
  {\spb{3}.{ 4} \spb{\ell_2}.4 \spb{\ell_2}.{\ell_1} \spa{\ell_2}.{ 5} \spa{\ell_2}.{\ell_1} \spab{5}.{ P_{61}}.{ 2}
   \spb{1}.{ 2} \spb{6}.{ 1} (P_{345}^2)}
}. }} In this shape, $C_{34}^{(2)}$ can be obtained from
$C_{34}^{(1)}$, as the following relation holds,
\eqn\cXXXIVrel{\eqalign{ P_{345}^2 \ C_{34}(2,0) = (-1) \ Tfm \ \{
P_{561}^2 \ C_{34}(1,0) \}
}} where $Tfm $ is the composition of three operations: i) Parity:
$\langle ~,~ \ra \leftrightarrow [~,~]$; ii) relabeling: $\{1,2,3,4,5,6\}
\to \{6,5,4,3,2,1\} $; iii) exchange: $\ell_1 \leftrightarrow
\ell_2$.
Therefore one can worry only about $C_{34}^{(1)}$ and recover
$C_{34}^{(2)}$ at the end through the above relation.

After the $t$-integration, the rational contribution coming from
$C_{34}^{(1)}$ can be read as \eqn\name{\eqalign{ C_{34}^{(1r)} &= {
{2 \dea \deb \spab{1}.{ P_{56}}.{ \ell}^3 \spb{3}.{ \ell} \spb{4}.{
3} P_{34}} \over
  {\spa{5}.{ 6} \spa{6}.{ 1} \spab{5}.{ P_{61}}.{ 2} \spab{\ell}.{ P_{34}}.{ \ell}^2 \spb{2}.{ \ell}
   \spb{4}.{ \ell}^3 (P_{561}^2)}  } \cr  &  - {  {2 \dea \deb \spab{1}.{ P_{56}}.{ 3}
   \spab{1}.{ P_{56}}.{ \ell}^2 \spb{3}.{ \ell} P_{34}} \over
  {\spa{5}.{ 6} \spa{6}.{ 1} \spab{5}.{ P_{61}}.{ 2} \spab{\ell}.{ P_{34}}.{ \ell}^2 \spb{2}.{ \ell}
   \spb{4}.{ \ell}^2 (P_{561}^2)}  } \cr  &  - {  {2 \dea \deb \spa{4}.{ \ell} \spab{1}.{ P_{56}}.{ \ell}^3
   \spb{3}.{ \ell} \spb{4}.{ 3} P_{34}} \over {\spa{5}.{ 6} \spa{6}.{ 1} \spab{5}.{ P_{61}}.{ 2}
   \spab{\ell}.{ P_{34}}.{ \ell}^3 \spb{2}.{ \ell} \spb{4}.{ \ell}^2 (P_{561}^2)}  } \cr  &  + {
 {2 \dea \deb \spa{4}.{ \ell} \spab{1}.{ P_{56}}.{ 3} \spab{1}.{ P_{56}}.{ \ell}^2 \spb{3}.{ \ell}
   P_{34}} \over {\spa{5}.{ 6} \spa{6}.{ 1} \spab{5}.{ P_{61}}.{ 2} \spab{\ell}.{ P_{34}}.{ \ell}^3
   \spb{2}.{ \ell} \spb{4}.{ \ell} (P_{561}^2)}  } \cr  &  + {
 {2 \dea \deb \spa{4}.{ \ell}^2 \spab{1}.{ P_{56}}.{ \ell}^3 \spb{3}.{ \ell} \spb{4}.{ 3}
   P_{34}} \over {\spa{5}.{ 6} \spa{6}.{ 1} \spab{5}.{ P_{61}}.{ 2} \spab{\ell}.{ P_{34}}.{ \ell}^4
   \spb{2}.{ \ell} \spb{4}.{ \ell} (P_{561}^2)}
}. }}
This can be written as a full derivative by choosing as reference
spinor $\eta = 4$:
\eqn\name{\eqalign{ C_{34}^{(1r)} &= { {2 \deb \dedea  \spa{4}.{
\ell} \spab{1}.{ P_{56}}.{ \ell}^3 \spb{4}.{ 3} P_{34}} \over
  {\spa{4}.{ 3} \spa{5}.{ 6} \spa{6}.{ 1} \spab{5}.{ P_{61}}.{ 2} \spab{\ell}.{ P_{34}}.{ \ell}
   \spb{2}.{ \ell} \spb{4}.{ \ell}^3 (P_{561}^2)}  } \cr  &  - {
 {2 \deb \dedea  \spa{4}.{ \ell} \spab{1}.{ P_{56}}.{ 3} \spab{1}.{ P_{56}}.{ \ell}^2 P_{34}} \over
  {\spa{4}.{ 3} \spa{5}.{ 6} \spa{6}.{ 1} \spab{5}.{ P_{61}}.{ 2} \spab{\ell}.{ P_{34}}.{ \ell}
   \spb{2}.{ \ell} \spb{4}.{ \ell}^2 (P_{561}^2)}  } \cr  &  - {
 {\deb \dedea  \spa{4}.{ \ell}^2 \spab{1}.{ P_{56}}.{ \ell}^3 \spb{4}.{ 3} P_{34}} \over
  {\spa{4}.{ 3} \spa{5}.{ 6} \spa{6}.{ 1} \spab{5}.{ P_{61}}.{ 2} \spab{\ell}.{ P_{34}}.{ \ell}^2
   \spb{2}.{ \ell} \spb{4}.{ \ell}^2 (P_{561}^2)}  } \cr  &  + {
 {\deb \dedea  \spa{4}.{ \ell}^2 \spab{1}.{ P_{56}}.{ 3} \spab{1}.{ P_{56}}.{ \ell}^2 P_{34}} \over
  {\spa{4}.{ 3} \spa{5}.{ 6} \spa{6}.{ 1} \spab{5}.{ P_{61}}.{ 2} \spab{\ell}.{ P_{34}}.{ \ell}^2
   \spb{2}.{ \ell} \spb{4}.{ \ell} (P_{561}^2)}  } \cr  &  + {
 {2 \deb \dedea  \spa{4}.{ \ell}^3 \spab{1}.{ P_{56}}.{ \ell}^3 \spb{4}.{ 3} P_{34}} \over
  {3 \spa{4}.{ 3} \spa{5}.{ 6} \spa{6}.{ 1} \spab{5}.{ P_{61}}.{ 2} \spab{\ell}.{ P_{34}}.{ \ell}^3
   \spb{2}.{ \ell} \spb{4}.{ \ell} (P_{561}^2)}
}. }}
As one can see, the higher pole $\spb{4}.{\ell}^n$ is neutralized,
and the only active pole is $\ell = 2$, whose residue reads as
follows: \eqn\cXXXIVresI{\eqalign{ C_{34}^{(1r:s)} &=
 - {{ 2 \spa{4}.{ 2} \spab{1}.{ P_{56}}.{ 2}^2 \spab{1}.{ P_{56}}.{ 3} P_{34}} \over
  {\spa{4}.{ 3} \spa{5}.{ 6} \spa{6}.{ 1} \spab{2}.{ P_{34}}.{ 2} \spab{5}.{ P_{61}}.{ 2}
   \spb{2}.{ 4}^2 (P_{561}^2)}  } \cr  &  - {  {\spa{4}.{ 2}^2 \spab{1}.{ P_{56}}.{ 2}^2
   \spab{1}.{ P_{56}}.{ 3} P_{34}} \over {\spa{4}.{ 3} \spa{5}.{ 6} \spa{6}.{ 1}
   \spab{2}.{ P_{34}}.{ 2}^2 \spab{5}.{ P_{61}}.{ 2} \spb{2}.{ 4} (P_{561}^2)}  } \cr  &  - {
 {2 \spa{4}.{ 2} \spab{1}.{ P_{56}}.{ 2}^3 \spb{4}.{ 3} P_{34}} \over
  {\spa{4}.{ 3} \spa{5}.{ 6} \spa{6}.{ 1} \spab{2}.{ P_{34}}.{ 2} \spab{5}.{ P_{61}}.{ 2}
   \spb{2}.{ 4}^3 (P_{561}^2)}  } \cr  &  - {  {\spa{4}.{ 2}^2 \spab{1}.{ P_{56}}.{ 2}^3 \spb{4}.{ 3}
   P_{34}} \over {\spa{4}.{ 3} \spa{5}.{ 6} \spa{6}.{ 1} \spab{2}.{ P_{34}}.{ 2}^2
   \spab{5}.{ P_{61}}.{ 2} \spb{2}.{ 4}^2 (P_{561}^2)}  } \cr  &  - {
 {2 \spa{4}.{ 2}^3 \spab{1}.{ P_{56}}.{ 2}^3 \spb{4}.{ 3} P_{34}} \over
  {3 \spa{4}.{ 3} \spa{5}.{ 6} \spa{6}.{ 1} \spab{2}.{ P_{34}}.{ 2}^3 \spab{5}.{ P_{61}}.{ 2}
   \spb{2}.{ 4} (P_{561}^2)}
}
}}

Through the relation in eq.\cXXXIVrel\ one can get the other term
contributing to the cut, \eqn\cXXXIVresII{\eqalign{ C_{34}^{(2r:s)}
&= { {2 \spab{4}.{ P_{12}}.{ 6} \spab{5}.{ P_{12}}.{ 6}^2 \spb{3}.{
5} P_{34}} \over
  {\spa{5}.{ 3}^2 \spab{5}.{ P_{34}}.{ 5} \spab{5}.{ P_{61}}.{ 2} \spb{1}.{ 6} \spb{2}.{ 1}
   \spb{3}.{ 4} (P_{612}^2)}  } \cr  &  + {  {2 \spa{3}.{ 4} \spab{5}.{ P_{12}}.{ 6}^3 \spb{3}.{ 5}
   P_{34}} \over {\spa{5}.{ 3}^3 \spab{5}.{ P_{34}}.{ 5} \spab{5}.{ P_{61}}.{ 2} \spb{1}.{ 6}
   \spb{2}.{ 1} \spb{3}.{ 4} (P_{612}^2)}  } \cr  &  + {
 {\spab{4}.{ P_{12}}.{ 6} \spab{5}.{ P_{12}}.{ 6}^2 \spb{3}.{ 5}^2 P_{34}} \over
  {\spa{5}.{ 3} \spab{5}.{ P_{34}}.{ 5}^2 \spab{5}.{ P_{61}}.{ 2} \spb{1}.{ 6} \spb{2}.{ 1}
   \spb{3}.{ 4} (P_{612}^2)}  } \cr  &  + {  {\spa{3}.{ 4} \spab{5}.{ P_{12}}.{ 6}^3 \spb{3}.{ 5}^2
   P_{34}} \over {\spa{5}.{ 3}^2 \spab{5}.{ P_{34}}.{ 5}^2 \spab{5}.{ P_{61}}.{ 2} \spb{1}.{ 6}
   \spb{2}.{ 1} \spb{3}.{ 4} (P_{612}^2)}  } \cr  &  + {
 {2 \spa{3}.{ 4} \spab{5}.{ P_{12}}.{ 6}^3 \spb{3}.{ 5}^3 P_{34}} \over
  {3 \spa{5}.{ 3} \spab{5}.{ P_{34}}.{ 5}^3 \spab{5}.{ P_{61}}.{ 2} \spb{1}.{ 6} \spb{2}.{ 1}
   \spb{3}.{ 4} (P_{612}^2)}
}.
}}

\bigskip

Finally, the coefficient of the bubble $I_{2:2;3}$ is obtained by
adding \cXXXIVresI\ and \cXXXIVresII: \eqn\name{\eqalign{ c_{2:2;3}
=   C_{34}^{(1r:s)}
            + C_{34}^{(2r:s)}
}}


\subsec{Cut $C_{23}$}

The cut in the $P_{23}$-channel receives two contributions:

\eqn\name{\eqalign{ C_{23} &=  \int d\mu \ \Big[
                   A(\ell_1^+, 2^-, 3^+, \ell_2^-)
                   A(\ell_2^+, 4^-, 5^+, 6^+, 1^-, \ell_1^-)
    \cr & \qquad \qquad     +  A(\ell_1^-, 2^-, 3^+, \ell_2^+)
                   A(\ell_2^-, 4^-, 5^+, 6^+, 1^-, \ell_1^+)
                \Big]
\cr & = 2 C_{23}^{(1)}, }} where \eqn\name{\eqalign{ C_{23}^{(1)} &=
\int d\mu \ \bigg\{ { {\spa{1}.{\ell_1} \spa{1}.{\ell_2}^2
\spa{2}.{\ell_1}^2 \spa{2}.{\ell_2}^2 \spb{5}.{ 6}^3} \over
  {\spa{2}.{ 3} \spa{3}.{\ell_2} \spa{\ell_1}.{ 2} \spa{\ell_1}.{\ell_2} \spa{\ell_2}.{\ell_1}
   \spab{1}.{ P_{456}}.{ 4} \spab{\ell_2}.{ P_{456}}.{ 6} \spb{4}.{ 5} P_{456}}  } \cr  &  + {
 {\spa{2}.{\ell_1}^2 \spa{2}.{\ell_2}^2 \spab{1}.{ P_{561}}.{\ell_1}^2
   \spab{1}.{ P_{561}}.{\ell_2}^2} \over {\spa{2}.{ 3} \spa{3}.{\ell_2} \spa{5}.{ 6} \spa{6}.{ 1}
   \spa{\ell_1}.{ 2} \spa{\ell_2}.{\ell_1} \spab{1}.{ P_{561}}.{ 4} \spab{5}.{ P_{561}}.{\ell_1}
   \spb{\ell_1}.{\ell_2} \spb{\ell_2}.{ 4} P_{561}}  } \cr  &  + {
 {\spa{2}.{\ell_1}^2 \spa{2}.{\ell_2}^2 \spa{\ell_2}.{ 4}^3 \spab{4}.{ P_{61\ell_1}}.{ 6}^2
   \spb{6}.{\ell_1}^4} \over {\spa{2}.{ 3} \spa{3}.{\ell_2} \spa{4}.{ 5} \spa{4}.{\ell_2}^2 \spa{\ell_1}.{ 2}
   \spa{\ell_2}.{\ell_1} \spab{5}.{ P_{61\ell_1}}.{\ell_1} \spab{\ell_2}.{ P_{61\ell_1}}.{ 6} \spb{1}.{\ell_1}
   \spb{6}.{ 1} \spb{\ell_1}.{ 6}^2 (P_{61\ell_1}^2)}
} \bigg\}. }}

Therefore we concentrate just on the term $C_{23}^{(1)}$, and
finally multiply by 2, in order to get the coefficients of the
proper functions. In particular, since this cut contains the
contributions to both bubbles and three-mass triangle coefficients,
we discuss these separately.

In the following formulas, {\sl for this cut only}, we define $$Q=
 (P_{61}^2/P_{23}^2) \ P_{23} + P_{61}$$ and $$|\omega_6] = P_{23}
P_{123} |6].$$

\subsubsec {{\bf Rational contribution from $C_{23}^{(1)}$}}

After the $t$-integration, the rational term which will contribute
to the bubble coefficients $I_{2:2;2}$ reads as a sum of six terms:

\eqn\name{\eqalign{ C_{23}^{(1,rat)} &=   C_{23}^{(1,r,1)}
                    + C_{23}^{(1,r,2)}
                    + C_{23}^{(1,r,3)}
\cr &                   + C_{23}^{(1,r,4)}
                    + C_{23}^{(1,r,5)}
                    + C_{23}^{(1,r,6)}.
}}
For each of these we give the expression after the $t-$integration,
the form as full derivative, and the residues.

\subsubsec{The term $C_{23}^{(1,r,1)} $}

i) $t$-integrated formula: \eqn\name{\eqalign{ C_{23}^{(1,r,1)} &= {
{\dea \deb \spab{1}.{ P_{56}}.{ 3}^2 \spab{1}.{ P_{56}}.{ \ell}^2
\spb{2}.{ 3} \spb{3}.{ \ell}} \over
  {\spa{5}.{ 6} \spa{6}.{ 1} \spab{1}.{ P_{56}}.{ 4} \spab{5}.{ P_{61}}.{ \ell}
   \spab{\ell}.{ P_{23}}.{ \ell}^2 \spb{2}.{ \ell} \spb{4}.{ \ell} (P_{561}^2)}  } \cr  &  + {
 {\dea \deb \spa{2}.{ 3}^2 \spab{1}.{ P_{56}}.{ \ell}^4 \spb{2}.{ 3}^3 \spb{3}.{ 4}^2
   \spb{3}.{ \ell}} \over {\spa{5}.{ 6} \spa{6}.{ 1} \spab{1}.{ P_{56}}.{ 4} \spab{5}.{ P_{61}}.{ \ell}
   \spab{\ell}.{ P_{23}}.{ \ell}^2 \spb{2}.{ \ell} \spb{4}.{ \ell}^3 (P_{23}^2)^2 (P_{561}^2)}  } \cr  &  - {
 {2 \dea \deb \spa{2}.{ 3} \spab{1}.{ P_{56}}.{ 3} \spab{1}.{ P_{56}}.{ \ell}^3 \spb{2}.{ 3}^2
   \spb{3}.{ 4} \spb{3}.{ \ell}} \over {\spa{5}.{ 6} \spa{6}.{ 1} \spab{1}.{ P_{56}}.{ 4}
   \spab{5}.{ P_{61}}.{ \ell} \spab{\ell}.{ P_{23}}.{ \ell}^2 \spb{2}.{ \ell} \spb{4}.{ \ell}^2
   (P_{23}^2) (P_{561}^2)}
} }}

ii) full derivative: \eqn\cXXIIIderI{\eqalign{ C_{23}^{(1,r,1)} &= {
{\deb \dedea  \spa{4}.{ \ell} \spab{1}.{ P_{56}}.{ 3}^2 \spab{1}.{
P_{56}}.{ \ell}^2 \spb{2}.{ 3}
   \spb{3}.{ \ell}} \over {\spa{5}.{ 6} \spa{6}.{ 1} \spab{1}.{ P_{56}}.{ 4} \spab{4}.{ P_{23}}.{ \ell}
   \spab{5}.{ P_{61}}.{ \ell} \spab{\ell}.{ P_{23}}.{ \ell} \spb{2}.{ \ell} \spb{4}.{ \ell}
   (P_{561}^2)}  } \cr  &  + {  {\deb \dedea  \spa{2}.{ 3}^2 \spa{4}.{ \ell} \spab{1}.{ P_{56}}.{ \ell}^4
   \spb{2}.{ 3}^3 \spb{3}.{ 4}^2 \spb{3}.{ \ell}} \over {\spa{5}.{ 6} \spa{6}.{ 1}
   \spab{1}.{ P_{56}}.{ 4} \spab{4}.{ P_{23}}.{ \ell} \spab{5}.{ P_{61}}.{ \ell}
   \spab{\ell}.{ P_{23}}.{ \ell} \spb{2}.{ \ell} \spb{4}.{ \ell}^3 (P_{23}^2)^2 (P_{561}^2)}  } \cr  &  - {
 {2 \deb \dedea  \spa{2}.{ 3} \spa{4}.{ \ell} \spab{1}.{ P_{56}}.{ 3} \spab{1}.{ P_{56}}.{ \ell}^3
   \spb{2}.{ 3}^2 \spb{3}.{ 4} \spb{3}.{ \ell}} \over {\spa{5}.{ 6} \spa{6}.{ 1} \spab{1}.{ P_{56}}.{ 4}
   \spab{4}.{ P_{23}}.{ \ell} \spab{5}.{ P_{61}}.{ \ell} \spab{\ell}.{ P_{23}}.{ \ell} \spb{2}.{ \ell}
   \spb{4}.{ \ell}^2 (P_{23}^2) (P_{561}^2)}
} }}
As one can see the higher pole $\spb{4}.{\ell}^n$ in neutralized by
the presence of $\spa{4}.{\ell}$ in the corresponding numerator.
Therefore, only single poles will give nonzero residues,
\eqn\cXXIIIresI{\eqalign{ C_{23}^{(1,r,1:s)} &= \ \sum_{i=1}^{4}
\lim_{\ell \to \ell_i} \spb{\ell}.{\ell_i} \bigg\{ \ { {  \spa{4}.{
\ell} \spab{1}.{ P_{56}}.{ 3}^2 \spab{1}.{ P_{56}}.{ \ell}^2
\spb{2}.{ 3}
   \spb{3}.{ \ell}} \over {\spa{5}.{ 6} \spa{6}.{ 1} \spab{1}.{ P_{56}}.{ 4} \spab{4}.{ P_{23}}.{ \ell}
   \spab{5}.{ P_{61}}.{ \ell} \spab{\ell}.{ P_{23}}.{ \ell} \spb{2}.{ \ell} \spb{4}.{ \ell}
   (P_{561}^2)}  } \cr  &  + {  {  \spa{2}.{ 3}^2 \spa{4}.{ \ell} \spab{1}.{ P_{56}}.{ \ell}^4
   \spb{2}.{ 3}^3 \spb{3}.{ 4}^2 \spb{3}.{ \ell}} \over {\spa{5}.{ 6} \spa{6}.{ 1}
   \spab{1}.{ P_{56}}.{ 4} \spab{4}.{ P_{23}}.{ \ell} \spab{5}.{ P_{61}}.{ \ell}
   \spab{\ell}.{ P_{23}}.{ \ell} \spb{2}.{ \ell} \spb{4}.{ \ell}^3 (P_{23}^2)^2 (P_{561}^2)}  } \cr  &  - {
 {2  \spa{2}.{ 3} \spa{4}.{ \ell} \spab{1}.{ P_{56}}.{ 3} \spab{1}.{ P_{56}}.{ \ell}^3
   \spb{2}.{ 3}^2 \spb{3}.{ 4} \spb{3}.{ \ell}} \over {\spa{5}.{ 6} \spa{6}.{ 1} \spab{1}.{ P_{56}}.{ 4}
   \spab{4}.{ P_{23}}.{ \ell} \spab{5}.{ P_{61}}.{ \ell} \spab{\ell}.{ P_{23}}.{ \ell} \spb{2}.{ \ell}
   \spb{4}.{ \ell}^2 (P_{23}^2) (P_{561}^2)}
} \bigg\} }}
with $|\ell_i] = P_{23}|4 \ra, P_{61}|5\ra, |2], |4]$ for
$(i=1,\ldots,4)$.

\subsubsec{The term $C_{23}^{(1,r,2)} $}

i) $t$-integrated formula: \eqn\name{\eqalign{ C_{23}^{(1,r,2)} &= {
{2 \dea \deb \spa{2}.{ \ell} \spab{1}.{ P_{56}}.{ 3} \spab{1}.{
P_{56}}.{ \ell}^3 \spb{2}.{ 3}^2
   \spb{3}.{ \ell}} \over {\spa{5}.{ 6} \spa{6}.{ 1} \spab{1}.{ P_{56}}.{ 4} \spab{5}.{ P_{61}}.{ \ell}
   \spab{\ell}.{ P_{23}}.{ \ell}^3 \spb{2}.{ \ell} \spb{4}.{ \ell} (P_{561}^2)}  } \cr  &  - {
 {\dea \deb \spa{2}.{ 3} \spa{2}.{ \ell} \spab{1}.{ P_{56}}.{ \ell}^4 \spb{2}.{ 3}^3 \spb{3}.{ 4}
   \spb{3}.{ \ell}} \over {\spa{5}.{ 6} \spa{6}.{ 1} \spab{1}.{ P_{56}}.{ 4} \spab{5}.{ P_{61}}.{ \ell}
   \spab{\ell}.{ P_{23}}.{ \ell}^3 \spb{2}.{ \ell} \spb{4}.{ \ell}^2 (P_{23}^2) (P_{561}^2)}
} }}

ii) full derivative: \eqn\cXXIIIderII{\eqalign{ C_{23}^{(1,r,2)} &=
{ {\deb \dedea \spa{2}.{ \ell}^2 \spab{1}.{ P_{56}}.{ 3} \spab{1}.{
P_{56}}.{ \ell}^3 \spb{2}.{ 3}^2} \over
  {\spa{2}.{ 3} \spa{5}.{ 6} \spa{6}.{ 1} \spab{1}.{ P_{56}}.{ 4} \spab{5}.{ P_{61}}.{ \ell}
   \spab{\ell}.{ P_{23}}.{ \ell}^2 \spb{2}.{ \ell} \spb{4}.{ \ell} (P_{561}^2)}  } \cr & - {
 {\deb \dedea \spa{2}.{ \ell}^2 \spab{1}.{ P_{56}}.{ \ell}^4 \spb{2}.{ 3}^3 \spb{3}.{ 4}} \over
  {2 \spa{5}.{ 6} \spa{6}.{ 1} \spab{1}.{ P_{56}}.{ 4} \spab{5}.{ P_{61}}.{ \ell}
   \spab{\ell}.{ P_{23}}.{ \ell}^2 \spb{2}.{ \ell} \spb{4}.{ \ell}^2 (P_{23}^2) (P_{561}^2)}
} }}
In this case we have both the residues of the simple poles,
\eqn\cXXIIIresIIs{\eqalign{ C_{23}^{(1,r,2:s)} &= { \spab{1}.{
P_{56}}.{ 3} \spb{2}.{ 3}^2 \over \spa{2}.{ 3} \spa{5}.{ 6}
\spa{6}.{ 1} (P_{561}^2) } \ \sum_{i=2}^{4} \lim_{\ell \to \ell_i}
\spb{\ell}.{\ell_i} { { \spa{2}.{ \ell}^2  \spab{1}.{ P_{56}}.{
\ell}^3 } \over
  { \spab{1}.{ P_{56}}.{ 4} \spab{5}.{ P_{61}}.{ \ell}
   \spab{\ell}.{ P_{23}}.{ \ell}^2 \spb{2}.{ \ell} \spb{4}.{ \ell} }  }
\cr & - { \spb{2}.{ 3}^3 \spb{3}.{ 4} \over 2 \spa{5}.{ 6} \spa{6}.{
1} \spab{1}.{ P_{56}}.{ 4} (P_{23}^2) (P_{561}^2) } \ \sum_{i=2}^{3}
\lim_{\ell \to \ell_i} \spb{\ell}.{\ell_i} {
 { \spa{2}.{ \ell}^2 \spab{1}.{ P_{56}}.{ \ell}^4 } \over
  { \spab{5}.{ P_{61}}.{ \ell}
   \spab{\ell}.{ P_{23}}.{ \ell}^2 \spb{2}.{ \ell} \spb{4}.{ \ell}^2 }
}, }}
and the residue from a double pole, \eqn\cXXIIIresIId{\eqalign{
C_{23}^{(1,r,2:d)} &= - { \spb{2}.{ 3}^3 \spb{3}.{ 4} \spa{2}.{ 4}^2
\over 2 \spa{5}.{ 6} \spa{6}.{ 1} \spab{1}.{ P_{56}}.{ 4} (P_{23}^2)
(P_{561}^2) } \ P_2\Big[ |4], L_1^{II:C_{23}},L_2^{II:C_{23}} \Big],
}}
with \eqn\name{\eqalign{ L_1^{II:C_{23}} &=  \{P_{561}|1\ra,
P_{561}|1\ra, P_{561}|1\ra, P_{561}|1\ra \} \cr L_2^{II:C_{23}} &=
\{P_{561}|5\ra, P_{23}|4\ra,  P_{23}|4\ra, |2] \}. }}

\subsubsec{The term $C_{23}^{(1,r,3)} $}

i) $t$-integrated formula: \eqn\name{\eqalign{ C_{23}^{(1,r,3)} &= {
{\dea \deb \spa{2}.{ \ell}^2 \spab{1}.{ P_{56}}.{ \ell}^4 \spb{2}.{
3}^3 \spb{3}.{ \ell}} \over
 {\spa{5}.{ 6} \spa{6}.{ 1} \spab{1}.{ P_{56}}.{ 4} \spab{5}.{ P_{61}}.{ \ell}
  \spab{\ell}.{ P_{23}}.{ \ell}^4 \spb{2}.{ \ell} \spb{4}.{ \ell} (P_{561}^2)}
} }}

ii) full derivative: \eqn\cXXIIIderIII{\eqalign{ C_{23}^{(1,r,3)} &=
{ {\deb \dedea  \spa{2}.{ \ell}^3 \spab{1}.{ P_{56}}.{ \ell}^4
\spb{2}.{ 3}^3} \over
 {3 \spa{2}.{ 3} \spa{5}.{ 6} \spa{6}.{ 1} \spab{1}.{ P_{56}}.{ 4} \spab{5}.{ P_{61}}.{ \ell}
  \spab{\ell}.{ P_{23}}.{ \ell}^3 \spb{2}.{ \ell} \spb{4}.{ \ell} (P_{561}^2)}
} }} There are only simple poles. Therefore,
\eqn\cXXIIIresIIIs{\eqalign{ C_{23}^{(1,r,3:s)} &= {\spb{2}.{ 3}^3
\over 3 \spa{2}.{ 3} \spa{5}.{ 6} \spa{6}.{ 1} \spab{1}.{ P_{56}}.{
4} (P_{561}^2)} \ \sum_{i=2}^{4} \lim_{\ell \to \ell_i}
\spb{\ell}.{\ell_i} \ { {\spa{2}.{ \ell}^3 \spab{1}.{ P_{56}}.{
\ell}^4 } \over
 {   \spab{5}.{ P_{61}}.{ \ell}
  \spab{\ell}.{ P_{23}}.{ \ell}^3 \spb{2}.{ \ell} \spb{4}.{ \ell} }
}. }}

\subsubsec{The term $C_{23}^{(1,r,4)} $}

i) $t$-integrated formula: \eqn\name{\eqalign{ C_{23}^{(1,r,4)} &=
  - { {{\dea \deb \spa{2}.{ 3} \spa{4}.{ 1}^2 \spab{4}.{ P_{23}}.{ \ell} \spb{3}.{ \ell}^3 \spb{6}.{ 1}
    \spb{6}.{ \ell}^2} \over {\spa{4}.{ 5} \spab{5}.{ P_{61}}.{ \ell} \spab{\ell}.{ P_{23}}.{ \ell}^2
    \spb{1}.{ \ell} \spb{2}.{ \ell} \spbb{\ell}.{ P_{23}}.{ Q}.{ \ell} \spb{\ell}.{\omega_6}}}  }
\cr  &  - {  {2 \dea \deb \spa{4}.{ 1} \spab{2}.{ Q}.{ \ell}
\spab{4}.{ P_{23}}.{ \ell}^2
   \spb{3}.{ \ell}^2 \spb{6}.{ \ell}^3 (P_{23}^2)} \over {\spa{4}.{ 5} \spab{5}.{ P_{61}}.{ \ell}
   \spab{\ell}.{ P_{23}}.{ \ell}^2 \spb{1}.{ \ell} \spb{2}.{ \ell} \spbb{\ell}.{ P_{23}}.{ Q}.{ \ell}^2
   \spb{\ell}.{\omega_6}}  } \cr  &  - {
 {\dea \deb \spab{2}.{ Q}.{ \ell} \spab{4}.{ Q}.{ \ell} \spab{4}.{ P_{23}}.{ \ell}^2 \spb{3}.{ \ell}^2
   \spb{6}.{ \ell}^4 (P_{23}^2)^2} \over {\spa{4}.{ 5} \spab{5}.{ P_{61}}.{ \ell}
   \spab{\ell}.{ P_{23}}.{ \ell}^2 \spb{1}.{ \ell} \spb{2}.{ \ell} \spb{6}.{ 1}
   \spbb{\ell}.{ P_{23}}.{ Q}.{ \ell}^3 \spb{\ell}.{\omega_6}}
} }}

ii) full derivative: \eqn\name{\eqalign{ C_{23}^{(1,r,4)} = \deb
\dedea \ {\cal I}^{(4)} }}

\eqn\cXXIIIderIV{\eqalign{
{\cal I}^{(4)} &=
  - { {{   \spa{2}.{ \ell} \spa{4}.{ 1}^2 \spab{4}.{ P_{23}}.{ \ell} \spb{3}.{ \ell}^2 \spb{6}.{ 1}
    \spb{6}.{ \ell}^2 \spb{a}.{ b}^2} \over {\spa{4}.{ 5} \spab{5}.{ P_{61}}.{ \ell}
    \spab{\ell}.{ P_{23}}.{ \ell} \spb{1}.{ \ell} \spb{2}.{ \ell} \spb{\eta_1}.{ \ell} \spb{\eta_2}.{ \ell}
    \spbb{b}.{ P_{23}}.{ Q}.{ b} \spb{\ell}.{\omega_6}}}  } \cr  &  - {
 {2    \spa{4}.{ 1} \spa{\eta_1}.{ \ell} \spab{2}.{ Q}.{ \ell} \spab{4}.{ P_{23}}.{ \ell}^2
   \spb{6}.{ \ell}^3 \spb{a}.{ b}^4 \spb{\eta_1}.{ 3}^2 (P_{23}^2)} \over
  {\spa{4}.{ 5} \spab{5}.{ P_{61}}.{ \ell} \spab{\ell}.{ P_{23}}.{ \ell} \spab{\eta_1}.{ P_{23}}.{ \ell}
   \spb{1}.{ \ell} \spb{2}.{ \ell} \spb{\eta_1}.{ \ell}^2 \spb{\eta_1}.{ \eta_2}^2
   \spbb{b}.{ P_{23}}.{ Q}.{ b}^2 \spb{\ell}.{\omega_6}}  } \cr  &  - {
 {2    \spa{4}.{ 1} \spa{\eta_2}.{ \ell} \spab{2}.{ Q}.{ \ell} \spab{4}.{ P_{23}}.{ \ell}^2
   \spb{3}.{ \eta_2}^2 \spb{6}.{ \ell}^3 \spb{a}.{ b}^4 (P_{23}^2)} \over
  {\spa{4}.{ 5} \spab{5}.{ P_{61}}.{ \ell} \spab{\ell}.{ P_{23}}.{ \ell} \spab{\eta_2}.{ P_{23}}.{ \ell}
   \spb{1}.{ \ell} \spb{2}.{ \ell} \spb{\eta_1}.{ \eta_2}^2 \spb{\eta_2}.{ \ell}^2
   \spbb{b}.{ P_{23}}.{ Q}.{ b}^2 \spb{\ell}.{\omega_6}}  } \cr  &  - {
 {4    \spa{1}.{ \ell} \spa{4}.{ 1} \spab{2}.{ Q}.{ \ell} \spab{4}.{ P_{23}}.{ \ell}^2
   \spb{3}.{ \eta_2} \spb{6}.{ \ell}^3 \spb{a}.{ b}^4 \spb{\eta_1}.{ 3} (P_{23}^2)} \over
  {\spa{4}.{ 5} \spab{1}.{ P_{23}}.{ \ell} \spab{5}.{ P_{61}}.{ \ell} \spab{\ell}.{ P_{23}}.{ \ell}
   \spb{1}.{ \ell} \spb{2}.{ \ell} \spb{\eta_1}.{ \ell} \spb{\eta_1}.{ \eta_2}^2 \spb{\eta_2}.{ \ell}
   \spbb{b}.{ P_{23}}.{ Q}.{ b}^2 \spb{\ell}.{\omega_6}}  } \cr  &  - {
 {   \spa{\eta_1}.{ \ell} \spab{2}.{ Q}.{ \ell} \spab{4}.{ Q}.{ \ell} \spab{4}.{ P_{23}}.{ \ell}^2
   \spb{3}.{ \ell}^2 \spb{6}.{ \ell} \spb{a}.{ b}^6 \spb{\eta_1}.{ 6}^3 (P_{23}^2)^2} \over
  {\spa{4}.{ 5} \spab{5}.{ P_{61}}.{ \ell} \spab{\ell}.{ P_{23}}.{ \ell} \spab{\eta_1}.{ P_{23}}.{ \ell}
   \spb{1}.{ \ell} \spb{2}.{ \ell} \spb{6}.{ 1} \spb{\eta_1}.{ \ell}^3 \spb{\eta_1}.{ \eta_2}^3
   \spbb{b}.{ P_{23}}.{ Q}.{ b}^3 \spb{\ell}.{\omega_6}}  } \cr  &  - {
 {   \spa{\eta_2}.{ \ell} \spab{2}.{ Q}.{ \ell} \spab{4}.{ Q}.{ \ell} \spab{4}.{ P_{23}}.{ \ell}^2
   \spb{3}.{ \ell}^2 \spb{6}.{ \ell} \spb{6}.{ \eta_2}^3 \spb{a}.{ b}^6 (P_{23}^2)^2} \over
  {\spa{4}.{ 5} \spab{5}.{ P_{61}}.{ \ell} \spab{\ell}.{ P_{23}}.{ \ell} \spab{\eta_2}.{ P_{23}}.{ \ell}
   \spb{1}.{ \ell} \spb{2}.{ \ell} \spb{6}.{ 1} \spb{\eta_1}.{ \eta_2}^3 \spb{\eta_2}.{ \ell}^3
   \spbb{b}.{ P_{23}}.{ Q}.{ b}^3 \spb{\ell}.{\omega_6}}  } \cr  &  - {
 {3    \spa{\eta_2}.{ \ell} \spab{2}.{ Q}.{ \ell} \spab{4}.{ Q}.{ \ell} \spab{4}.{ P_{23}}.{ \ell}^2
   \spb{3}.{ \ell}^2 \spb{6}.{ \ell} \spb{6}.{ \eta_2}^2 \spb{a}.{ b}^6 \spb{\eta_1}.{ 6}
   (P_{23}^2)^2} \over {\spa{4}.{ 5} \spab{5}.{ P_{61}}.{ \ell} \spab{\ell}.{ P_{23}}.{ \ell}
   \spab{\eta_2}.{ P_{23}}.{ \ell} \spb{1}.{ \ell} \spb{2}.{ \ell} \spb{6}.{ 1} \spb{\eta_1}.{ \ell}
   \spb{\eta_1}.{ \eta_2}^3 \spb{\eta_2}.{ \ell}^2 \spbb{b}.{ P_{23}}.{ Q}.{ b}^3
   \spb{\ell}.{\omega_6}}  } \cr  &  - {
 {3    \spa{\eta_1}.{ \ell} \spab{2}.{ Q}.{ \ell} \spab{4}.{ Q}.{ \ell} \spab{4}.{ P_{23}}.{ \ell}^2
   \spb{3}.{ \ell}^2 \spb{6}.{ \ell} \spb{6}.{ \eta_2} \spb{a}.{ b}^6 \spb{\eta_1}.{ 6}^2
   (P_{23}^2)^2} \over {\spa{4}.{ 5} \spab{5}.{ P_{61}}.{ \ell} \spab{\ell}.{ P_{23}}.{ \ell}
   \spab{\eta_1}.{ P_{23}}.{ \ell} \spb{1}.{ \ell} \spb{2}.{ \ell} \spb{6}.{ 1} \spb{\eta_1}.{ \ell}^2
   \spb{\eta_1}.{ \eta_2}^3 \spb{\eta_2}.{ \ell} \spbb{b}.{ P_{23}}.{ Q}.{ b}^3
   \spb{\ell}.{\omega_6}}
}. }}

\subsubsec{The term $C_{23}^{(1,r,5)} $}

i) $t$-integrated formula: \eqn\name{\eqalign{ C_{23}^{(1,r,5)} &=
 - 2 {{ \dea \deb \spa{2}.{ 3} \spa{4}.{ 1} \spa{4}.{ \ell} \spab{4}.{ P_{23}}.{ \ell} \spb{3}.{ \ell}^3
   \spb{6}.{ \ell}^3 (P_{23}^2)} \over {\spa{4}.{ 5} \spab{5}.{ P_{61}}.{ \ell}
   \spab{\ell}.{ P_{23}}.{ \ell}^3 \spb{1}.{ \ell} \spb{2}.{ \ell} \spbb{\ell}.{ P_{23}}.{ Q}.{ \ell}
   \spb{\ell}.{\omega_6}}  } \cr  &  - {
 {\dea \deb \spa{4}.{ \ell} \spab{2}.{ Q}.{ \ell} \spab{4}.{ P_{23}}.{ \ell}^2 \spb{3}.{ \ell}^2
   \spb{6}.{ \ell}^4 (P_{23}^2)^2} \over {\spa{4}.{ 5} \spab{5}.{ P_{61}}.{ \ell}
   \spab{\ell}.{ P_{23}}.{ \ell}^3 \spb{1}.{ \ell} \spb{2}.{ \ell} \spb{6}.{ 1}
   \spbb{\ell}.{ P_{23}}.{ Q}.{ \ell}^2 \spb{\ell}.{\omega_6}}  } \cr  &  + {
 {\dea \deb \spa{1}.{ \ell} \spa{2}.{ 1} \spab{1}.{ P_{23}}.{ \ell} \spb{3}.{ \ell}^2 \spb{5}.{ 6}^3
   (P_{23}^2)} \over {\spab{1}.{ P_{456}}.{ 4} \spab{\ell}.{ P_{23}}.{ \ell}^3 \spb{2}.{ \ell}
   \spb{4}.{ 5} \spb{\ell}.{\omega_6} (P_{123}^2)}
} }}

ii) full derivative:
\eqn\name{\eqalign{ C_{23}^{(1,r,5)} =
\deb \dedea \ {\cal I}^{(5)} }} \eqn\cXXIIIderV{\eqalign{
{\cal I}^{(5)} &=
  - { {{   \spa{2}.{ 3} \spa{4}.{ 1} \spa{4}.{ \ell}^2 \spb{3}.{ \ell}^3 \spb{6}.{ \ell}^3 \spb{a}.{ b}^2
    (P_{23}^2)} \over {\spa{4}.{ 5} \spab{5}.{ P_{61}}.{ \ell} \spab{\ell}.{ P_{23}}.{ \ell}^2
    \spb{1}.{ \ell} \spb{2}.{ \ell} \spb{\eta_1}.{ \ell} \spb{\eta_2}.{ \ell} \spbb{b}.{ P_{23}}.{ Q}.{ b}
    \spb{\ell}.{\omega_6}}}  } \cr  &  - {
 {   \spa{\eta_1}.{ \ell}^2 \spab{2}.{ Q}.{ \ell} \spab{4}.{ P_{23}}.{ \ell}^3 \spb{6}.{ \ell}^4
   \spb{a}.{ b}^4 \spb{\eta_1}.{ 3}^2 (P_{23}^2)^2} \over {2 \spa{4}.{ 5} \spab{5}.{ P_{61}}.{ \ell}
   \spab{\ell}.{ P_{23}}.{ \ell}^2 \spab{\eta_1}.{ P_{23}}.{ \ell}^2 \spb{1}.{ \ell} \spb{2}.{ \ell}
   \spb{6}.{ 1} \spb{\eta_1}.{ \ell}^2 \spb{\eta_1}.{ \eta_2}^2 \spbb{b}.{ P_{23}}.{ Q}.{ b}^2
   \spb{\ell}.{\omega_6}}  } \cr  &  - {
 {   \spa{4}.{ \eta_1} \spa{\eta_1}.{ \ell} \spab{2}.{ Q}.{ \ell} \spab{4}.{ P_{23}}.{ \ell}^2
   \spb{6}.{ \ell}^4 \spb{a}.{ b}^4 \spb{\eta_1}.{ 3}^2 (P_{23}^2)^2} \over
  {\spa{4}.{ 5} \spab{5}.{ P_{61}}.{ \ell} \spab{\ell}.{ P_{23}}.{ \ell}
   \spab{\eta_1}.{ P_{23}}.{ \ell}^2 \spb{1}.{ \ell} \spb{2}.{ \ell} \spb{6}.{ 1} \spb{\eta_1}.{ \ell}^2
   \spb{\eta_1}.{ \eta_2}^2 \spbb{b}.{ P_{23}}.{ Q}.{ b}^2 \spb{\ell}.{\omega_6}  }} \cr  &  - {  {   \spa{\eta_2}.{ \ell}^2 \spab{2}.{ Q}.{ \ell} \spab{4}.{ P_{23}}.{ \ell}^3
   \spb{3}.{ \eta_2}^2 \spb{6}.{ \ell}^4 \spb{a}.{ b}^4 (P_{23}^2)^2} \over
  {2 \spa{4}.{ 5} \spab{5}.{ P_{61}}.{ \ell} \spab{\ell}.{ P_{23}}.{ \ell}^2
   \spab{\eta_2}.{ P_{23}}.{ \ell}^2 \spb{1}.{ \ell} \spb{2}.{ \ell} \spb{6}.{ 1} \spb{\eta_1}.{ \eta_2}^2
   \spb{\eta_2}.{ \ell}^2 \spbb{b}.{ P_{23}}.{ Q}.{ b}^2 \spb{\ell}.{\omega_6}}  } \cr  &  - {
 {   \spa{4}.{ \eta_2} \spa{\eta_2}.{ \ell} \spab{2}.{ Q}.{ \ell} \spab{4}.{ P_{23}}.{ \ell}^2
   \spb{3}.{ \eta_2}^2 \spb{6}.{ \ell}^4 \spb{a}.{ b}^4 (P_{23}^2)^2} \over
  {\spa{4}.{ 5} \spab{5}.{ P_{61}}.{ \ell} \spab{\ell}.{ P_{23}}.{ \ell}
   \spab{\eta_2}.{ P_{23}}.{ \ell}^2 \spb{1}.{ \ell} \spb{2}.{ \ell} \spb{6}.{ 1} \spb{\eta_1}.{ \eta_2}^2
   \spb{\eta_2}.{ \ell}^2 \spbb{b}.{ P_{23}}.{ Q}.{ b}^2 \spb{\ell}.{\omega_6}}  } \cr  &  - {
 {   \spa{4}.{ \ell}^2 \spab{2}.{ Q}.{ \ell} \spab{4}.{ P_{23}}.{ \ell} \spb{3}.{ \eta_2}
   \spb{6}.{ \ell}^4 \spb{a}.{ b}^4 \spb{\eta_1}.{ 3} (P_{23}^2)^2} \over
  {\spa{4}.{ 5} \spab{5}.{ P_{61}}.{ \ell} \spab{\ell}.{ P_{23}}.{ \ell}^2 \spb{1}.{ \ell} \spb{2}.{ \ell}
   \spb{6}.{ 1} \spb{\eta_1}.{ \ell} \spb{\eta_1}.{ \eta_2}^2 \spb{\eta_2}.{ \ell}
   \spbb{b}.{ P_{23}}.{ Q}.{ b}^2 \spb{\ell}.{\omega_6}}  } \cr  &  + {
 {   \spa{1}.{ \ell}^2 \spa{2}.{ 1} \spb{3}.{ \ell}^2 \spb{5}.{ 6}^3 (P_{23}^2)} \over
  {2 \spab{1}.{ P_{456}}.{ 4} \spab{\ell}.{ P_{23}}.{ \ell}^2 \spb{2}.{ \ell} \spb{4}.{ 5}
   \spb{\ell}.{\omega_6} (P_{123}^2)}
}. }}

The above expression contains single and double poles.
The residues of the single poles will be read off later; for the total
rational contribution, we consider here only the terms having double
poles, $\spab{\eta_1}.{ P_{23}}.{ \ell}^2$ and $\spab{\eta_2}.{ P_{23}}.{ \ell}^2$:

\eqn\name{\eqalign{ C_{23}^{(1,r,5:d)} = \deb \dedea \ {\cal I}^{(5:d)} }}
\eqn\cXXIIIderV{\eqalign{
{\cal I}^{(5:d)} =&
- {
 {   \spa{\eta_1}.{ \ell}^2 \spab{2}.{ Q}.{ \ell} \spab{4}.{ P_{23}}.{ \ell}^3 \spb{6}.{ \ell}^4
   \spb{a}.{ b}^4 \spb{\eta_1}.{ 3}^2 (P_{23}^2)^2} \over {2 \spa{4}.{ 5} \spab{5}.{ P_{61}}.{ \ell}
   \spab{\ell}.{ P_{23}}.{ \ell}^2 \spab{\eta_1}.{ P_{23}}.{ \ell}^2 \spb{1}.{ \ell} \spb{2}.{ \ell}
   \spb{6}.{ 1} \spb{\eta_1}.{ \ell}^2 \spb{\eta_1}.{ \eta_2}^2 \spbb{b}.{ P_{23}}.{ Q}.{ b}^2
   \spb{\ell}.{\omega_6}}  } \cr  &  - {
 {   \spa{4}.{ \eta_1} \spa{\eta_1}.{ \ell} \spab{2}.{ Q}.{ \ell} \spab{4}.{ P_{23}}.{ \ell}^2
   \spb{6}.{ \ell}^4 \spb{a}.{ b}^4 \spb{\eta_1}.{ 3}^2 (P_{23}^2)^2} \over
  {\spa{4}.{ 5} \spab{5}.{ P_{61}}.{ \ell} \spab{\ell}.{ P_{23}}.{ \ell}
   \spab{\eta_1}.{ P_{23}}.{ \ell}^2 \spb{1}.{ \ell} \spb{2}.{ \ell} \spb{6}.{ 1} \spb{\eta_1}.{ \ell}^2
   \spb{\eta_1}.{ \eta_2}^2 \spbb{b}.{ P_{23}}.{ Q}.{ b}^2
   \spb{\ell}.{\omega_6}  }} \cr
& + \{\eta_1 \to \eta_2 \}
} }

The residues in this case give

\eqn\cXXIIIresVd{\eqalign{ C_{23}^{(1,r,5:d)} =&
{ \spb{a}.{ b}^4  \spb{\eta_1}.{ 3}^2
  \spab{\eta_1}.{P_{23}}.{ \eta_1}^2
  \over
    2 \spa{4}.{ 5}
    \spb{6}.{ 1}
    \spb{\eta_1}.{ \eta_2}^2
    \spbb{b}.{ P_{23}}.{ Q}.{ b}^2
}
\ {\tilde P}_2 \Big[P_{23}|\eta_1\ra ,L_1^{II:C_{23}},L_2^{II:C_{23}} \Big] \cr
&
+
{
   \spa{4}.{ \eta_1}
   \spb{a}.{ b}^4
   \spb{\eta_1}.{ 3}^2
   \spab{\eta_1}.{P_{23}}.{ \eta_1}
  (P_{23}^2)
\over
   \spa{4}.{ 5}
   \spb{6}.{ 1}
   \spb{\eta_1}.{ \eta_2}^2
   \spbb{b}.{ P_{23}}.{ Q}.{ b}^2
}
\ {\tilde P}_2 \Big[P_{23}|\eta_1\ra ,M_1^{II:C_{23}},M_2^{II:C_{23}} \Big] \cr
& + \{\eta_1 \leftrightarrow \eta_2 \},
} }
with \eqn\name{\eqalign{
L_1^{II:C_{23}} &=  \{Q|2\ra, P_{23}|4\ra,P_{23}|4\ra,P_{23}|4\ra,|6],|6],|6],|6] \} \cr
L_2^{II:C_{23}} &=  \{P_{61}|5\ra, |1], |2],|\eta_1],|\eta_1],|\eta_1],|\eta_1], |\omega_6] \} \cr
M_1^{II:C_{23}} &=  \{Q|2\ra,P_{23}|4\ra,P_{23}|4\ra,|6],|6],|6],|6] \} \cr
M_2^{II:C_{23}} &=  \{P_{61}|5\ra, |1], |2] ,|\eta_1],|\eta_1],|\eta_1], |\omega_6] \}
}}
since we used $|\ell\ra = P_{23}|\eta_1]$.

\subsubsec{The term $C_{23}^{(1,r,6)} $}

i) $t$-integrated formula: \eqn\name{\eqalign{ C_{23}^{(1,r,6)} &=
 - { {{\dea \deb \spa{2}.{ 3} \spa{4}.{ \ell}^2 \spab{4}.{ P_{23}}.{ \ell} \spb{3}.{ \ell}^3
    \spb{6}.{ \ell}^4 (P_{23}^2)^2} \over {\spa{4}.{ 5} \spab{5}.{ P_{61}}.{ \ell}
    \spab{\ell}.{ P_{23}}.{ \ell}^4 \spb{1}.{ \ell} \spb{2}.{ \ell} \spb{6}.{ 1}
    \spbb{\ell}.{ P_{23}}.{ Q}.{ \ell} \spb{\ell}.{\omega_6}}}  } \cr  &  + {
 {\dea \deb \spa{1}.{ \ell}^2 \spa{2}.{ 3} \spab{1}.{ P_{23}}.{ \ell} \spb{3}.{ \ell}^3 \spb{5}.{ 6}^3
   (P_{23}^2)} \over {\spab{1}.{ P_{456}}.{ 4} \spab{\ell}.{ P_{23}}.{ \ell}^4 \spb{2}.{ \ell}
   \spb{4}.{ 5} \spb{\ell}.{\omega_6} (P_{123}^2)}
} }}

ii) full derivative: \eqn\name{\eqalign{ C_{23}^{(1,r,6)} = \deb
\dedea \ {\cal I}^{(6)} }} \eqn\cXXIIIderVI{\eqalign{
{\cal I}^{(6)} &=
  - { {  \spa{2}.{ 3} \spa{4}.{ \ell}^3 \spb{3}.{ \ell}^3 \spb{6}.{ \ell}^4 \spb{a}.{ b}^2
    (P_{23}^2)^2} \over {3 \spa{4}.{ 5} \spab{5}.{ P_{61}}.{ \ell} \spab{\ell}.{ P_{23}}.{ \ell}^3
   \spb{1}.{ \ell} \spb{2}.{ \ell} \spb{6}.{ 1} \spb{\eta_1}.{ \ell} \spb{\eta_2}.{ \ell}
   \spbb{b}.{ P_{23}}.{ Q}.{ b} \spb{\ell}.{\omega_6}}  } \cr  &  + {
 {  \spa{1}.{ \ell}^3 \spa{2}.{ 3} \spb{3}.{ \ell}^3 \spb{5}.{ 6}^3 (P_{23}^2)} \over
  {3 \spab{1}.{ P_{456}}.{ 4} \spab{\ell}.{ P_{23}}.{ \ell}^3 \spb{2}.{ \ell} \spb{4}.{ 5}
   \spb{\ell}.{\omega_6} (P_{123}^2)}
} }}

Given the expressions of $C_{23}^{(1,r,4)}, C_{23}^{(1,r,5)}$ and
$C_{23}^{(1,r,6)}$, in eqs.\cXXIIIderIV, \cXXIIIderV, and
\cXXIIIderVI, their combined contribution can be written as,
\eqn\name{\eqalign{ C_{23}^{(1,r,4,5,6)} &= \deb \dedea \Big\{
{\cal{I}}^{(4)} + {\cal{I}}^{(5)} + {\cal{I}}^{(6)} \Big\} }}
Therefore, the sum of residues of their single poles will
give
\eqn\cXXIIIresCOMBO{\eqalign{ C_{23}^{(1,r,4,5,6:s)} &= \
\sum_{j=1}^{10} \lim_{\ell \to \ell_j} \spb{\ell}.{\ell_j} \Big\{
{\cal{I}}^{(4)} + {\cal{I}}^{(5)} + {\cal{I}}^{(6)} \Big\},
}}
with \eqn\name{\eqalign{ |\ell_j] =  P_{61}|5\ra, |1], |2], |4],
|\eta_1], |\eta_2], |\omega_6], P_{23}|1 \ra, P_{23}|\eta_1 \ra,
P_{23}|\eta_2 \ra \qquad (j=1,\ldots,10). }}

The coefficient of the bubble $I_{2:2;2}$ is given by adding eqs.
\cXXIIIresI, \cXXIIIresIIs, \cXXIIIresIId, \cXXIIIresIIIs, \cXXIIIresCOMBO,
\cXXIIIresVd, and multiplying by 2:
\eqn\name{\eqalign{ c_{2:2;2}
= 2 \Big(
              C_{23}^{(1,r,1:s)}
            + C_{23}^{(1,r,2:s)}
            + C_{23}^{(1,r,2:d)}
            + C_{23}^{(1,r,3:s)}
            + C_{23}^{(1,r,3,4,5,6:s)}
            + C_{23}^{(1,r,5:d)}
          \Big).
}}

\subsubsec {{\bf 3-Mass-Triangle contribution from $C_{23}^{(1)}$}}

After the $t$-integration the contribution to the
three-mass-triangle coefficient reads: \eqn\name{\eqalign{
C_{23}^{(1,3m)} &=
                       C_{23}^{(1,3m,1)}
                     + C_{23}^{(1,3m,2)}
                     + C_{23}^{(1,3m,3)}.
}} Unlike in the case presented in Section 4, here we make use of the functions defined
in Appendix C to write down answers directly and compactly.

\subsubsec{The term $C_{23}^{(1,3m,1)} $}

i) after $t$-integration \eqn\name{\eqalign{ C_{23}^{(1,3m,1)} &= -
{ \spa{4}.{ 1}^2 \spb{6}.{ 1} \over \spa{4}.{ 5} } {\dea \deb \
{\spab{2}.{ Q}.{ \ell} \spab{4}.{ P_{23}}.{ \ell} \spb{3}.{ \ell}^2
   \spb{6}.{ \ell}^2} \over
{
  \spab{\ell}.{ Q}.{ \ell}
  \spab{\ell}.{ P_{23}}.{ \ell} \
  \spbb{\ell}.{ P_{23}}.{ Q}.{ \ell} \
  \spab{5}.{ P_{61}}.{ \ell}
  \spb{1}.{ \ell} \spb{2}.{ \ell}
  \spb{\omega_6}.{\ell}
} } }}

ii) triangle coefficient:
 \eqn\cXXIIImmmI{\eqalign{
C_{23}^{(1,3m,1)} &= - { \spa{4}.{ 1}^2 \spb{6}.{ 1} \over \spa{4}.{
5} } \ \tilde{C}_3^{II}[L_a,L_b^{II},P_{23},Q] \cr }}
\eqn\name{\eqalign{ L_a &= \{ Q|2\ra, P_{23}|4\ra, |3],|3],|6],|6]
       \} \cr
L_b^{II} &= \{ P_{61}|5\ra, |1], |2], |\omega_6], |\eta]
           \}
}}

\subsubsec{The term $C_{23}^{(1,3m,2)} $}

i) after $t$-integration \eqn\name{\eqalign{ C_{23}^{(1,3m,2)} &= -
{2 \spa{4}.{ 1} (P_{23}^2) \over \spa{4}.{ 5}} { \dea \deb \ {
\spab{2}.{ Q}.{ \ell} \spab{4}.{ Q}.{ \ell} \spab{4}.{ P_{23}}.{
\ell}
  \spb{3}.{ \ell}^2 \spb{6}.{ \ell}^3 }
\over {
  \spab{\ell}.{ Q}.{ \ell} \spab{\ell}.{ P_{23}}.{ \ell} \
  \spbb{\ell}.{ P_{23}}.{ Q}.{ \ell}^2 \
  \spab{5}.{ P_{61}}.{ \ell}
  \spb{1}.{ \ell} \spb{2}.{ \ell}
  \spb{\omega_6}.{\ell}
} } }}

ii) triangle coefficient:
 \eqn\cXXIIImmmII{\eqalign{
C_{23}^{(1,3m,2)} &= - {2 \spa{4}.{ 1} (P_{23}^2) \over \spa{4}.{
5}} \ \tilde{C}_3^{III}[L_a,L_{b,1}^{III},L_{b,2}^{III},P_{23},Q]
\cr }}
\eqn\name{\eqalign{ L_a &= \{ Q|2\ra, Q|4\ra, P_{23}|4\ra,
|3],|3],|6],|6],|6]
       \} \cr
L_{b,1}^{III} &= \{ P_{61}|5\ra, |1], |2], |\omega_6],
                   |\eta], |\eta_2], |\eta_2]
           \} \cr
L_{b,2}^{III} &= \{ P_{61}|5\ra, |1], |2], |\omega_6],
                   |\eta], |\eta_1], |\eta_1]
           \} \cr
}}

\subsubsec{The term $C_{23}^{(1,3m,3)} $}

i) after $t$-integration \eqn\name{\eqalign{ C_{23}^{(1,3m,3)} &= -
{ (P_{23}^2)^2 \over \spa{4}.{ 5} \spb{6}.{ 1} } {\dea \deb \ {
\spab{2}.{ Q}.{ \ell} \spab{4}.{ Q}.{ \ell}^2 \spab{4}.{ P_{23}}.{
\ell} \spb{3}.{ \ell}^2
  \spb{6}.{ \ell}^4 } \over
{
  \spab{\ell}.{ Q}.{ \ell}
  \spab{\ell}.{ P_{23}}.{ \ell} \
  \spbb{\ell}.{ P_{23}}.{ Q}.{ \ell}^3 \
  \spab{5}.{ P_{61}}.{ \ell}
  \spb{1}.{ \ell} \spb{2}.{ \ell}
  \spb{\omega_6}.{\ell}
} } }}

ii) triangle coefficient:
 \eqn\cXXIIImmmIII{\eqalign{
C_{23}^{(1,3m,3)} &= - { (P_{23}^2)^2 \over \spa{4}.{ 5} \spb{6}.{
1} } \tilde{C}_3^{IV}[L_a,L_{b,1}^{IV},L_{b,2}^{IV},P_{23},Q] \cr }}
\eqn\name{\eqalign{ L_a &= \{ Q|2\ra, Q|4\ra, Q|4\ra, P_{23}|4\ra,
|3],|3],|6],|6],|6],|6]
       \} \cr
L_{b,1}^{IV} &= \{ P_{61}|5\ra, |1], |2], |\omega_6],
                   |\eta], |\eta_2], |\eta_2], |\eta_2]
           \} \cr
L_{b,2}^{IV} &= \{ P_{61}|5\ra, |1], |2], |\omega_6],
                   |\eta], |\eta_1], |\eta_1], |\eta_1]
           \} \cr
}}

\bigskip

Finally, the coefficient of the thee-mass triangle $I_{3:2:2;2}$ can
be obtained by taking the sum of \cXXIIImmmI, \cXXIIImmmII, and
\cXXIIImmmIII\ and multiplying the result by 2: \eqn\name{\eqalign{
c_{3:2:2;2} = 2 \Big(
               C_{23}^{(1,3m,1)}
             + C_{23}^{(1,3m,2)}
             + C_{23}^{(1,3m,3)}
          \Big).
}}


\subsec{Cut $C_{61}$}

This cut has contributions from three-mass triangle, so the result
will be more complicated. Also since it is the same three-mass
triangle function as in cut $C_{23}$, we can use it as an
independent check for this coefficient.

The cut is given by  \eqn\name{\eqalign{ C_{61} &=  \int d\mu \
\Big[
                   A(\ell_1^+, 6^+, 1^-, \ell_2^-)
                   A(\ell_2^+, 2^-, 3^+, 4^-, 5^+, \ell_1^-)
  \cr & \qquad \qquad +  A(\ell_1^-, 6^+, 1^-, \ell_2^+)
                   A(\ell_2^-, 2^-, 3^+, 4^-, 5^+, \ell_1^+)
                \Big]
\cr &=
 - {{ 2 \spa{1}.{\ell_1}^2 \spa{1}.{\ell_2} \spa{4}.{ 2}^4 \spb{5}.{\ell_1} \spb{5}.{\ell_2}^2} \over
  {\spa{2}.{ 3} \spa{3}.{ 4} \spa{6}.{ 1} \spa{6}.{\ell_1} \spa{\ell_2}.{\ell_1} \spab{2}.{ P_{234}}.{ 5}
   \spab{4}.{ P_{234}}.{\ell_2} \spb{\ell_1}.{\ell_2} (P_{234}^2)}  } \cr  &  + {
 {2 \spa{1}.{\ell_1}^2 \spa{1}.{\ell_2} \spa{2}.{\ell_1}^2 \spa{2}.{\ell_2} \spb{3}.{ 5}^4} \over
  {\spa{6}.{ 1} \spa{6}.{\ell_1} \spa{\ell_1}.{\ell_2} \spa{\ell_2}.{\ell_1}
    \spab{2}.{ P_{\ell_1 \ell_2  2}}.{ 5}
   \spab{\ell_1}.{ P_{\ell_1 \ell_2  2}}.{ 3} \spb{3}.{ 4} \spb{4}.{ 5}
   (P_{\ell_1 \ell_2  2}^2)}  } \cr  &  - {
 {2 \spa{1}.{\ell_1}^2 \spa{1}.{\ell_2} \spa{4}.{\ell_1}^2 \spab{4}.{P_{\ell_2 23}}.{ 3}^2
   \spb{\ell_2}.{ 3}^2} \over {\spa{4}.{ 5} \spa{5}.{\ell_1} \spa{6}.{ 1} \spa{6}.{\ell_1} \spa{\ell_2}.{\ell_1}
   \spab{4}.{ P_{\ell_2 2 3}}.{\ell_2} \spab{\ell_1}.{ P_{\ell_2  2 3}}.{ 3} \spb{2}.{ 3} \spb{\ell_2}.{ 2}
   (P_{\ell_2  2  3}^2)}
} }}

In the following formulas,  {\sl for this cut only}, we define
$$| \omega_5 \ra = P_{61} P_{23} |4 \ra$$
and $$Q=  (P_{23}^2/P_{61}^2) \ P_{61} + P_{23}.$$

\subsubsec {{\bf Rational contribution from $C_{61}$}}

\eqn\name{\eqalign{ C_{61}^{rat} &=
                      C_{61}^{(r,1)}
                    + C_{61}^{(r,2)}
                    + C_{61}^{(r,3)}
                    + C_{61}^{(r,4)}
                    + C_{61}^{(r,5)}.
}}

\subsubsec{The term $C_{61}^{(r,1)} $}

i) $t$-integrated formula: \eqn\name{\eqalign{ C_{61}^{(r,1)} &= {
{2 \dea \deb \spa{1}.{ \ell}^2 \spa{2}.{ 1} \spa{2}.{ \ell}^2
\spb{1}.{ \ell} \spb{3}.{ 5}^4
  (P_{61}^2)} \over {\spa{6}.{ \ell}^2 \spab{2}.{ P_{612}}.{ 5} \spab{\ell}.{ P_{61}}.{ \ell}^3
  \spab{\ell}.{ P_{612}}.{ 3} \spb{3}.{ 4} \spb{4}.{ 5} (P_{612}^2)}
} }}

We describe in detail how to write this term as a full derivative
to show a technical point
which allows us to get rid of higher poles---when possible.

We have seen in many examples that
a suitable choice of the reference spinor
$\eta$ can simplify writing an integrand as a full derivative, when
using the integration-by-parts identity eq.\speziell.
In the above expression there is a double pole $\spa{6}.{\ell}^2$,
but the presence in the numerator of $\spb{1}.{\ell}$
seems to force us to pick up $\eta=1$. By doing so, one would end up
with an expression containing a triple pole $\spa{\ell}.{6}^3$, to be dealt with afterwards.

Alternatively, one can multiply $C_{61}^{(r,1)}$ by
$1=\spb{6}.{\ell}/\spb{6}.{\ell}$ and use the following identity,

\eqn\name{\eqalign{
{\spb{1}.{ \ell} \over {\spab{\ell}.{ P_{61}}.{ \ell} \spb{6}.{ \ell}}}
=
{1 \over \spab{\ell}.{ P_{61}}.{ 6}} \Bigg(
{\spab{\ell}.{ P_{61}}.{ 1} \over \spab{\ell}.{ P_{61}}.{ \ell}  }
+ { \spb{1}.{ 6} \over \spb{6}.{ \ell}} \Bigg)
}}
to write $C_{61}^{(r,1)}$ as a sum of two terms,
\eqn\name{\eqalign{
C_{61}^{(r,1)} &=
 - {{ 2 \dea \deb \spa{1}.{ \ell} \spa{2}.{ 1} \spa{2}.{ \ell}^2 \spb{3}.{ 5}^4 P_{61}} \over
  {\spa{6}.{ \ell}^2 \spab{2}.{ P_{612}}.{ 5} \spab{\ell}.{ P_{61}}.{ \ell}^2
   \spab{\ell}.{ P_{612}}.{ 3} \spb{3}.{ 4} \spb{4}.{ 5} (P_{612}^2)}  }
\cr  &  + {
 {2 \dea \deb \spa{1}.{ \ell} \spa{2}.{ 1} \spa{2}.{ \ell}^2 \spb{3}.{ 5}^4 \spb{6}.{ 1}
   \spb{6}.{ \ell} P_{61}} \over {\spa{6}.{ \ell} \spab{2}.{ P_{612}}.{ 5}
   \spab{\ell}.{ P_{61}}.{ \ell}^3 \spab{\ell}.{ P_{612}}.{ 3} \spb{1}.{ 6} \spb{3}.{ 4}
   \spb{4}.{ 5} (P_{612}^2)}
}
}}
each of which can be integrated by parts by using $\eta=6$, to neutralize the
double pole.
In fact its expression can be written as,

ii) full derivative: \eqn\name{\eqalign{ C_{61}^{(r,1)} = \dea
\dedeb \ {\cal I}^{(1)} }} \eqn\cLXIderI{\eqalign{
{\cal I}^{(1)} &=
 - {{ 2    \spa{2}.{ 1} \spa{2}.{ \ell}^2 \spb{3}.{ 5}^4 \spb{6}.{ \ell} (P_{61}^2)} \over
  {\spa{6}.{ \ell}^2 \spab{2}.{ P_{612}}.{ 5} \spab{\ell}.{ P_{61}}.{ \ell}
   \spab{\ell}.{ P_{612}}.{ 3} \spb{1}.{ 6} \spb{3}.{ 4} \spb{4}.{ 5} (P_{612}^2)}  } \cr  &  + {
 {  \spa{2}.{ 1} \spa{2}.{ \ell}^2 \spb{3}.{ 5}^4 \spb{6}.{ 1} \spb{6}.{ \ell}^2 (P_{61}^2)} \over
  {\spa{6}.{ \ell} \spab{2}.{ P_{612}}.{ 5} \spab{\ell}.{ P_{61}}.{ \ell}^2
   \spab{\ell}.{ P_{612}}.{ 3} \spb{1}.{ 6}^2 \spb{3}.{ 4} \spb{4}.{ 5} (P_{612}^2)}
} }}
where the term with $1/\spa{6}.{\ell}^2$ 
has a factor in the numerator of 
$\spb{6}.{\ell}$, which annihilates its residue.
Therefore the contribution of this term to the corresponding bubble
coefficients will be given by the sum of residues of only {\it simple} poles.

\subsubsec{The term $C_{61}^{(r,2)} $}

i) $t$-integrated formula: \eqn\name{\eqalign{ C_{61}^{(r,2)} &=
 - {{ 2 \dea \deb \spa{1}.{ \ell}^3 \spa{2}.{ 1} \spa{2}.{ \ell}^2 \spb{1}.{ 6} \spb{1}.{ \ell}^2
   \spb{3}.{ 5}^4 (P_{61}^2)} \over {\spa{6}.{ \ell}^2 \spab{2}.{ P_{612}}.{ 5}
   \spab{\ell}.{ P_{61}}.{ \ell}^4 \spab{\ell}.{ P_{612}}.{ 3} \spb{3}.{ 4} \spb{4}.{ 5}
   \spb{6}.{ 1} (P_{612}^2)}  } \cr  &  - {  {2 \dea \deb \spa{1}.{ \ell}^2 \spa{2}.{ 6} \spa{2}.{ \ell}^2
   \spb{3}.{ 5}^4 \spb{6}.{ \ell}^2 (P_{61}^2)} \over {\spa{6}.{ \ell} \spab{2}.{ P_{612}}.{ 5}
   \spab{\ell}.{ P_{61}}.{ \ell}^4 \spab{\ell}.{ P_{612}}.{ 3} \spb{3}.{ 4} \spb{4}.{ 5}
   (P_{612}^2)}
} }}

ii) full derivative: \eqn\name{\eqalign{ C_{61}^{(r,2)} = \dea
\dedeb \ {\cal I}^{(2)} }} \eqn\cLXIderII{\eqalign{
{\cal I}^{(2)}&= { {2    \spa{2}.{ 1} \spa{2}.{ \ell}^2 \spb{3}.{
5}^4 \spb{6}.{ \ell} (P_{61}^2)} \over
  {\spa{6}.{ \ell}^2 \spab{2}.{ P_{612}}.{ 5} \spab{\ell}.{ P_{61}}.{ \ell}
   \spab{\ell}.{ P_{612}}.{ 3} \spb{3}.{ 4} \spb{4}.{ 5} \spb{6}.{ 1} (P_{612}^2)}  } \cr  &  - {
 {2    \spa{2}.{ 1} \spa{2}.{ \ell}^2 \spb{3}.{ 5}^4 \spb{6}.{ \ell}^2 (P_{61}^2)} \over
  {\spa{6}.{ \ell} \spab{2}.{ P_{612}}.{ 5} \spab{\ell}.{ P_{61}}.{ \ell}^2
   \spab{\ell}.{ P_{612}}.{ 3} \spb{1}.{ 6} \spb{3}.{ 4} \spb{4}.{ 5} (P_{612}^2)}  } \cr  &  - {
 {2    \spa{1}.{ \ell} \spa{2}.{ 6} \spa{2}.{ \ell}^2 \spb{3}.{ 5}^4 \spb{6}.{ \ell}^3 (P_{61}^2)} \over
  {3 \spa{6}.{ \ell} \spab{2}.{ P_{612}}.{ 5} \spab{\ell}.{ P_{61}}.{ \ell}^3
   \spab{\ell}.{ P_{612}}.{ 3} \spb{1}.{ 6} \spb{3}.{ 4} \spb{4}.{ 5} (P_{612}^2)}  } \cr  &  + {
 {2    \spa{2}.{ 1} \spa{2}.{ \ell}^2 \spb{3}.{ 5}^4 \spb{6}.{ 1} \spb{6}.{ \ell}^3 (P_{61}^2)} \over
  {3 \spab{2}.{ P_{612}}.{ 5} \spab{\ell}.{ P_{61}}.{ \ell}^3 \spab{\ell}.{ P_{612}}.{ 3}
   \spb{1}.{ 6}^2 \spb{3}.{ 4} \spb{4}.{ 5} (P_{612}^2)}
} }}

\subsubsec{The term $C_{61}^{(r,3)} $}

i) $t$-integrated formula: \eqn\name{\eqalign{ C_{61}^{(r,3)} &=
 - {{ 2 \dea \deb \spa{1}.{ \ell}^3 \spa{4}.{ 6}^2 \spa{4}.{ \ell}^2 \spab{\ell}.{ Q}.{ 6}^2
   \spab{\ell}.{ P_{61}}.{ 3}^4 \spb{1}.{ 6}} \over {\spa{4}.{ 5} \spa{5}.{ \ell} \spa{6}.{ \ell}
   \spa{\omega_5}.{ \ell} \spaa{\ell}.{ P_{61}}.{ Q}.{ \ell}^3 \spab{\ell}.{ P_{61}}.{ 2}
   \spab{\ell}.{ P_{61}}.{ \ell}^2 \spab{\ell}.{ P_{612}}.{ 3} \spb{2}.{ 3}}  } \cr   &  - {
 {2 \dea \deb \spa{1}.{ \ell}^3 \spa{4}.{ 2}^2 \spa{4}.{ \ell}^2 \spab{\ell}.{ P_{61}}.{ 3}^2
   \spb{1}.{ 6} \spb{2}.{ 3}} \over {\spa{4}.{ 5} \spa{5}.{ \ell} \spa{6}.{ \ell} \spa{\omega_5}.{ \ell}
   \spaa{\ell}.{ P_{61}}.{ Q}.{ \ell} \spab{\ell}.{ P_{61}}.{ 2} \spab{\ell}.{ P_{61}}.{ \ell}^2
   \spab{\ell}.{ P_{612}}.{ 3}}  } \cr   &  - {  {4 \dea \deb \spa{1}.{ \ell}^3 \spa{4}.{ 1} \spa{4}.{ 2}
   \spa{4}.{ \ell}^2 \spab{\ell}.{ Q}.{ 1} \spab{\ell}.{ P_{61}}.{ 3}^3 \spb{6}.{ 1}} \over
  {\spa{4}.{ 5} \spa{5}.{ \ell} \spa{6}.{ \ell} \spa{\omega_5}.{ \ell} \spaa{\ell}.{ P_{61}}.{ Q}.{ \ell}^2
   \spab{\ell}.{ P_{61}}.{ 2} \spab{\ell}.{ P_{61}}.{ \ell}^2 \spab{\ell}.{ P_{612}}.{ 3}}  } \cr   &  - {
 {4 \dea \deb \spa{1}.{ \ell}^3 \spa{4}.{ 2} \spa{4}.{ 6} \spa{4}.{ \ell}^2 \spab{\ell}.{ Q}.{ 6}
   \spab{\ell}.{ P_{61}}.{ 3}^3 \spb{6}.{ 1}} \over {\spa{4}.{ 5} \spa{5}.{ \ell} \spa{6}.{ \ell}
   \spa{\omega_5}.{ \ell} \spaa{\ell}.{ P_{61}}.{ Q}.{ \ell}^2 \spab{\ell}.{ P_{61}}.{ 2}
   \spab{\ell}.{ P_{61}}.{ \ell}^2 \spab{\ell}.{ P_{612}}.{ 3}}  } \cr   &  - {
 {2 \dea \deb \spa{1}.{ \ell}^2 \spa{4}.{ 1}^2 \spa{4}.{ \ell}^2 \spab{\ell}.{ Q}.{ 1}
   \spab{\ell}.{ Q}.{ 6} \spab{\ell}.{ P_{61}}.{ 3}^4 \spb{6}.{ 1}} \over
  {\spa{4}.{ 5} \spa{5}.{ \ell} \spa{\omega_5}.{ \ell} \spaa{\ell}.{ P_{61}}.{ Q}.{ \ell}^3
   \spab{\ell}.{ P_{61}}.{ 2} \spab{\ell}.{ P_{61}}.{ \ell}^2 \spab{\ell}.{ P_{612}}.{ 3}
   \spb{2}.{ 3}}  } \cr   &  - {  {4 \dea \deb \spa{1}.{ \ell}^2 \spa{4}.{ 1} \spa{4}.{ 6} \spa{4}.{ \ell}^2
   \spab{\ell}.{ Q}.{ 6}^2 \spab{\ell}.{ P_{61}}.{ 3}^4 \spb{6}.{ 1}} \over
  {\spa{4}.{ 5} \spa{5}.{ \ell} \spa{\omega_5}.{ \ell} \spaa{\ell}.{ P_{61}}.{ Q}.{ \ell}^3
   \spab{\ell}.{ P_{61}}.{ 2} \spab{\ell}.{ P_{61}}.{ \ell}^2 \spab{\ell}.{ P_{612}}.{ 3}
   \spb{2}.{ 3}}
} }}

ii) full derivative:
\eqn\name{\eqalign{ C_{61}^{(r,3)} =
\dea\dedeb \ {\cal I}^{(3)}
}}
\eqn\cLXIderIII{\eqalign{
\! \!\! \!\! \!\! \!\! \!
{\cal I}^{(3)}
&=
 - {{ 2   \spa{1}.{ \ell}^2 \spa{4}.{ 2}^2 \spa{4}.{ \ell}^2 \spab{\ell}.{ P_{61}}.{ 3}^2 \spb{2}.{ 3}
   \spb{6}.{ \ell}} \over {\spa{4}.{ 5} \spa{5}.{ \ell} \spa{6}.{ \ell} \spa{\omega_5}.{ \ell}
   \spaa{\ell}.{ P_{61}}.{ Q}.{ \ell} \spab{\ell}.{ P_{61}}.{ 2} \spab{\ell}.{ P_{61}}.{ \ell}
   \spab{\ell}.{ P_{612}}.{ 3}}  } \cr   &  + {  {2   \spa{1}.{ \ell}^3 \spa{4}.{ 6}^2 \spa{a}.{ b}^6
   \spa{\eta_1}.{ 4}^3 \spab{\ell}.{ Q}.{ 6}^2 \spab{\ell}.{ P_{61}}.{ 3}^4 \spb{1}.{ 6}
   \spb{\eta_1}.{ \ell}} \over {\spa{4}.{ 5} \spa{4}.{ \ell} \spa{5}.{ \ell} \spa{6}.{ \ell} \spa{\eta_1}.{ \ell}^3
   \spa{\eta_1}.{ \eta_2}^3 \spa{\omega_5}.{ \ell} \spaa{b}.{ P_{61}}.{ Q}.{ b}^3
   \spab{\ell}.{ P_{61}}.{ 2} \spab{\ell}.{ P_{61}}.{ \ell} \spab{\ell}.{ P_{61}}.{ \eta_1}
   \spab{\ell}.{ P_{612}}.{ 3} \spb{2}.{ 3}}  } \cr   &  + {
 {6   \spa{1}.{ \ell}^3 \spa{4}.{ 6}^2 \spa{4}.{ \eta_2} \spa{a}.{ b}^6 \spa{\eta_1}.{ 4}^2
   \spab{\ell}.{ Q}.{ 6}^2 \spab{\ell}.{ P_{61}}.{ 3}^4 \spb{1}.{ 6} \spb{\eta_1}.{ \ell}} \over
  {\spa{4}.{ 5} \spa{4}.{ \ell} \spa{5}.{ \ell} \spa{6}.{ \ell} \spa{\eta_1}.{ \ell}^2
   \spa{\eta_1}.{ \eta_2}^3 \spa{\eta_2}.{ \ell} \spa{\omega_5}.{ \ell} \spaa{b}.{ P_{61}}.{ Q}.{ b}^3
   \spab{\ell}.{ P_{61}}.{ 2} \spab{\ell}.{ P_{61}}.{ \ell} \spab{\ell}.{ P_{61}}.{ \eta_1}
   \spab{\ell}.{ P_{612}}.{ 3} \spb{2}.{ 3}}  } \cr   &  + {
 {4   \spa{1}.{ \ell}^3 \spa{4}.{ 1} \spa{4}.{ 2} \spa{a}.{ b}^4 \spa{\eta_1}.{ 4}^2
   \spab{\ell}.{ Q}.{ 1} \spab{\ell}.{ P_{61}}.{ 3}^3 \spb{6}.{ 1} \spb{\eta_1}.{ \ell}} \over
  {\spa{4}.{ 5} \spa{5}.{ \ell} \spa{6}.{ \ell} \spa{\eta_1}.{ \ell}^2 \spa{\eta_1}.{ \eta_2}^2
   \spa{\omega_5}.{ \ell} \spaa{b}.{ P_{61}}.{ Q}.{ b}^2 \spab{\ell}.{ P_{61}}.{ 2}
   \spab{\ell}.{ P_{61}}.{ \ell} \spab{\ell}.{ P_{61}}.{ \eta_1} \spab{\ell}.{ P_{612}}.{ 3}}  } \cr   &  + {
 {8   \spa{1}.{ \ell}^3 \spa{4}.{ 1} \spa{4}.{ 2} \spa{4}.{ \eta_2} \spa{a}.{ b}^4
   \spa{\eta_1}.{ 4} \spab{\ell}.{ Q}.{ 1} \spab{\ell}.{ P_{61}}.{ 3}^3 \spb{6}.{ 1} \spb{\eta_1}.{ \ell}} \over
  {\spa{4}.{ 5} \spa{5}.{ \ell} \spa{6}.{ \ell} \spa{\eta_1}.{ \ell} \spa{\eta_1}.{ \eta_2}^2
   \spa{\eta_2}.{ \ell} \spa{\omega_5}.{ \ell} \spaa{b}.{ P_{61}}.{ Q}.{ b}^2 \spab{\ell}.{ P_{61}}.{ 2}
   \spab{\ell}.{ P_{61}}.{ \ell} \spab{\ell}.{ P_{61}}.{ \eta_1} \spab{\ell}.{ P_{612}}.{ 3}}  } \cr   &  + {
 {4   \spa{1}.{ \ell}^3 \spa{4}.{ 2} \spa{4}.{ 6} \spa{a}.{ b}^4 \spa{\eta_1}.{ 4}^2
   \spab{\ell}.{ Q}.{ 6} \spab{\ell}.{ P_{61}}.{ 3}^3 \spb{6}.{ 1} \spb{\eta_1}.{ \ell}} \over
  {\spa{4}.{ 5} \spa{5}.{ \ell} \spa{6}.{ \ell} \spa{\eta_1}.{ \ell}^2 \spa{\eta_1}.{ \eta_2}^2
   \spa{\omega_5}.{ \ell} \spaa{b}.{ P_{61}}.{ Q}.{ b}^2 \spab{\ell}.{ P_{61}}.{ 2}
   \spab{\ell}.{ P_{61}}.{ \ell} \spab{\ell}.{ P_{61}}.{ \eta_1} \spab{\ell}.{ P_{612}}.{ 3}}  } \cr   &  + {
 {8   \spa{1}.{ \ell}^3 \spa{4}.{ 2} \spa{4}.{ 6} \spa{4}.{ \eta_2} \spa{a}.{ b}^4
   \spa{\eta_1}.{ 4} \spab{\ell}.{ Q}.{ 6} \spab{\ell}.{ P_{61}}.{ 3}^3 \spb{6}.{ 1} \spb{\eta_1}.{ \ell}} \over
  {\spa{4}.{ 5} \spa{5}.{ \ell} \spa{6}.{ \ell} \spa{\eta_1}.{ \ell} \spa{\eta_1}.{ \eta_2}^2
   \spa{\eta_2}.{ \ell} \spa{\omega_5}.{ \ell} \spaa{b}.{ P_{61}}.{ Q}.{ b}^2 \spab{\ell}.{ P_{61}}.{ 2}
   \spab{\ell}.{ P_{61}}.{ \ell} \spab{\ell}.{ P_{61}}.{ \eta_1} \spab{\ell}.{ P_{612}}.{ 3}}  } \cr   &  + {
 {2   \spa{1}.{ \ell}^2 \spa{4}.{ 1}^2 \spa{a}.{ b}^6 \spa{\eta_1}.{ 4}^3 \spab{\ell}.{ Q}.{ 1}
   \spab{\ell}.{ Q}.{ 6} \spab{\ell}.{ P_{61}}.{ 3}^4 \spb{6}.{ 1} \spb{\eta_1}.{ \ell}} \over
  {\spa{4}.{ 5} \spa{4}.{ \ell} \spa{5}.{ \ell} \spa{\eta_1}.{ \ell}^3 \spa{\eta_1}.{ \eta_2}^3
   \spa{\omega_5}.{ \ell} \spaa{b}.{ P_{61}}.{ Q}.{ b}^3 \spab{\ell}.{ P_{61}}.{ 2}
   \spab{\ell}.{ P_{61}}.{ \ell} \spab{\ell}.{ P_{61}}.{ \eta_1} \spab{\ell}.{ P_{612}}.{ 3}
   \spb{2}.{ 3}}  } \cr   &  + {  {6   \spa{1}.{ \ell}^2 \spa{4}.{ 1}^2 \spa{4}.{ \eta_2} \spa{a}.{ b}^6
   \spa{\eta_1}.{ 4}^2 \spab{\ell}.{ Q}.{ 1} \spab{\ell}.{ Q}.{ 6} \spab{\ell}.{ P_{61}}.{ 3}^4
   \spb{6}.{ 1} \spb{\eta_1}.{ \ell}} \over {\spa{4}.{ 5} \spa{4}.{ \ell} \spa{5}.{ \ell} \spa{\eta_1}.{ \ell}^2
   \spa{\eta_1}.{ \eta_2}^3 \spa{\eta_2}.{ \ell} \spa{\omega_5}.{ \ell} \spaa{b}.{ P_{61}}.{ Q}.{ b}^3
   \spab{\ell}.{ P_{61}}.{ 2} \spab{\ell}.{ P_{61}}.{ \ell} \spab{\ell}.{ P_{61}}.{ \eta_1}
   \spab{\ell}.{ P_{612}}.{ 3} \spb{2}.{ 3}}  } \cr   &  + {
 {4   \spa{1}.{ \ell}^2 \spa{4}.{ 1} \spa{4}.{ 6} \spa{a}.{ b}^6 \spa{\eta_1}.{ 4}^3
   \spab{\ell}.{ Q}.{ 6}^2 \spab{\ell}.{ P_{61}}.{ 3}^4 \spb{6}.{ 1} \spb{\eta_1}.{ \ell}} \over
  {\spa{4}.{ 5} \spa{4}.{ \ell} \spa{5}.{ \ell} \spa{\eta_1}.{ \ell}^3 \spa{\eta_1}.{ \eta_2}^3
   \spa{\omega_5}.{ \ell} \spaa{b}.{ P_{61}}.{ Q}.{ b}^3 \spab{\ell}.{ P_{61}}.{ 2}
   \spab{\ell}.{ P_{61}}.{ \ell} \spab{\ell}.{ P_{61}}.{ \eta_1} \spab{\ell}.{ P_{612}}.{ 3}
   \spb{2}.{ 3}}  } \cr   &  + {  {12   \spa{1}.{ \ell}^2 \spa{4}.{ 1} \spa{4}.{ 6} \spa{4}.{ \eta_2}
   \spa{a}.{ b}^6 \spa{\eta_1}.{ 4}^2 \spab{\ell}.{ Q}.{ 6}^2 \spab{\ell}.{ P_{61}}.{ 3}^4
   \spb{6}.{ 1} \spb{\eta_1}.{ \ell}} \over {\spa{4}.{ 5} \spa{4}.{ \ell} \spa{5}.{ \ell} \spa{\eta_1}.{ \ell}^2
   \spa{\eta_1}.{ \eta_2}^3 \spa{\eta_2}.{ \ell} \spa{\omega_5}.{ \ell} \spaa{b}.{ P_{61}}.{ Q}.{ b}^3
   \spab{\ell}.{ P_{61}}.{ 2} \spab{\ell}.{ P_{61}}.{ \ell} \spab{\ell}.{ P_{61}}.{ \eta_1}
   \spab{\ell}.{ P_{612}}.{ 3} \spb{2}.{ 3}}  } \cr   &  + {
 {2   \spa{1}.{ \ell}^3 \spa{4}.{ 6}^2 \spa{4}.{ \eta_2}^3 \spa{a}.{ b}^6 \spab{\ell}.{ Q}.{ 6}^2
   \spab{\ell}.{ P_{61}}.{ 3}^4 \spb{1}.{ 6} \spb{\eta_2}.{ \ell}} \over
  {\spa{4}.{ 5} \spa{4}.{ \ell} \spa{5}.{ \ell} \spa{6}.{ \ell} \spa{\eta_1}.{ \eta_2}^3
   \spa{\eta_2}.{ \ell}^3 \spa{\omega_5}.{ \ell} \spaa{b}.{ P_{61}}.{ Q}.{ b}^3
   \spab{\ell}.{ P_{61}}.{ 2} \spab{\ell}.{ P_{61}}.{ \ell} \spab{\ell}.{ P_{61}}.{ \eta_2}
   \spab{\ell}.{ P_{612}}.{ 3} \spb{2}.{ 3}}  } \cr   &  + {
 {6   \spa{1}.{ \ell}^3 \spa{4}.{ 6}^2 \spa{4}.{ \eta_2}^2 \spa{a}.{ b}^6 \spa{\eta_1}.{ 4}
   \spab{\ell}.{ Q}.{ 6}^2 \spab{\ell}.{ P_{61}}.{ 3}^4 \spb{1}.{ 6} \spb{\eta_2}.{ \ell}} \over
  {\spa{4}.{ 5} \spa{4}.{ \ell} \spa{5}.{ \ell} \spa{6}.{ \ell} \spa{\eta_1}.{ \ell} \spa{\eta_1}.{ \eta_2}^3
   \spa{\eta_2}.{ \ell}^2 \spa{\omega_5}.{ \ell} \spaa{b}.{ P_{61}}.{ Q}.{ b}^3
   \spab{\ell}.{ P_{61}}.{ 2} \spab{\ell}.{ P_{61}}.{ \ell} \spab{\ell}.{ P_{61}}.{ \eta_2}
   \spab{\ell}.{ P_{612}}.{ 3} \spb{2}.{ 3}}  } \cr   &  + {
 {4   \spa{1}.{ \ell}^3 \spa{4}.{ 1} \spa{4}.{ 2} \spa{4}.{ \eta_2}^2 \spa{a}.{ b}^4
   \spab{\ell}.{ Q}.{ 1} \spab{\ell}.{ P_{61}}.{ 3}^3 \spb{6}.{ 1} \spb{\eta_2}.{ \ell}} \over
  {\spa{4}.{ 5} \spa{5}.{ \ell} \spa{6}.{ \ell} \spa{\eta_1}.{ \eta_2}^2 \spa{\eta_2}.{ \ell}^2
   \spa{\omega_5}.{ \ell} \spaa{b}.{ P_{61}}.{ Q}.{ b}^2 \spab{\ell}.{ P_{61}}.{ 2}
   \spab{\ell}.{ P_{61}}.{ \ell} \spab{\ell}.{ P_{61}}.{ \eta_2} \spab{\ell}.{ P_{612}}.{ 3}}  } \cr   &  + {
 {4   \spa{1}.{ \ell}^3 \spa{4}.{ 2} \spa{4}.{ 6} \spa{4}.{ \eta_2}^2 \spa{a}.{ b}^4
   \spab{\ell}.{ Q}.{ 6} \spab{\ell}.{ P_{61}}.{ 3}^3 \spb{6}.{ 1} \spb{\eta_2}.{ \ell}} \over
  {\spa{4}.{ 5} \spa{5}.{ \ell} \spa{6}.{ \ell} \spa{\eta_1}.{ \eta_2}^2 \spa{\eta_2}.{ \ell}^2
   \spa{\omega_5}.{ \ell} \spaa{b}.{ P_{61}}.{ Q}.{ b}^2 \spab{\ell}.{ P_{61}}.{ 2}
   \spab{\ell}.{ P_{61}}.{ \ell} \spab{\ell}.{ P_{61}}.{ \eta_2} \spab{\ell}.{ P_{612}}.{ 3}}  } \cr   &  + {
 {2   \spa{1}.{ \ell}^2 \spa{4}.{ 1}^2 \spa{4}.{ \eta_2}^3 \spa{a}.{ b}^6 \spab{\ell}.{ Q}.{ 1}
   \spab{\ell}.{ Q}.{ 6} \spab{\ell}.{ P_{61}}.{ 3}^4 \spb{6}.{ 1} \spb{\eta_2}.{ \ell}} \over
  {\spa{4}.{ 5} \spa{4}.{ \ell} \spa{5}.{ \ell} \spa{\eta_1}.{ \eta_2}^3 \spa{\eta_2}.{ \ell}^3
   \spa{\omega_5}.{ \ell} \spaa{b}.{ P_{61}}.{ Q}.{ b}^3 \spab{\ell}.{ P_{61}}.{ 2}
   \spab{\ell}.{ P_{61}}.{ \ell} \spab{\ell}.{ P_{61}}.{ \eta_2} \spab{\ell}.{ P_{612}}.{ 3}
   \spb{2}.{ 3}}  } \cr   &  + {  {6   \spa{1}.{ \ell}^2 \spa{4}.{ 1}^2 \spa{4}.{ \eta_2}^2 \spa{a}.{ b}^6
   \spa{\eta_1}.{ 4} \spab{\ell}.{ Q}.{ 1} \spab{\ell}.{ Q}.{ 6} \spab{\ell}.{ P_{61}}.{ 3}^4 \spb{6}.{ 1}
   \spb{\eta_2}.{ \ell}} \over {\spa{4}.{ 5} \spa{4}.{ \ell} \spa{5}.{ \ell} \spa{\eta_1}.{ \ell}
   \spa{\eta_1}.{ \eta_2}^3 \spa{\eta_2}.{ \ell}^2 \spa{\omega_5}.{ \ell} \spaa{b}.{ P_{61}}.{ Q}.{ b}^3
   \spab{\ell}.{ P_{61}}.{ 2} \spab{\ell}.{ P_{61}}.{ \ell} \spab{\ell}.{ P_{61}}.{ \eta_2}
   \spab{\ell}.{ P_{612}}.{ 3} \spb{2}.{ 3}}  } \cr   &  + {
 {4   \spa{1}.{ \ell}^2 \spa{4}.{ 1} \spa{4}.{ 6} \spa{4}.{ \eta_2}^3 \spa{a}.{ b}^6
   \spab{\ell}.{ Q}.{ 6}^2 \spab{\ell}.{ P_{61}}.{ 3}^4 \spb{6}.{ 1} \spb{\eta_2}.{ \ell}} \over
  {\spa{4}.{ 5} \spa{4}.{ \ell} \spa{5}.{ \ell} \spa{\eta_1}.{ \eta_2}^3 \spa{\eta_2}.{ \ell}^3
   \spa{\omega_5}.{ \ell} \spaa{b}.{ P_{61}}.{ Q}.{ b}^3 \spab{\ell}.{ P_{61}}.{ 2}
   \spab{\ell}.{ P_{61}}.{ \ell} \spab{\ell}.{ P_{61}}.{ \eta_2} \spab{\ell}.{ P_{612}}.{ 3}
   \spb{2}.{ 3}}  } \cr   &  + {  {12   \spa{1}.{ \ell}^2 \spa{4}.{ 1} \spa{4}.{ 6} \spa{4}.{ \eta_2}^2
   \spa{a}.{ b}^6 \spa{\eta_1}.{ 4} \spab{\ell}.{ Q}.{ 6}^2 \spab{\ell}.{ P_{61}}.{ 3}^4 \spb{6}.{ 1}
   \spb{\eta_2}.{ \ell}} \over {\spa{4}.{ 5} \spa{4}.{ \ell} \spa{5}.{ \ell} \spa{\eta_1}.{ \ell}
   \spa{\eta_1}.{ \eta_2}^3 \spa{\eta_2}.{ \ell}^2 \spa{\omega_5}.{ \ell} \spaa{b}.{ P_{61}}.{ Q}.{ b}^3
   \spab{\ell}.{ P_{61}}.{ 2} \spab{\ell}.{ P_{61}}.{ \ell} \spab{\ell}.{ P_{61}}.{ \eta_2}
   \spab{\ell}.{ P_{612}}.{ 3} \spb{2}.{ 3}}
} }}

\subsubsec{The term $C_{61}^{(r,4)} $}

i) $t$-integrated formula: \eqn\name{\eqalign{ C_{61}^{(r,4)} &= {
{4 \dea \deb \spa{1}.{ \ell}^3 \spa{4}.{ 1} \spa{4}.{ 6} \spa{4}.{
\ell}^2
   \spab{\ell}.{ P_{61}}.{ 3}^4 \spb{1}.{ 6} \spb{1}.{ \ell}} \over
  {\spa{4}.{ 5} \spa{5}.{ \ell} \spa{6}.{ \ell}^2 \spa{\omega_5}.{ \ell} \spaa{\ell}.{ P_{61}}.{ Q}.{ \ell}
   \spab{\ell}.{ P_{61}}.{ 2} \spab{\ell}.{ P_{61}}.{ \ell}^3 \spab{\ell}.{ P_{612}}.{ 3}
   \spb{2}.{ 3}}  } \cr   &  + {  {4 \dea \deb \spa{1}.{ \ell}^3 \spa{4}.{ 1} \spa{4}.{ 6} \spa{4}.{ \ell}^2
   \spab{\ell}.{ Q}.{ 6} \spab{\ell}.{ P_{61}}.{ 3}^4 \spb{1}.{ 6} \spb{1}.{ \ell}} \over
  {\spa{4}.{ 5} \spa{5}.{ \ell} \spa{6}.{ \ell} \spa{\omega_5}.{ \ell} \spaa{\ell}.{ P_{61}}.{ Q}.{ \ell}^2
   \spab{\ell}.{ P_{61}}.{ 2} \spab{\ell}.{ P_{61}}.{ \ell}^3 \spab{\ell}.{ P_{612}}.{ 3}
   \spb{2}.{ 3}}  } \cr   &  + {  {2 \dea \deb \spa{1}.{ \ell}^2 \spa{4}.{ 1}^2 \spa{4}.{ \ell}^2
   \spab{\ell}.{ Q}.{ 6} \spab{\ell}.{ P_{61}}.{ 3}^4 \spb{1}.{ \ell} \spb{6}.{ 1}} \over
  {\spa{4}.{ 5} \spa{5}.{ \ell} \spa{\omega_5}.{ \ell} \spaa{\ell}.{ P_{61}}.{ Q}.{ \ell}^2
   \spab{\ell}.{ P_{61}}.{ 2} \spab{\ell}.{ P_{61}}.{ \ell}^3 \spab{\ell}.{ P_{612}}.{ 3}
   \spb{2}.{ 3}}  } \cr   &  + {  {2 \dea \deb \spa{1}.{ \ell}^3 \spa{4}.{ 6}^2 \spa{4}.{ \ell}^2
   \spab{\ell}.{ Q}.{ 6} \spab{\ell}.{ P_{61}}.{ 3}^4 \spb{1}.{ 6} \spb{6}.{ \ell}} \over
  {\spa{4}.{ 5} \spa{5}.{ \ell} \spa{6}.{ \ell} \spa{\omega_5}.{ \ell} \spaa{\ell}.{ P_{61}}.{ Q}.{ \ell}^2
   \spab{\ell}.{ P_{61}}.{ 2} \spab{\ell}.{ P_{61}}.{ \ell}^3 \spab{\ell}.{ P_{612}}.{ 3}
   \spb{2}.{ 3}}  } \cr   &  - {  {4 \dea \deb \spa{1}.{ \ell}^2 \spa{4}.{ 1} \spa{4}.{ 2} \spa{4}.{ \ell}^2
   \spab{\ell}.{ P_{61}}.{ 3}^3 \spb{6}.{ 1} \spb{6}.{ \ell}} \over
  {\spa{4}.{ 5} \spa{5}.{ \ell} \spa{\omega_5}.{ \ell} \spaa{\ell}.{ P_{61}}.{ Q}.{ \ell}
   \spab{\ell}.{ P_{61}}.{ 2} \spab{\ell}.{ P_{61}}.{ \ell}^3 \spab{\ell}.{ P_{612}}.{ 3}}  } \cr   &  + {
 {4 \dea \deb \spa{1}.{ \ell}^3 \spa{4}.{ 2} \spa{4}.{ 6} \spa{4}.{ \ell}^2
   \spab{\ell}.{ P_{61}}.{ 3}^3 \spb{6}.{ 1} \spb{6}.{ \ell}} \over
  {\spa{4}.{ 5} \spa{5}.{ \ell} \spa{6}.{ \ell} \spa{\omega_5}.{ \ell} \spaa{\ell}.{ P_{61}}.{ Q}.{ \ell}
   \spab{\ell}.{ P_{61}}.{ 2} \spab{\ell}.{ P_{61}}.{ \ell}^3 \spab{\ell}.{ P_{612}}.{ 3}}  } \cr   &  + {
 {2 \dea \deb \spa{1}.{ \ell} \spa{4}.{ 2}^4 \spab{\ell}.{ P_{61}}.{ 5}^2 \spb{5}.{ 6} \spb{6}.{ \ell}
   (P_{61}^2)} \over {\spa{2}.{ 3} \spa{3}.{ 4} \spa{6}.{ \ell} \spa{\omega_5}.{ \ell}
   \spab{2}.{ P_{234}}.{ 5} \spab{\ell}.{ P_{61}}.{ \ell}^3 \spb{1}.{ 6} (P_{234}^2)}
} }}

ii) full derivative:
\eqn\name{\eqalign{ C_{61}^{(r,4)} = \dea \dedeb \ {\cal I}^{(4)} }}
\eqn\cLXIderIV{\eqalign{
 {\cal I}^{(4)} &=
{ {2   \spa{1}.{ \ell}^2 \spa{4}.{ 1}^2 \spa{4}.{ \eta_2} \spa{a}.{
b}^4 \spa{\eta_1}.{ 4}
   \spab{\ell}.{ Q}.{ 6} \spab{\ell}.{ P_{61}}.{ 3}^4 \spb{1}.{ \ell}^2} \over
  {\spa{4}.{ 5} \spa{5}.{ \ell} \spa{6}.{ \ell} \spa{\eta_1}.{ \ell} \spa{\eta_1}.{ \eta_2}^2
   \spa{\eta_2}.{ \ell} \spa{\omega_5}.{ \ell} \spaa{b}.{ P_{61}}.{ Q}.{ b}^2 \spab{\ell}.{ P_{61}}.{ 2}
   \spab{\ell}.{ P_{61}}.{ \ell}^2 \spab{\ell}.{ P_{612}}.{ 3} \spb{2}.{ 3}}  } \cr   &  - {
 {4   \spa{1}.{ \ell} \spa{4}.{ 1} \spa{4}.{ 6} \spa{4}.{ \ell}^2 \spa{a}.{ b}^2
   \spab{\ell}.{ P_{61}}.{ 3}^4 \spb{6}.{ \ell}} \over {\spa{4}.{ 5} \spa{5}.{ \ell} \spa{6}.{ \ell}^2
   \spa{\eta_1}.{ \ell} \spa{\eta_2}.{ \ell} \spa{\omega_5}.{ \ell} \spaa{b}.{ P_{61}}.{ Q}.{ b}
   \spab{\ell}.{ P_{61}}.{ 2} \spab{\ell}.{ P_{61}}.{ \ell} \spab{\ell}.{ P_{612}}.{ 3}
   \spb{2}.{ 3}}  } \cr   &  - {  {8   \spa{1}.{ \ell} \spa{4}.{ 1} \spa{4}.{ 6} \spa{4}.{ \eta_2}
   \spa{a}.{ b}^4 \spa{\eta_1}.{ 4} \spab{\ell}.{ Q}.{ 6} \spab{\ell}.{ P_{61}}.{ 3}^4 \spb{6}.{ \ell}} \over
  {\spa{4}.{ 5} \spa{5}.{ \ell} \spa{6}.{ \ell} \spa{\eta_1}.{ \ell} \spa{\eta_1}.{ \eta_2}^2
   \spa{\eta_2}.{ \ell} \spa{\omega_5}.{ \ell} \spaa{b}.{ P_{61}}.{ Q}.{ b}^2 \spab{\ell}.{ P_{61}}.{ 2}
   \spab{\ell}.{ P_{61}}.{ \ell} \spab{\ell}.{ P_{612}}.{ 3} \spb{2}.{ 3}}  } \cr   &  + {
 {2   \spa{1}.{ \ell}^2 \spa{4}.{ 6}^2 \spa{4}.{ \eta_2} \spa{a}.{ b}^4 \spa{\eta_1}.{ 4}
   \spab{\ell}.{ Q}.{ 6} \spab{\ell}.{ P_{61}}.{ 3}^4 \spb{6}.{ \ell}^2} \over
  {\spa{4}.{ 5} \spa{5}.{ \ell} \spa{6}.{ \ell} \spa{\eta_1}.{ \ell} \spa{\eta_1}.{ \eta_2}^2
   \spa{\eta_2}.{ \ell} \spa{\omega_5}.{ \ell} \spaa{b}.{ P_{61}}.{ Q}.{ b}^2 \spab{\ell}.{ P_{61}}.{ 2}
   \spab{\ell}.{ P_{61}}.{ \ell}^2 \spab{\ell}.{ P_{612}}.{ 3} \spb{2}.{ 3}}  } \cr   &  - {
 {2   \spa{1}.{ \ell} \spa{4}.{ 1} \spa{4}.{ 2} \spa{4}.{ \ell}^2 \spa{a}.{ b}^2
   \spab{\ell}.{ P_{61}}.{ 3}^3 \spb{6}.{ 1} \spb{6}.{ \ell}^2} \over
  {\spa{4}.{ 5} \spa{5}.{ \ell} \spa{\eta_1}.{ \ell} \spa{\eta_2}.{ \ell} \spa{\omega_5}.{ \ell}
   \spaa{b}.{ P_{61}}.{ Q}.{ b} \spab{\ell}.{ P_{61}}.{ 2} \spab{\ell}.{ P_{61}}.{ \ell}^2
   \spab{\ell}.{ P_{612}}.{ 3} \spb{1}.{ 6}}  } \cr   &  + {
 {2   \spa{1}.{ \ell}^2 \spa{4}.{ 2} \spa{4}.{ 6} \spa{4}.{ \ell}^2 \spa{a}.{ b}^2
   \spab{\ell}.{ P_{61}}.{ 3}^3 \spb{6}.{ 1} \spb{6}.{ \ell}^2} \over
  {\spa{4}.{ 5} \spa{5}.{ \ell} \spa{6}.{ \ell} \spa{\eta_1}.{ \ell} \spa{\eta_2}.{ \ell}
   \spa{\omega_5}.{ \ell} \spaa{b}.{ P_{61}}.{ Q}.{ b} \spab{\ell}.{ P_{61}}.{ 2}
   \spab{\ell}.{ P_{61}}.{ \ell}^2 \spab{\ell}.{ P_{612}}.{ 3} \spb{1}.{ 6}}  } \cr   &  + {
 {2   \spa{1}.{ \ell} \spa{4}.{ 1} \spa{4}.{ 6} \spa{4}.{ \ell}^2 \spa{a}.{ b}^2
   \spab{\ell}.{ P_{61}}.{ 3}^4 \spb{6}.{ 1} \spb{6}.{ \ell}^2} \over
  {\spa{4}.{ 5} \spa{5}.{ \ell} \spa{6}.{ \ell} \spa{\eta_1}.{ \ell} \spa{\eta_2}.{ \ell}
   \spa{\omega_5}.{ \ell} \spaa{b}.{ P_{61}}.{ Q}.{ b} \spab{\ell}.{ P_{61}}.{ 2}
   \spab{\ell}.{ P_{61}}.{ \ell}^2 \spab{\ell}.{ P_{612}}.{ 3} \spb{1}.{ 6}
   \spb{2}.{ 3}}  } \cr   &  + {
 {4   \spa{1}.{ \ell} \spa{4}.{ 1} \spa{4}.{ 6} \spa{4}.{ \eta_2} \spa{a}.{ b}^4 \spa{\eta_1}.{ 4}
   \spab{\ell}.{ Q}.{ 6} \spab{\ell}.{ P_{61}}.{ 3}^4 \spb{6}.{ 1} \spb{6}.{ \ell}^2} \over
  {\spa{4}.{ 5} \spa{5}.{ \ell} \spa{\eta_1}.{ \ell} \spa{\eta_1}.{ \eta_2}^2 \spa{\eta_2}.{ \ell}
   \spa{\omega_5}.{ \ell} \spaa{b}.{ P_{61}}.{ Q}.{ b}^2 \spab{\ell}.{ P_{61}}.{ 2}
   \spab{\ell}.{ P_{61}}.{ \ell}^2 \spab{\ell}.{ P_{612}}.{ 3} \spb{1}.{ 6} \spb{2}.{ 3}}  } \cr   &  - {
 {4   \spa{1}.{ \ell}^3 \spa{4}.{ 1} \spa{4}.{ 6} \spa{a}.{ b}^4 \spa{\eta_1}.{ 4}^2
   \spab{\ell}.{ Q}.{ 6} \spab{\ell}.{ P_{61}}.{ 3}^4 \spb{1}.{ 6} \spb{1}.{ \eta_1} \spb{\eta_1}.{ \ell}} \over
  {\spa{4}.{ 5} \spa{5}.{ \ell} \spa{6}.{ \ell} \spa{\eta_1}.{ \ell}^2 \spa{\eta_1}.{ \eta_2}^2
   \spa{\omega_5}.{ \ell} \spaa{b}.{ P_{61}}.{ Q}.{ b}^2 \spab{\ell}.{ P_{61}}.{ 2}
   \spab{\ell}.{ P_{61}}.{ \ell} \spab{\ell}.{ P_{61}}.{ \eta_1}^2 \spab{\ell}.{ P_{612}}.{ 3}
   \spb{2}.{ 3}}  } \cr   &  - {  {2   \spa{1}.{ \ell}^2 \spa{4}.{ 1}^2 \spa{a}.{ b}^4 \spa{\eta_1}.{ 4}^2
   \spab{\ell}.{ Q}.{ 6} \spab{\ell}.{ P_{61}}.{ 3}^4 \spb{1}.{ \eta_1} \spb{6}.{ 1} \spb{\eta_1}.{ \ell}} \over
  {\spa{4}.{ 5} \spa{5}.{ \ell} \spa{\eta_1}.{ \ell}^2 \spa{\eta_1}.{ \eta_2}^2 \spa{\omega_5}.{ \ell}
   \spaa{b}.{ P_{61}}.{ Q}.{ b}^2 \spab{\ell}.{ P_{61}}.{ 2} \spab{\ell}.{ P_{61}}.{ \ell}
   \spab{\ell}.{ P_{61}}.{ \eta_1}^2 \spab{\ell}.{ P_{612}}.{ 3} \spb{2}.{ 3}}  } \cr   &  - {
 {2   \spa{1}.{ \ell}^3 \spa{4}.{ 6}^2 \spa{a}.{ b}^4 \spa{\eta_1}.{ 4}^2 \spab{\ell}.{ Q}.{ 6}
   \spab{\ell}.{ P_{61}}.{ 3}^4 \spb{1}.{ 6} \spb{6}.{ \eta_1} \spb{\eta_1}.{ \ell}} \over
  {\spa{4}.{ 5} \spa{5}.{ \ell} \spa{6}.{ \ell} \spa{\eta_1}.{ \ell}^2 \spa{\eta_1}.{ \eta_2}^2
   \spa{\omega_5}.{ \ell} \spaa{b}.{ P_{61}}.{ Q}.{ b}^2 \spab{\ell}.{ P_{61}}.{ 2}
   \spab{\ell}.{ P_{61}}.{ \ell} \spab{\ell}.{ P_{61}}.{ \eta_1}^2 \spab{\ell}.{ P_{612}}.{ 3}
   \spb{2}.{ 3}}  } \cr   &  + {  {  \spa{1}.{ \ell}^4 \spa{4}.{ 6}^2 \spa{a}.{ b}^4 \spa{\eta_1}.{ 4}^2
   \spab{\ell}.{ Q}.{ 6} \spab{\ell}.{ P_{61}}.{ 3}^4 \spb{1}.{ 6}^2 \spb{\eta_1}.{ \ell}^2} \over
  {\spa{4}.{ 5} \spa{5}.{ \ell} \spa{6}.{ \ell} \spa{\eta_1}.{ \ell}^2 \spa{\eta_1}.{ \eta_2}^2
   \spa{\omega_5}.{ \ell} \spaa{b}.{ P_{61}}.{ Q}.{ b}^2 \spab{\ell}.{ P_{61}}.{ 2}
   \spab{\ell}.{ P_{61}}.{ \ell}^2 \spab{\ell}.{ P_{61}}.{ \eta_1}^2 \spab{\ell}.{ P_{612}}.{ 3}
   \spb{2}.{ 3}}  } \cr   &  - {  {2   \spa{1}.{ \ell}^3 \spa{4}.{ 1} \spa{4}.{ 6} \spa{a}.{ b}^4
   \spa{\eta_1}.{ 4}^2 \spab{\ell}.{ Q}.{ 6} \spab{\ell}.{ P_{61}}.{ 3}^4 \spb{6}.{ 1}^2
   \spb{\eta_1}.{ \ell}^2} \over {\spa{4}.{ 5} \spa{5}.{ \ell} \spa{\eta_1}.{ \ell}^2 \spa{\eta_1}.{ \eta_2}^2
   \spa{\omega_5}.{ \ell} \spaa{b}.{ P_{61}}.{ Q}.{ b}^2 \spab{\ell}.{ P_{61}}.{ 2}
   \spab{\ell}.{ P_{61}}.{ \ell}^2 \spab{\ell}.{ P_{61}}.{ \eta_1}^2 \spab{\ell}.{ P_{612}}.{ 3}
   \spb{2}.{ 3}}  } \cr   &  + {  {  \spa{1}.{ \ell}^2 \spa{4}.{ 1}^2 \spa{6}.{ \ell} \spa{a}.{ b}^4
   \spa{\eta_1}.{ 4}^2 \spab{\ell}.{ Q}.{ 6} \spab{\ell}.{ P_{61}}.{ 3}^4 \spb{6}.{ 1}^2
   \spb{\eta_1}.{ \ell}^2} \over {\spa{4}.{ 5} \spa{5}.{ \ell} \spa{\eta_1}.{ \ell}^2 \spa{\eta_1}.{ \eta_2}^2
   \spa{\omega_5}.{ \ell} \spaa{b}.{ P_{61}}.{ Q}.{ b}^2 \spab{\ell}.{ P_{61}}.{ 2}
   \spab{\ell}.{ P_{61}}.{ \ell}^2 \spab{\ell}.{ P_{61}}.{ \eta_1}^2 \spab{\ell}.{ P_{612}}.{ 3}
   \spb{2}.{ 3}}  } \cr   &  - {  {4   \spa{1}.{ \ell}^3 \spa{4}.{ 1} \spa{4}.{ 6} \spa{a}.{ b}^4
   \spa{\eta_2}.{ 4}^2 \spab{\ell}.{ Q}.{ 6} \spab{\ell}.{ P_{61}}.{ 3}^4 \spb{1}.{ 6} \spb{1}.{ \eta_2}
   \spb{\eta_2}.{ \ell}} \over {\spa{4}.{ 5} \spa{5}.{ \ell} \spa{6}.{ \ell} \spa{\eta_2}.{ \ell}^2
   \spa{\eta_2}.{ \eta_1}^2 \spa{\omega_5}.{ \ell} \spaa{b}.{ P_{61}}.{ Q}.{ b}^2
   \spab{\ell}.{ P_{61}}.{ 2} \spab{\ell}.{ P_{61}}.{ \ell} \spab{\ell}.{ P_{61}}.{ \eta_2}^2
   \spab{\ell}.{ P_{612}}.{ 3} \spb{2}.{ 3}}  } \cr   &  - {
 {2   \spa{1}.{ \ell}^2 \spa{4}.{ 1}^2 \spa{a}.{ b}^4 \spa{\eta_2}.{ 4}^2 \spab{\ell}.{ Q}.{ 6}
   \spab{\ell}.{ P_{61}}.{ 3}^4 \spb{1}.{ \eta_2} \spb{6}.{ 1} \spb{\eta_2}.{ \ell}} \over
  {\spa{4}.{ 5} \spa{5}.{ \ell} \spa{\eta_2}.{ \ell}^2 \spa{\eta_2}.{ \eta_1}^2 \spa{\omega_5}.{ \ell}
   \spaa{b}.{ P_{61}}.{ Q}.{ b}^2 \spab{\ell}.{ P_{61}}.{ 2} \spab{\ell}.{ P_{61}}.{ \ell}
   \spab{\ell}.{ P_{61}}.{ \eta_2}^2 \spab{\ell}.{ P_{612}}.{ 3} \spb{2}.{ 3}}  } \cr   &  - {
 {2   \spa{1}.{ \ell}^3 \spa{4}.{ 6}^2 \spa{a}.{ b}^4 \spa{\eta_2}.{ 4}^2 \spab{\ell}.{ Q}.{ 6}
   \spab{\ell}.{ P_{61}}.{ 3}^4 \spb{1}.{ 6} \spb{6}.{ \eta_2} \spb{\eta_2}.{ \ell}} \over
  {\spa{4}.{ 5} \spa{5}.{ \ell} \spa{6}.{ \ell} \spa{\eta_2}.{ \ell}^2 \spa{\eta_2}.{ \eta_1}^2
   \spa{\omega_5}.{ \ell} \spaa{b}.{ P_{61}}.{ Q}.{ b}^2 \spab{\ell}.{ P_{61}}.{ 2}
   \spab{\ell}.{ P_{61}}.{ \ell} \spab{\ell}.{ P_{61}}.{ \eta_2}^2 \spab{\ell}.{ P_{612}}.{ 3}
   \spb{2}.{ 3}}  } \cr
 & (\rm continues \ to \ the \ next \ page)
} }

\eqn\name{\eqalign{
  & (\rm continues \ from \ the \ previous \ page) \cr
  &  + {  {  \spa{1}.{ \ell}^4 \spa{4}.{ 6}^2 \spa{a}.{ b}^4 \spa{\eta_2}.{ 4}^2
   \spab{\ell}.{ Q}.{ 6} \spab{\ell}.{ P_{61}}.{ 3}^4 \spb{1}.{ 6}^2 \spb{\eta_2}.{ \ell}^2} \over
  {\spa{4}.{ 5} \spa{5}.{ \ell} \spa{6}.{ \ell} \spa{\eta_2}.{ \ell}^2 \spa{\eta_2}.{ \eta_1}^2
   \spa{\omega_5}.{ \ell} \spaa{b}.{ P_{61}}.{ Q}.{ b}^2 \spab{\ell}.{ P_{61}}.{ 2}
   \spab{\ell}.{ P_{61}}.{ \ell}^2 \spab{\ell}.{ P_{61}}.{ \eta_2}^2 \spab{\ell}.{ P_{612}}.{ 3}
   \spb{2}.{ 3}}  } \cr   &  - {  {2   \spa{1}.{ \ell}^3 \spa{4}.{ 1} \spa{4}.{ 6} \spa{a}.{ b}^4
   \spa{\eta_2}.{ 4}^2 \spab{\ell}.{ Q}.{ 6} \spab{\ell}.{ P_{61}}.{ 3}^4 \spb{6}.{ 1}^2
   \spb{\eta_2}.{ \ell}^2} \over {\spa{4}.{ 5} \spa{5}.{ \ell} \spa{\eta_2}.{ \ell}^2 \spa{\eta_2}.{ \eta_1}^2
   \spa{\omega_5}.{ \ell} \spaa{b}.{ P_{61}}.{ Q}.{ b}^2 \spab{\ell}.{ P_{61}}.{ 2}
   \spab{\ell}.{ P_{61}}.{ \ell}^2 \spab{\ell}.{ P_{61}}.{ \eta_2}^2 \spab{\ell}.{ P_{612}}.{ 3}
   \spb{2}.{ 3}}  } \cr   &  + {  {  \spa{1}.{ \ell}^2 \spa{4}.{ 1}^2 \spa{6}.{ \ell} \spa{a}.{ b}^4
   \spa{\eta_2}.{ 4}^2 \spab{\ell}.{ Q}.{ 6} \spab{\ell}.{ P_{61}}.{ 3}^4 \spb{6}.{ 1}^2
   \spb{\eta_2}.{ \ell}^2} \over {\spa{4}.{ 5} \spa{5}.{ \ell} \spa{\eta_2}.{ \ell}^2 \spa{\eta_2}.{ \eta_1}^2
   \spa{\omega_5}.{ \ell} \spaa{b}.{ P_{61}}.{ Q}.{ b}^2 \spab{\ell}.{ P_{61}}.{ 2}
   \spab{\ell}.{ P_{61}}.{ \ell}^2 \spab{\ell}.{ P_{61}}.{ \eta_2}^2 \spab{\ell}.{ P_{612}}.{ 3}
   \spb{2}.{ 3}}  } \cr   &  + {  {  \spa{4}.{ 2}^4 \spab{\ell}.{ P_{61}}.{ 5}^2 \spb{5}.{ 6} \spb{6}.{ \ell}^2
   (P_{61}^2)} \over {\spa{2}.{ 3} \spa{3}.{ 4} \spa{6}.{ \ell} \spa{\omega_5}.{ \ell}
   \spab{2}.{ P_{234}}.{ 5} \spab{\ell}.{ P_{61}}.{ \ell}^2 \spb{1}.{ 6}^2 (P_{234}^2)}
} }}
The above expression contains both single and double poles.
Here we give only the expression for the residues of the double poles
$\spab{\ell}.{ P_{61}}.{ \eta_1}^2$ and $\spab{\ell}.{ P_{61}}.{ \eta_2}^2$,

\eqn\name{\eqalign{ C_{61}^{(r,4:d)} = \dea \dedeb \ {\cal I}^{(4:d)} }}
\eqn\name{\eqalign{
 {\cal I}^{(4:d)} &=
- {
 {4   \spa{1}.{ \ell}^3 \spa{4}.{ 1} \spa{4}.{ 6} \spa{a}.{ b}^4 \spa{\eta_1}.{ 4}^2
   \spab{\ell}.{ Q}.{ 6} \spab{\ell}.{ P_{61}}.{ 3}^4 \spb{1}.{ 6} \spb{1}.{
   \eta_1} \spb{\eta_1}.{ \ell}}
\over
  {\spa{4}.{ 5} \spa{5}.{ \ell} \spa{6}.{ \ell} \spa{\eta_1}.{ \ell}^2 \spa{\eta_1}.{ \eta_2}^2
   \spa{\omega_5}.{ \ell} \spaa{b}.{ P_{61}}.{ Q}.{ b}^2 \spab{\ell}.{ P_{61}}.{ 2}
   \spab{\ell}.{ P_{61}}.{ \ell} \spab{\ell}.{ P_{61}}.{ \eta_1}^2 \spab{\ell}.{ P_{612}}.{ 3}
   \spb{2}.{ 3}}  }
\cr   &  - {  {2   \spa{1}.{ \ell}^2 \spa{4}.{ 1}^2 \spa{a}.{ b}^4 \spa{\eta_1}.{ 4}^2
   \spab{\ell}.{ Q}.{ 6} \spab{\ell}.{ P_{61}}.{ 3}^4 \spb{1}.{ \eta_1}
   \spb{6}.{ 1} \spb{\eta_1}.{ \ell}}
\over
  {\spa{4}.{ 5} \spa{5}.{ \ell} \spa{\eta_1}.{ \ell}^2 \spa{\eta_1}.{ \eta_2}^2 \spa{\omega_5}.{ \ell}
   \spaa{b}.{ P_{61}}.{ Q}.{ b}^2 \spab{\ell}.{ P_{61}}.{ 2} \spab{\ell}.{ P_{61}}.{ \ell}
   \spab{\ell}.{ P_{61}}.{ \eta_1}^2 \spab{\ell}.{ P_{612}}.{ 3} \spb{2}.{ 3}}
  }
\cr   &  - {
 {2   \spa{1}.{ \ell}^3 \spa{4}.{ 6}^2 \spa{a}.{ b}^4 \spa{\eta_1}.{ 4}^2 \spab{\ell}.{ Q}.{ 6}
   \spab{\ell}.{ P_{61}}.{ 3}^4 \spb{1}.{ 6} \spb{6}.{ \eta_1} \spb{\eta_1}.{
   \ell}}
\over
  {\spa{4}.{ 5} \spa{5}.{ \ell} \spa{6}.{ \ell} \spa{\eta_1}.{ \ell}^2 \spa{\eta_1}.{ \eta_2}^2
   \spa{\omega_5}.{ \ell} \spaa{b}.{ P_{61}}.{ Q}.{ b}^2 \spab{\ell}.{ P_{61}}.{ 2}
   \spab{\ell}.{ P_{61}}.{ \ell} \spab{\ell}.{ P_{61}}.{ \eta_1}^2 \spab{\ell}.{ P_{612}}.{ 3}
   \spb{2}.{ 3}}  }
\cr   &  + {  {  \spa{1}.{ \ell}^4 \spa{4}.{ 6}^2 \spa{a}.{ b}^4 \spa{\eta_1}.{ 4}^2
   \spab{\ell}.{ Q}.{ 6} \spab{\ell}.{ P_{61}}.{ 3}^4 \spb{1}.{ 6}^2
   \spb{\eta_1}.{ \ell}^2}
\over
  {\spa{4}.{ 5} \spa{5}.{ \ell} \spa{6}.{ \ell} \spa{\eta_1}.{ \ell}^2 \spa{\eta_1}.{ \eta_2}^2
   \spa{\omega_5}.{ \ell} \spaa{b}.{ P_{61}}.{ Q}.{ b}^2 \spab{\ell}.{ P_{61}}.{ 2}
   \spab{\ell}.{ P_{61}}.{ \ell}^2 \spab{\ell}.{ P_{61}}.{ \eta_1}^2 \spab{\ell}.{ P_{612}}.{ 3}
   \spb{2}.{ 3}}  }
\cr   &  - {  {2   \spa{1}.{ \ell}^3 \spa{4}.{ 1} \spa{4}.{ 6} \spa{a}.{ b}^4
   \spa{\eta_1}.{ 4}^2 \spab{\ell}.{ Q}.{ 6} \spab{\ell}.{ P_{61}}.{ 3}^4 \spb{6}.{ 1}^2
   \spb{\eta_1}.{ \ell}^2}
\over {\spa{4}.{ 5} \spa{5}.{ \ell} \spa{\eta_1}.{ \ell}^2 \spa{\eta_1}.{ \eta_2}^2
   \spa{\omega_5}.{ \ell} \spaa{b}.{ P_{61}}.{ Q}.{ b}^2 \spab{\ell}.{ P_{61}}.{ 2}
   \spab{\ell}.{ P_{61}}.{ \ell}^2 \spab{\ell}.{ P_{61}}.{ \eta_1}^2 \spab{\ell}.{ P_{612}}.{ 3}
   \spb{2}.{ 3}}  }
\cr &  + {  {  \spa{1}.{ \ell}^2 \spa{4}.{ 1}^2 \spa{6}.{ \ell} \spa{a}.{ b}^4
   \spa{\eta_1}.{ 4}^2 \spab{\ell}.{ Q}.{ 6} \spab{\ell}.{ P_{61}}.{ 3}^4 \spb{6}.{ 1}^2
   \spb{\eta_1}.{ \ell}^2}
\over {\spa{4}.{ 5} \spa{5}.{ \ell} \spa{\eta_1}.{ \ell}^2 \spa{\eta_1}.{ \eta_2}^2
   \spa{\omega_5}.{ \ell} \spaa{b}.{ P_{61}}.{ Q}.{ b}^2 \spab{\ell}.{ P_{61}}.{ 2}
   \spab{\ell}.{ P_{61}}.{ \ell}^2 \spab{\ell}.{ P_{61}}.{ \eta_1}^2 \spab{\ell}.{ P_{612}}.{ 3}
   \spb{2}.{ 3}}  }
\cr
 & + \{\eta_1 \leftrightarrow \eta_2 \}
}}
The sum of residue of the double poles reads,
\eqn\cLXIresIVd{\eqalign{ C_{61}^{(r,4:d)} =&
- { 4
   \spa{4}.{ 1}
   \spa{4}.{ 6}
   \spa{a}.{ b}^4
   \spa{\eta_1}.{ 4}^2
   \spb{1}.{ 6}
   \spb{1}.{\eta_1}
   \spab{\eta_1}.{P_{61}}.{ \eta_1}
\over
   \spa{4}.{ 5}
   \spa{\eta_1}.{ \eta_2}^2
   \spaa{b}.{ P_{61}}.{ Q}.{ b}^2
   \spb{2}.{ 3} (P_{61}^2)
}
\ P_2 \Big[P_{61}|\eta_1] ,L_1^{II:C_{61}},L_2^{II:C_{61}} \Big]
\cr   &
- { 2
   \spa{4}.{ 1}^2
   \spa{a}.{ b}^4
   \spb{1}.{ \eta_1}
   \spb{6}.{ 1}
   \spa{\eta_1}.{ 4}^2
   \spab{\eta_1}.P_{61}.{ \eta_1}
\over
   \spa{4}.{ 5}
   \spa{\eta_1}.{ \eta_2}^2
   \spaa{b}.{ P_{61}}.{ Q}.{ b}^2
   \spb{2}.{ 3}
   (P_{61}^2)
}
\ P_2 \Big[P_{61}|\eta_1] ,M_1^{II:C_{61}},M_2^{II:C_{61}} \Big]
\cr   &
- {2
   \spa{4}.{ 6}^2
   \spa{a}.{ b}^4
   \spa{\eta_1}.{ 4}^2
   \spb{1}.{ 6}
   \spb{6}.{ \eta_1}
   \spab{\eta_1}.P_{61}.{\eta_1}
\over
   \spa{4}.{ 5}
   \spa{\eta_1}.{ \eta_2}^2
   \spaa{b}.{ P_{61}}.{ Q}.{ b}^2
   \spb{2}.{ 3}
  (P_{61}^2)
}
\ P_2 \Big[P_{61}|\eta_1] ,L_1^{II:C_{61}},L_2^{II:C_{61}} \Big]
\cr   &
-
{
   \spa{4}.{ 6}^2
   \spa{a}.{ b}^4
   \spa{\eta_1}.{ 4}^2
   \spb{1}.{ 6}^2
   \spab{\eta_1}.P_{61}.{ \eta_1}^2
\over
   \spa{4}.{ 5}
   \spa{\eta_1}.{ \eta_2}^2
   \spaa{b}.{ P_{61}}.{ Q}.{ b}^2
   \spb{2}.{ 3}
   (P_{61}^2)^2
}
\ P_2 \Big[P_{61}|\eta_1] ,N_1^{II:C_{61}},N_2^{II:C_{61}} \Big]
\cr   &
+
{2
   \spa{4}.{ 1}
   \spa{4}.{ 6}
   \spa{a}.{ b}^4
   \spa{\eta_1}.{ 4}^2
   \spb{6}.{ 1}^2
   \spab{\eta_1}.P_{61}.{ \eta_1}^2
\over
   \spa{4}.{ 5}
   \spa{\eta_1}.{ \eta_2}^2
   \spaa{b}.{ P_{61}}.{ Q}.{ b}^2
   \spb{2}.{ 3}
   (P_{61}^2)^2
}
\ P_2 \Big[P_{61}|\eta_1] ,L_1^{II:C_{61}},O_2^{II:C_{61}} \Big]
\cr &
-
{
   \spa{4}.{ 1}^2
   \spa{a}.{ b}^4
   \spa{\eta_1}.{ 4}^2
   \spb{6}.{ 1}^2
   \spab{\eta_1}.P_{61}.{ \eta_1}^2
\over
   \spa{4}.{ 5}
   \spa{\eta_1}.{ \eta_2}^2
   \spaa{b}.{ P_{61}}.{ Q}.{ b}^2
   \spb{2}.{ 3}
   (P_{61}^2)^2
}
\ P_2 \Big[P_{61}|\eta_1] ,O_1^{II:C_{61}},O_2^{II:C_{61}} \Big]
\cr
 & + \{\eta_1 \leftrightarrow \eta_2 \}
}}
with
\eqn\name{\eqalign{
L_1^{II:C_{61}} &=  \{|1\ra,|1\ra,|1\ra, Q|6], P_{61}|3],P_{61}|3],P_{61}|3],P_{61}|3] \} \cr
L_2^{II:C_{61}} &=  \{|5\ra,|6\ra,|\eta_1\ra,|\eta_1\ra,|\eta_1\ra,|\omega_5\ra,P_{61}|2],P_{612}|3]  \} \cr
M_1^{II:C_{61}} &=  \{|1\ra,|1\ra, Q|6], P_{61}|3],P_{61}|3],P_{61}|3],P_{61}|3] \} \cr
M_2^{II:C_{61}} &=  \{|5\ra,|\eta_1\ra,|\eta_1\ra,|\eta_1\ra,|\omega_5\ra,P_{61}|2],P_{612}|3]  \} \cr
N_1^{II:C_{61}} &=  \{|1\ra,|1\ra,|1\ra,|1\ra, Q|6], P_{61}|3],P_{61}|3],P_{61}|3],P_{61}|3] \} \cr
N_2^{II:C_{61}} &=  \{|5\ra,|6\ra,|\eta_1\ra,|\eta_1\ra,|\eta_1\ra,|\eta_1\ra,|\omega_5\ra,P_{61}|2],P_{612}|3]  \} \cr
O_1^{II:C_{61}} &=  \{|1\ra,|1\ra,|6\ra, Q|6], P_{61}|3],P_{61}|3],P_{61}|3],P_{61}|3] \} \cr
O_2^{II:C_{61}} &=  \{|5\ra,|\eta_1\ra,|\eta_1\ra,|\eta_1\ra,|\eta_1\ra,|\omega_5\ra,P_{61}|2],P_{612}|3]  \} \cr
}}
since we used $|\ell] = P_{61}|\eta_1\ra$.

\subsubsec{The term $C_{61}^{(r,5)} $}

i) $t$-integrated formula:
\eqn\name{\eqalign{ C_{61}^{(r,5)} &=
 - {{ 2 \dea \deb \spa{1}.{ \ell}^3 \spa{4}.{ 1}^2 \spa{4}.{ \ell}^2 \spab{\ell}.{ P_{61}}.{ 3}^4
   \spb{1}.{ 6} \spb{1}.{ \ell}^2} \over {\spa{4}.{ 5} \spa{5}.{ \ell} \spa{6}.{ \ell} \spa{\omega_5}.{ \ell}
   \spaa{\ell}.{ P_{61}}.{ Q}.{ \ell} \spab{\ell}.{ P_{61}}.{ 2} \spab{\ell}.{ P_{61}}.{ \ell}^4
   \spab{\ell}.{ P_{612}}.{ 3} \spb{2}.{ 3}}  } \cr   &  - {
 {4 \dea \deb \spa{1}.{ \ell}^4 \spa{4}.{ 1} \spa{4}.{ 6} \spa{4}.{ \ell}^2
   \spab{\ell}.{ P_{61}}.{ 3}^4 \spb{1}.{ 6}^2 \spb{1}.{ \ell}^2} \over
  {\spa{4}.{ 5} \spa{5}.{ \ell} \spa{6}.{ \ell}^2 \spa{\omega_5}.{ \ell} \spaa{\ell}.{ P_{61}}.{ Q}.{ \ell}
   \spab{\ell}.{ P_{61}}.{ 2} \spab{\ell}.{ P_{61}}.{ \ell}^4 \spab{\ell}.{ P_{612}}.{ 3}
   \spb{2}.{ 3} \spb{6}.{ 1}}  } \cr   &  - {  {2 \dea \deb \spa{1}.{ \ell}^3 \spa{4}.{ 6}^2 \spa{4}.{ \ell}^2
   \spab{\ell}.{ P_{61}}.{ 3}^4 \spb{1}.{ 6} \spb{6}.{ \ell}^2} \over
  {\spa{4}.{ 5} \spa{5}.{ \ell} \spa{6}.{ \ell} \spa{\omega_5}.{ \ell} \spaa{\ell}.{ P_{61}}.{ Q}.{ \ell}
   \spab{\ell}.{ P_{61}}.{ 2} \spab{\ell}.{ P_{61}}.{ \ell}^4 \spab{\ell}.{ P_{612}}.{ 3}
   \spb{2}.{ 3}}  } \cr   &  + {  {2 \dea \deb \spa{1}.{ \ell} \spa{4}.{ 2}^4 \spab{\ell}.{ P_{61}}.{ 5}^3
   \spb{6}.{ \ell}^2 (P_{61}^2)} \over {\spa{2}.{ 3} \spa{3}.{ 4} \spa{6}.{ \ell} \spa{\omega_5}.{ \ell}
   \spab{2}.{ P_{234}}.{ 5} \spab{\ell}.{ P_{61}}.{ \ell}^4 \spb{1}.{ 6} (P_{234}^2)}
} }}

ii) full derivative: \eqn\name{\eqalign{ C_{61}^{(r,5)} = \dea
\dedeb \ {\cal I}^{(5)} }} \eqn\cLXIderV{\eqalign{
{\cal I}^{(5)} &=
 - {{ 2   \spa{4}.{ 1}^2 \spa{4}.{ \ell}^2 \spab{\ell}.{ P_{61}}.{ 3}^4 \spb{6}.{ \ell}} \over
  {\spa{4}.{ 5} \spa{5}.{ \ell} \spa{6}.{ \ell} \spa{\omega_5}.{ \ell} \spaa{\ell}.{ P_{61}}.{ Q}.{ \ell}
   \spab{\ell}.{ P_{61}}.{ 2} \spab{\ell}.{ P_{61}}.{ \ell} \spab{\ell}.{ P_{612}}.{ 3}
   \spb{2}.{ 3}}  } \cr   &  + {  {2   \spa{4}.{ 1}^2 \spa{4}.{ \ell}^2 \spab{\ell}.{ P_{61}}.{ 3}^4
   \spb{6}.{ 1} \spb{6}.{ \ell}^2} \over {\spa{4}.{ 5} \spa{5}.{ \ell} \spa{\omega_5}.{ \ell}
   \spaa{\ell}.{ P_{61}}.{ Q}.{ \ell} \spab{\ell}.{ P_{61}}.{ 2} \spab{\ell}.{ P_{61}}.{ \ell}^2
   \spab{\ell}.{ P_{612}}.{ 3} \spb{1}.{ 6} \spb{2}.{ 3}}  } \cr   &  - {
 {2   \spa{1}.{ \ell}^2 \spa{4}.{ 6}^2 \spa{4}.{ \ell}^2 \spab{\ell}.{ P_{61}}.{ 3}^4
   \spb{6}.{ \ell}^3} \over {3 \spa{4}.{ 5} \spa{5}.{ \ell} \spa{6}.{ \ell} \spa{\omega_5}.{ \ell}
   \spaa{\ell}.{ P_{61}}.{ Q}.{ \ell} \spab{\ell}.{ P_{61}}.{ 2} \spab{\ell}.{ P_{61}}.{ \ell}^3
   \spab{\ell}.{ P_{612}}.{ 3} \spb{2}.{ 3}}  } \cr   &  - {
 {2   \spa{4}.{ 1}^2 \spa{4}.{ \ell}^2 \spa{6}.{ \ell} \spab{\ell}.{ P_{61}}.{ 3}^4 \spb{6}.{ 1}^2
   \spb{6}.{ \ell}^3} \over {3 \spa{4}.{ 5} \spa{5}.{ \ell} \spa{\omega_5}.{ \ell}
   \spaa{\ell}.{ P_{61}}.{ Q}.{ \ell} \spab{\ell}.{ P_{61}}.{ 2} \spab{\ell}.{ P_{61}}.{ \ell}^3
   \spab{\ell}.{ P_{612}}.{ 3} \spb{1}.{ 6}^2 \spb{2}.{ 3}}  } \cr   &  + {
 {2   \spa{4}.{ 2}^4 \spab{\ell}.{ P_{61}}.{ 5}^3 \spb{6}.{ \ell}^3 (P_{61}^2)} \over
  {3 \spa{2}.{ 3} \spa{3}.{ 4} \spa{6}.{ \ell} \spa{\omega_5}.{ \ell} \spab{2}.{ P_{234}}.{ 5}
   \spab{\ell}.{ P_{61}}.{ \ell}^3 \spb{1}.{ 6}^2 (P_{234}^2)}
} }}

\bigskip

Finally, the coefficient of the bubble $I_{2:2;6}$ can be written
as a sum of the residues of all the poles,
by adding eqs. \cLXIderI, \cLXIderII, \cLXIderIII,
\cLXIderIV, \cLXIderV, and \cLXIresIVd:

\eqn\name{\eqalign{ c_{2:2;6} = \sum_{j=1}^{10} \lim_{\ell \to
\ell_j} \spa{\ell}.{\ell_j}
 \sum_{i=1}^{5}{\cal I}^{(i)}
 + C_{61}^{(4:d)},
}}
where the single poles are
\eqn\name{\eqalign{ |\ell_j\ra = |4\ra, |5\ra, |6\ra,
P_{612}|3], |\omega_5\ra, P_{61}|2],
 |\eta_1\ra, |\eta_2\ra, P_{61}|\eta_1], P_{61}|\eta_2], \qquad (j=1,\ldots,10).
}}

\subsubsec {{\bf 3-Mass-Triangle contribution from $C_{61}$}}

\eqn\name{\eqalign{ C_{61}^{(3m)} &=
                       C_{61}^{(3m,1)}
                     + C_{61}^{(3m,2)}
                     + C_{61}^{(3m,3)}
}}

\subsubsec{The term $C_{61}^{(3m,1)} $}

i) after $t$-integration: \eqn\name{\eqalign{ C_{61}^{(3m,1)} &=
  { 2 \spa{4}.{ 2}^2 \spb{2}.{ 3} \over \spa{4}.{ 5}}
 {{ \dea \deb \ \spa{ \ell}.{1}^2  \spa{ \ell}.{4}^2 \spab{\ell}.{ Q}.{ 6}
  \spab{\ell}.{ P_{61}}.{ 3}^2 }
 \over
{
   \spab{\ell}.{ Q}.{ \ell}
   \spab{\ell}.{ P_{61}}.{ \ell} \
   \spaa{\ell}.{ P_{61}}.{ Q}.{ \ell} \
   \spa{ \ell}.5
   \spa{ \ell}.6
   \spa{ \ell}.{\omega_5}
   \spab{\ell}.{ P_{61}}.{ 2}
   \spab{\ell}.{ P_{612}}.{ 3}}
} }}

ii) triangle coefficient:
 \eqn\cLXImmmI{\eqalign{
C_{61}^{(3m,1)} &=
  { 2 \spa{4}.{ 2}^2 \spb{2}.{ 3} \over \spa{4}.{ 5}} \
C_3^{II}[L_a,L_b^{II},P_{61},Q] \cr }}
\eqn\name{\eqalign{ L_a &= \{|1\ra,|1\ra,|4\ra,|4\ra,  Q|6],
P_{61}|3], P_{61}|3]
       \} \cr
L_b^{II} &= \{ |5\ra,|6\ra,|\omega_5\ra, P_{61}|2],P_{612}|3],
|\eta\ra
           \}
}}

\subsubsec{The term $C_{61}^{(3m,2)} $}

i) after $t$-integration: \eqn\name{\eqalign{ C_{61}^{(3m,2)} =& -
{4 \spa{4}.{ 1} \spa{4}.{ 2} \over \spa{4}.{ 5}} { { \dea \deb \
   \spa{ \ell}.{ 1}^2
   \spa{ \ell}.{ 4}^2
   \spab{\ell}.{ Q}.{ 1}
   \spab{\ell}.{ Q}.{ 6}
   \spab{\ell}.{ P_{61}}.{ 3}^3}
\over {
   \spab{\ell}.{ Q}.{ \ell}
   \spab{\ell}.{ P_{61}}.{ \ell} \
   \spaa{\ell}.{ P_{61}}.{ Q}.{ \ell}^2 \
   \spa{ \ell}.5
   \spa{ \ell}.6
   \spa{ \ell}.{\omega_5}
   \spab{\ell}.{ P_{61}}.{ 2}
   \spab{\ell}.{P_{612}}.{ 3}}
} \cr
& - {4 \spa{4}.{ 2} \spa{4}.{ 6} \over \spa{4}.{ 5} } {
 { \dea \deb \
   \spa{ \ell}.1^2
   \spa{ \ell}.4^2
   \spab{\ell}.{ Q}.{ 6}^2
   \spab{\ell}.{ P_{61}}.{ 3}^3}
  \over
 {
   \spab{\ell}.{ Q}.{ \ell}
   \spab{\ell}.{ P_{61}}.{ \ell} \
   \spaa{\ell}.{ P_{61}}.{ Q}.{ \ell}^2 \
   \spa{ \ell}.5
   \spa{ \ell}.6
   \spa{ \ell}.{\omega_5}
   \spab{\ell}.{ P_{61}}.{ 2}
   \spab{\ell}.{ P_{612}}.{ 3}}
} }}

ii) triangle coefficient:
 \eqn\cLXImmmII{\eqalign{
C_{61}^{(3m,2)} =& - {4 \spa{4}.{ 1} \spa{4}.{ 2} \over \spa{4}.{
5}} \ C_3^{III}[L_a,L_{b,1}^{III},L_{b,2}^{III},P_{61},Q] \cr &
- {4 \spa{4}.{ 2} \spa{4}.{ 6} \over \spa{4}.{ 5} }
C_3^{III}[M_a,L_{b,1}^{III},L_{b,2}^{III},P_{61},Q]
\cr }}
\eqn\name{\eqalign{ L_a &= \{ |1\ra,|1\ra,|4\ra,|4\ra,
          Q|1], Q|6], P_{61}|3], P_{61}|3], P_{61}|3]
       \} \cr
M_a &= \{ |1\ra,|1\ra,|4\ra,|4\ra,
          Q|6], Q|6], P_{61}|3], P_{61}|3], P_{61}|3]
       \} \cr
L_{b,1}^{III} &= \{|5\ra,|6\ra,|\omega_5\ra, P_{61}|2],P_{612}|3],
                   |\eta\ra, |\eta_2\ra, |\eta_2\ra
           \} \cr
L_{b,2}^{III} &= \{|5\ra,|6\ra,|\omega_5\ra, P_{61}|2],P_{612}|3],
                   |\eta\ra, |\eta_1\ra, |\eta_1\ra
           \} \cr
}}

\subsubsec{The term $C_{61}^{(3m,3)} $}

i) after $t$-integration: \eqn\name{\eqalign{ C_{61}^{(3m,3)}  =& \
 {2 \spa{4}.{ 1}^2 \over \spa{4}.{ 5} \spb{2}.{ 3}  }
{{ \dea \deb \
   \spa{ \ell}.{1}^2
   \spa{ \ell}.{4}^2
   \spab{\ell}.{ Q}.{ 1}^2
   \spab{\ell}.{ Q}.{ 6}
   \spab{\ell}.{ P_{61}}.{ 3}^4
  }
   \over
 {
   \spab{\ell}.{ Q}.{ \ell}
   \spab{\ell}.{ P_{61}}.{ \ell} \
   \spaa{\ell}.{ P_{61}}.{ Q}.{ \ell}^3 \
   \spa{ \ell}.5
   \spa{ \ell}.6
   \spa{ \ell}.{\omega_5}
   \spab{\ell}.{ P_{61}}.{ 2}
   \spab{\ell}.{ P_{612}}.{ 3}
 }  } \cr
&  + {4  \spa{4}.{ 1} \spa{4}.{ 6} \over \spa{4}.{ 5} \spb{2}.{ 3}}
{  { \dea \deb \
   \spa{ \ell}.{1}^2
   \spa{ \ell}.{4}^2
   \spab{\ell}.{ Q}.{ 1}
   \spab{\ell}.{ Q}.{ 6}^2
   \spab{\ell}.{ P_{61}}.{ 3}^4}
\over
  {
   \spab{\ell}.{ Q}.{ \ell}
   \spab{\ell}.{ P_{61}}.{ \ell} \
   \spaa{\ell}.{ P_{61}}.{ Q}.{ \ell}^3 \
   \spa{ \ell}.5
   \spa{ \ell}.6
   \spa{ \ell}.{\omega_5}
   \spab{\ell}.{ P_{61}}.{ 2}
   \spab{\ell}.{ P_{612}}.{ 3}
}  } \cr
& + {2 \spa{4}.{ 6}^2  \over  \spa{4}.{ 5} \spb{2}.{ 3}}
 {
 { \dea \deb \
   \spa{ \ell}.{1}^2
   \spa{ \ell}.{4}^2
   \spab{\ell}.{ Q}.{ 6}^3
   \spab{\ell}.{ P_{61}}.{ 3}^4}
\over {
   \spab{\ell}.{ Q}.{ \ell}
   \spab{\ell}.{ P_{61}}.{ \ell} \
   \spaa{\ell}.{ P_{61}}.{ Q}.{ \ell}^3 \
   \spa{ \ell}.5
   \spa{ \ell}.6
   \spa{ \ell}.{\omega_5}
   \spab{\ell}.{ P_{61}}.{ 2}
   \spab{\ell}.{ P_{612}}.{ 3}
} } }}

ii) triangle coefficient:
 \eqn\cLXImmmIII{\eqalign{
C_{61}^{(3m,3)}  =& \
 {2 \spa{4}.{ 1}^2 \over \spa{4}.{ 5} \spb{2}.{ 3}  } \
C_3^{IV}[L_a,L_{b,1}^{IV},L_{b,2}^{IV},P_{61},Q] \cr
& + {4  \spa{4}.{ 1} \spa{4}.{ 6} \over \spa{4}.{ 5} \spb{2}.{ 3}} \
C_3^{IV}[M_a,L_{b,1}^{IV},L_{b,2}^{IV},P_{61},Q] \cr
& + {2 \spa{4}.{ 6}^2  \over  \spa{4}.{ 5} \spb{2}.{ 3}} \
C_3^{IV}[N_a,L_{b,1}^{IV},L_{b,2}^{IV},P_{61},Q] }}
\eqn\name{\eqalign{ L_a &= \{ |1\ra,|1\ra,|4\ra,|4\ra,
          Q|1], Q|1], Q|6], P_{61}|3], P_{61}|3], P_{61}|3], P_{61}|3]
       \} \cr
M_a &= \{ |1\ra,|1\ra,|4\ra,|4\ra,
          Q|1], Q|6], Q|6], P_{61}|3], P_{61}|3], P_{61}|3], P_{61}|3]
       \} \cr
N_a &= \{ |1\ra,|1\ra,|4\ra,|4\ra,
          Q|6], Q|6], Q|6], P_{61}|3], P_{61}|3], P_{61}|3], P_{61}|3]
       \} \cr
L_{b,1}^{IV} &= \{ |5\ra,|6\ra,|\omega_5\ra, P_{61}|2],P_{612}|3],
                   |\eta\ra, |\eta_2\ra, |\eta_2\ra, |\eta_2\ra
           \} \cr
L_{b,2}^{IV} &= \{ |5\ra,|6\ra,|\omega_5\ra, P_{61}|2],P_{612}|3],
                   |\eta\ra, |\eta_1\ra, |\eta_1\ra, |\eta_1\ra
           \} \cr
}}


\bigskip

Finally, the coefficient of the three-mass triangle $I_{3:2:2;2}$ is
given by the sum of \cLXImmmI, \cLXImmmII, and \cLXImmmIII:
\eqn\name{\eqalign{ c_{3:2:2;2} =
               C_{61}^{(1,3m,1)}
             + C_{61}^{(1,3m,2)}
             + C_{61}^{(1,3m,3)}.
}}

\newsec{Conclusions}

In this work we completed the cut-constructible part of
$\CA^{\rm scalar}$
for the one-loop six-gluon amplitude in QCD. This completes the calculation
of the cut-constructible component of the one-loop six-gluon amplitude in QCD
as whole.

The method we adopted from \BrittoHA\ relies on the combination of ideas belonging to
{\it generalized unitarity}, to build the cuts out of
tree-level amplitudes, and {\it complex spinor algebra}, to carry
on the phase space integration.
This method was already successfully applied in the computation of the
$\CA^{\N=1}$ partner of the six-gluon amplitude.
For the current task, we extended its features nontrivially to
deal with the more general features of amplitudes
belonging to
less supersymmetric
theories, such as QCD.

The background knowledge of two properties of one-loop amplitudes, namely
{\it supersymmetric decomposition}, and {\it integral reduction} to a linear
combination of analytically known scalar functions, associated to two-, three-,
and four-point topologies,
allowed us to concentrate mainly on the phase space integration.

By exploiting the finiteness of the amplitude, we chose an integral
basis involving only boxes, three-mass triangles, and bubbles.
Therefore, the
problem of computing the (cut-constructible part of the) amplitude was shifted
to the calculation of the corresponding coefficients -- or rather to their
{\it extraction} from the cut integrals.

Our goal was to reduce the complexity of the calculation as much as
possible to trivial spinor algebra manipulations, and to minimize the
number of actual integrations to be finally performed.

The coefficients of the box functions could be
computed via the quadruple-cut method without any integration whatsoever.
Indeed, these have all appeared previously in the literature.

The known analytic properties of the bubble and three-mass
triangle functions enabled us to distinguish unequivocally among them.
To be more explicit,
their branch cuts represented a specific
signature to identify the coefficients of bubbles and three-mass triangles
separately.
The former multiply
a term that after the
integration would have generated a rational term;
the latter multiply a term that after the
integration would have generated logarithms with square roots in the  arguments.

After this {\it a priori} analysis on the properties of the expected results,
we could apply our optimized algorithm for the phase space integration.
We wrote the `twistor-motivated' Lorentz invariant phase space measure and the cut
integrands in spinorial formalism, with the loop momentum
written in its two components of opposite helicity, namely holomorphic and an antiholomorphic spinor variables.
We used trivial spinor algebra to disentangle the dependence on the two
variables and to write the integrands as a spinor derivative with respect to
one of the two integration variables.
We found that the expression of the result, at that stage of the
calculation, contained only four classes of integrands.

The integration was finally performed by combining the
{\it holomorphic anomaly}, 
which is an adaptation of the Cauchy residue theorem,
 and {\it Feynman parametrization}.
That required the development of a technique for dealing with spinorial
integrands carrying {\it multiple poles}, which constitutes the novel
and the most powerful feature of our algorithm.

\bigskip
In view of the recent progress in understanding  the recursive behaviour
of scattering amplitudes, the results here obtained can, in principle, be used
to drive the recursion relations for constructing the left-over
rational piece of the six-gluon amplitude \BernCQ.
Moreover, due to its polylogarithmic structure, the amplitude we
computed could represent a bootstrap point for the calculation of
one-loop amplitudes with more external legs and different helicity
configuration, once the one-loop recursive behaviour is fully
sorted out.

\smallskip
The numerical implementation of the results here presented, and their
crosschecks, is left to future work. 

\bigskip
The method we have developed may be used to compute the cut-constructible part
of a generic one-loop $n-$point amplitude with different particle content.


\bigskip
\bigskip
\centerline{\bf Acknowledgments}

We thank all the participants and organizers of the
workshop  ``From Twistors To Amplitudes'', and especially Babis Anastasiou and Nigel Glover, for the fruitful environment out of 
which this project started.

We would like to thank Freddy Cachazo and Thomas Gehrmann for 
feedback on 
the manuscript and Evgeny Buchbinder for conversations.
P. M. wishes to thank Daniel Maitre for discussing useful features of
Mathematica and Ettore Remiddi for an enjoyable discussion on complex
analysis.

R. B. is supported by Stichting FOM.  
B. F. is supported by the Marie Curie Research Training Network 
under contract number MRTN-CT-2004-005104.  
P. M. has been partially supported by the US DOE grant
DE-FG03-91ER40662 Task J, and by the European Commission Marie Curie
Fellowship under contract number MEIF-CT-2006-024178.

\appendix{A}{Tree-Level Amplitudes}

MHV tree amplitudes with fermions and scalars may be derived from supersymmetric Ward identities \refs{\grisaru, \mangparke} applied to the Parke-Taylor formula \parke.  See \DixonCF\ for a review.

Here we summarize some NMHV tree-level amplitudes which are useful for
our calculations. A similar list was  given in Appendix B of \BrittoHA.
However there is a technical point in these results, so we recall
them again.

Each unitarity cut integral has two terms, from the two possible helicity assignments for the internal propagators.
It is found that to simplify results,
we would like the expressions for
the tree amplitudes in each of these two terms to be related in a simple, symmetric way.
For example, if we derive them by on-shell recursion relations
\refs{\BrittoAP,\BrittoFQ,\LuoMY,\LuoRX}, we should take the same
reference momenta for each pair at every step.
Under this convention we have
following results where $a=2$ is for  scalars and $a=1$ is for
fermions.  Taking $a=0$ reproduces the results for all-gluon
amplitudes; these may be used to check the relation \nzerosing.

\eqn\oppny{\eqalign{
    A( 4^+_{F/S}, 5^+, 6^+,1^-,2^-,3^-_{F/S}) = &
{ \gb{3|1+2|6}^3\over [6~1][1~2]\braket{3~4}\braket{4~5} P_{345}^2
\gb{5|6+1|2}} \left( -{ \gb{4|1+2|6}\over \gb{3|1+2|6}}\right)^a \cr
& +   { \gb{1|5+6|4}^3\over [2~3][3~4]\braket{5~6}\braket{6~1}
P_{561}^2 \gb{5|6+1|2}} \left({ \gb{1|5+6|3}\over
\gb{1|5+6|4}}\right)^a }}

\eqn\manif{\eqalign{
 A( 4^-_{F/S}, 5^+, 6^+,1^-,2^-,3^+_{F/S})=
&    { \gb{4|1+2|6}^4\over [6~1][1~2]\braket{3~4}\braket{4~5}
P_{345}^2 \gb{5|6+1|2}\gb{3|1+2|6}} \left({ \gb{3|1+2|6} \over
\gb{4|1+2|6}}\right)^a \cr &  +   { \gb{1|5+6|3}^4\over
[2~3][3~4]\braket{5~6}\braket{6~1} P_{561}^2
\gb{5|6+1|2}\gb{1|5+6|4}} \left(-{\gb{1|5+6|4}\over \gb{1|5+6|3}}
\right)^a }}

\eqn\sncf{\eqalign{
 A(1^+_{F/S},2^-,3^+,4^+, 5^-, 6^-_{F/S})
& =   { [3~4]^4 \braket{5~6}^4 \over [2~3][3~4]
\braket{5~6}\braket{6~1} P_{234}^2 \gb{1| P_{234}|4} \gb{5|
P_{234}|2}} \left( - {\braket{5~1}\over \braket{5~6}}\right)^a \cr &
+  { \gb{2|P_{456}|4}^4 \over \braket{1~2}\braket{2~3} [4~5][5~6]
\gb{3|P_{456}|6} \gb{1|P_{456}|4} P_{456}^2}
\left(-{\braket{2~1}[6~4] \over \gb{2|P_{456}|4}}\right)^a \cr &  +
{ \gb{5| P_{345}|1}^4 \over \braket{3~4}\braket{4~5} [6~1][1~2]
P_{345}^2 \gb{5| P_{345}|2}\gb{3| P_{345}|6}} \left( { \gb{5|
P_{345}|6} \over \gb{5| P_{345}|1}}\right)^a }}

\eqn\nvbv{\eqalign{
 A(1^-_{F/S},2^-,3^+,4^+, 5^-, 6^+_{F/S})
& =   { [3~4]^4 \braket{5~1}^4 \over [2~3][3~4]
\braket{5~6}\braket{6~1} P_{234}^2 \gb{1| P_{234}|4} \gb{5|
P_{234}|2}} \left( {\braket{5~6}\over \braket{5~1}}\right)^a \cr &
+  { \braket{1~2}^4 [4~6]^4 \over \braket{1~2}\braket{2~3}
[4~5][5~6] \gb{3|P_{456}|6} \gb{1|P_{456}|4} P_{456}^2} \left( {
\gb{2|P_{456}|4} \over \braket{2~1}[6~4]}\right)^a \cr &  + { \gb{5|
P_{345}|6}^4 \over \braket{3~4}\braket{4~5} [6~1][1~2] P_{345}^2
\gb{5| P_{345}|2}\gb{3| P_{345}|6}} \left( -{ \gb{5| P_{345}|1}
\over \gb{5| P_{345}|6}}\right)^a }}

\eqn\neufl{\eqalign{ A(1^-_{F/S},2^-,3^+,4^-, 5^+, 6^+_{F/S}) & =
{ \gb{4|P_{123}|3}^4 \over [1~2][2~3]\braket{4~5}\braket{5~6}
P_{123}^2 \gb{4|P_{123}|1} \gb{6|P_{123}|3}} \left( {
[1~3]\braket{4~6}\over \gb{4|P_{123}|3}}\right)^a \cr &  +  {
\braket{1~2}^4 [3~5]^4 \over \braket{6~1} \braket{1~2}[3~4][4~5]
\gb{2|P_{612}|5} \gb{6|P_{612}|3} P_{612}^2} \left( {
\braket{2~6}\over \braket{2~1}}\right)^a \cr &  +   { [5~6]^4
\braket{4~2}^4 \over \braket{2~3}\braket{3~4}[5~6][6~1] P_{234}^2
\gb{4|P_{234}|1} \gb{2|P_{234}|5}} \left( - {[5~1]\over [5~6] }
\right)^a }}

\eqn\linkon{\eqalign{ A(1^+_{F/S},2^-,3^+,4^-, 5^+, 6^-_{F/S}) & =
{[1~3]^4\braket{4~6}^4 \over [1~2][2~3]\braket{4~5}\braket{5~6}
P_{123}^2 \gb{4|P_{123}|1} \gb{6|P_{123}|3}} \left(- {
\gb{4|P_{123}|3}\over [1~3]\braket{4~6}}\right)^a\cr &  + {
\braket{6~2}^4 [3~5]^4 \over \braket{6~1} \braket{1~2}[3~4][4~5]
\gb{2|P_{612}|5} \gb{6|P_{612}|3} P_{612}^2} \left( -{
\braket{2~1}\over \braket{2~6}}\right)^a \cr &  +   { [5~1]^4
\braket{4~2}^4 \over \braket{2~3}\braket{3~4}[5~6][6~1] P_{234}^2
\gb{4|P_{234}|1} \gb{2|P_{234}|5}} \left( {[5~6]\over [5~1] }
\right)^a }}

Comparing each pair (\oppny,\manif), (\sncf,\nvbv),
(\neufl,\linkon), we see the pattern relating the amplitudes with opposite helicities for the fermions or scalars. In fact, this is also the
reason why a factor of  two appears in  each cut in our scalar loops.

The expression \sncf\ is different from the
one given in \BrittoHA. The reason is that we have used $(2,3)$ as
reference momenta in the recursion relations to derive the formula in \BrittoHA\ but used $(3,4)$ as reference momenta
here to match the ones used in \nvbv.

\appendix{B}{Feynman Parametrization and its Integration}

Here we demonstrate in detail how to integrate terms of types (3)
and (4) in the categorization \fourtypes.  These are the integrands
that give precisely the logarithmic contributions and require a
Feynman parameter.

We define certain functions that we find useful.
 Where these appear in the paper with a tilde, this means to take the complex conjugate.
We also use $\eta_{1,2}$ to denote the two solutions of the equation $\vev{\ell|P
Q|\ell}=0$ with the proper $P,Q$ momenta, both in this and the following appendix.

\subsec{Type (3)}

Let us start from the following integral: \eqn\typethr{ T_3  =  \int
\vev{\ell~d\ell}[\ell~d\ell]{F(\la)\vev{a~\ell}\over
\gb{\ell|P_{cut}|a}^{n-1}}{1\over
\gb{\ell|P_{cut}|\ell}\gb{\ell|P_a|\ell}}. } Here we have multiplied
numerator and denominator by a factor of $\vev{a~\ell}$. To do the
integration, first we introduce a Feynman parameter to rewrite $T_3$
as \eqn\typethrone{ T_3  =  \int_0^1 dz \int
\vev{\ell~d\ell}[\ell~d\ell]{F(\la)\vev{a~\ell}\over
\gb{\ell|P_{cut}|a}^{n-1}}{1\over \gb{\ell|P|\ell}^2},~~~P=z
P_{cut}+(1-z) P_a. } Next we use \speziell\ to write the integrand
as a derivative: \eqn\typethrtwo{ T_3  =  \int_0^1 dz \int
\vev{\ell~d\ell}[d\ell~\partial_\ell]\left({F(\la)\vev{a~\ell}[\W\eta~\ell]\over
\gb{\ell|P_{cut}|a}^{n-1}}{1\over
\gb{\ell|P|\ell}\gb{\ell|P|\W\eta}}\right), } where  $\W\eta$ is an
arbitrary but fixed spinor of negative chirality. A convenient
choice is $|\W\eta]=|P|a\rangle$.\foot{ There is one subtlety in the
choice of $\W\eta$. In \typethrtwo\ we must avoid choosing
$|\eta]\sim |a]$. The reason is that  the starting point \typethr\
already has a pole from the factor$[\ell~a]$. In fact, if we choose
$|\eta]\sim |a]$, then the integral $\int_0^1 dz$ will diverge. }
With this choice we find that \eqn\typethrthree{ T_3  =
 \int_0^1 dz \int
\vev{\ell~d\ell}[d\ell~\partial_\ell]\left(-{F(\la)\gb{a|P_{cut}|\ell}\over
\gb{\ell|P_{cut}|a}^{n-1}}{1\over \gb{\ell|P|\ell}(z P_{cut}^2-(1-z)
\gb{a|P_{cut}|a})}\right). }

Now we read out the residues of the poles. There are two kinds of
poles: single poles from $F(\la)$ (as explained in Section 2) and
the multiple pole from $\gb{\ell|P_{cut}|a}$. For residues from
a
single pole, the $z$-integration takes the form 
\eqn\inttypeone{
 Z_1 =\int_0^1 dz {1\over (a_1+ b_1 z)(c_1+d_1 z)}={1\over
a_1 d_1-b_1 c_1}\log\left( {a_1(c_1+d_1)\over c_1(a_1+b_1)}\right),
} where $a_1,b_1,c_1,d_1$ are rational functions (for example,
$a_1=\gb{\ell|P_a|\ell}$).\foot{There is a subtle point in this
expression: we have assumed that there is no pole $\vev{\ell~a}$. If
this pole exists, it is easy to see that $a_1=0$ at pole
$\ket{\ell}=\ket{a}$ and the $z$-integration diverges. We have no
general argument why this is true, except to notice that  there is a
factor ${1\over [\ell~a]}$ from the anti-holomorphic part. It may be
that for tree-level amplitudes, the two factors $\vev{\ell~a}$ and
$[\ell~a]$ have the property that  if one exists as a pole, then the
other cannot.} 
This is always a logarithmic function. Furthermore,
since it is rational, it does not have any square root which is the
signature of three-mass triangle and four-mass box functions. Thus
it contributes only to one-mass, two-mass and three-mass box
functions.

Now we discuss the residue from the multiple pole
$\gb{\ell|P_{cut}|a}^{n-1}$. One important feature of the form
\typethrthree\ is the factor $\gb{a|P_{cut}|\ell}$:\foot{Notice that
since $F(\la)$ is a function of $\la$ only, the factor
$\gb{a|P_{cut}|\ell}$ cannot be canceled.} {\sl it is zero
precisely at the location of the pole from $\gb{\ell|P_{cut}|a}$.}
In other words, there is no residue contribution from the multiple
pole $\gb{\ell|P_{cut}|a}$ at all.\foot{If for some reason we have
other multiple poles, we must be careful because in this case, the
$z$-integration may give a rational contribution. Fortunately, in
our decomposition no other multiple poles show up.}

\subsec{Type (4)}

The integral of type (4) is given by \eqn\typefour{\eqalign{ T_4  &
=  \int \vev{\ell~d\ell}[\ell~d\ell]{F(\la)\over
\vev{\ell|P_{cut}Q|\ell}^{n-1}}{1\over
\gb{\ell|P_{cut}|\ell}\gb{\ell|Q|\ell}} \cr & =  \int_0^1 dz \int
\vev{\ell~d\ell}[\ell~d\ell]{F(\la)\over
\vev{\ell|P_{cut}Q|\ell}^{n-1}}{1\over \gb{\ell|R|\ell}^2},~~~~R=z
Q+(1-z) P_{cut}. }} Similarly as for type (3), we arrive at
\eqn\typefourone{
 T_4  = \int_0^1 dz \int
\vev{\ell~d\ell}[d\ell~\partial_\ell]\left( {F(\la)\over
\vev{\ell|P_{cut}Q|\ell}^{n-1}}{\gb{\eta|R|\ell}\over
\vev{\ell~\eta} R^2\gb{\ell|R|\ell}}\right), } where we have chosen
the auxiliary spinor $|\W\eta]=|R|\eta\rangle$.

Again we have two kinds of poles: the single poles from $F(\la)$ and
multiple poles from the factor $\vev{\ell|P_{cut}Q|\ell}^{n-1}$. Let
us discuss them one by one.

\subsubsec{The Contribution from Single Poles}

The $z$-dependence comes from the momentum vector $R$, defined in
\typefour\ and appearing in the last factor of \typefourone. It may
be expressed as \eqn\ztwoform{
 Z_2= \int_0^1 dz { ( z c_1+c_2)\over
(a_0 z^2+ a_1 z+ a_2)
 (z b_1+b_2)},
} where we have defined \eqn\nocc{\eqalign{ a_0 & =
(Q-P_{cut})^2,~~~~a_1=2 P_{cut}\cdot (Q-P_{cut}),~~~~a_2= P_{cut}^2,
\cr b_1 & =
\gb{\ell|(Q-P_{cut})|\ell},~~~~b_2=\gb{\ell|P_{cut}|\ell},\cr c_1 &
=  \gb{\eta|(Q-P_{cut})|\ell},~~~~c_2=\gb{\eta|P_{cut}|\ell}. }}
To understand the various contributions, we split the integral as
follows: \eqn\also{\eqalign{ Z_2& =  \int_0^1 dz { ( z c_1 +c_2)
\over ( a_0 z^2 + a_1 z+ a_2) ( z b_1+ b_2)}
\cr & =
\int_0^1 dz \bigg\{ {
b_1( -b_2 c_1+ b_1 c_2) \over ( a_2 b_1^2 - a_1 b_1 b_2+ a_0 b_2^2)}
{1\over ( z b_1+ b_2)}\cr &  + { ( b_2 c_1-b_1 c_2)\over  2( a_2
b_1^2 - a_1 b_1 b_2+ a_0 b_2^2)} { (2 z a_0 + a_1) \over ( a_0 z^2 +
a_1 z+ a_2)} \cr &  + {   (2 a_2 b_1 c_1-a_1 b_2 c_1-a_1 b_1 c_2+ 2
a_0 b_2 c_2) \over 2( a_2 b_1^2 - a_1 b_1 b_2+ a_0 b_2^2)} {1\over (
a_0 z^2 + a_1 z+ a_2)} \bigg\}. }}
 Among these three terms, the first two will give
logarithmic functions with rational parameters, so these contribute
to one-mass, two-mass and three-mass box functions. The third term
will be \eqn\threemg{ \int_0^1 dz {1\over ( a_0 z^2 + a_1 z+ a_2)}
={1\over \sqrt{\Delta}} \log\left({ 2a_0 z +a_1-\sqrt{\Delta} \over
2a_0 z +a_1+\sqrt{\Delta}} \right)|_{0}^{1},~~~~\Delta= a_1^2-4 a_0
a_2, }
 where $\Delta$ is not a complete square,  so this
is not a rational function. In fact, the $\Delta$ is the
characteristic signature 
(Gram determinant)
which identifies the contribution with a particular
 three-mass triangle or four-mass box function. Specifically,
expression \threemg\ is the same as \threemdelta.

The above splitting of $Z_2$ makes sense if and only if $a_2 b_1^2 -
a_1 b_1 b_2+ a_0 b_2^2\neq 0$. From \nocc\ we find that
\eqn\xiku{\eqalign{
 a_2 b_1^2 -
a_1 b_1 b_2+ a_0 b_2^2 & =
-\vev{\ell|P_{cut}Q|\ell}[\ell|P_{cut}Q|\ell] \cr 2 a_2 b_1 c_1-a_1
b_2 c_1-a_1 b_1 c_2+ 2 a_0 b_2 c_2 & = -\vev{\ell|P_{cut} Q-
QP_{cut}|\eta}[\ell|P_{cut} Q|\ell] }} We see that $(a_2 b_1^2 - a_1
b_1 b_2+ a_0 b_2^2)$ is zero exactly for the  pole $\vev{\ell|P
Q|\ell}$, so our manipulation is safe for single poles other than
this one.

For future use, we define the following function: \eqn\rone{
R_1[\ell,\eta, P_{cut}, Q]={(2 a_2 b_1 c_1-a_1 b_2 c_1-a_1 b_1 c_2+
2 a_0 b_2 c_2) \over 2( a_2 b_1^2 - a_1 b_1 b_2+ a_0
b_2^2)}={\vev{\ell|P_{cut} Q- QP_{cut}|\eta}\over 2
\vev{\ell|P_{cut}Q|\ell}}. }

\subsubsec{The Contribution from  Poles in $\vev{\ell|P Q|\ell}$}

For the poles in $\vev{\ell|P_{cut} Q|\ell}^{n-1}$, things become much
more complicated. There are two facts we need to take into account.
The first is that, as seen in \xiku, \eqn\clotho{ a_2 b_1^2 - a_1
b_1 b_2+ a_0 b_2^2=0, }
 so we need to be careful when we do the splitting. The
second is that   the residue will have terms like
$$
{\gb{\eta|R|\eta_{1,2}}\gb{\xi|R|\eta_{1,2}}^{m-1}\over
R^2\gb{\eta_{1,2}|R|\eta_{1,2}}^{m}}
$$
with $m$ ranging from $1$ to $(n-1)$, where $\eta_{1,2}$ are the two
solutions for the poles in $\vev{\ell|P_{cut} Q|\ell}^{n-1}$. In other
words, we have more patterns for the $z$-integration.

Now we discuss the residues in detail for the cases $n=2,3,4,$ which
are the only ones needed in this paper.

\bigskip

{\bf The case of $n=2$:}

\bigskip

In this case, it is a single  pole. Using \clotho\  we can solve for
$a_0$:
$$a_0= {a_1
b_1 b_2-a_2 b_1^2\over b_2^2}.$$

Therefore we can split the integral as follows: \eqn\rarin{\eqalign{
 &  \int_0^1 dz { ( z c_1 +c_2)
\over ( a_0 z^2 + a_1 z+ a_2) ( z b_1+ b_2)} =\int_0^1 dz { b_2^2 (
z c_1 +c_2) \over(a_1 b_2 z+ a_2b_2- a_2 b_1 z)  ( z b_1+ b_2)^2}
\cr & =  \int_0^1 dz \bigg\{ {  (a_2 b_2 c_1+a_2 b_1 c_2-a_1 b_2 c_2)\over
b_2(2 a_2 b_1-a_1 b_2)}{b_2^2 \over (b_1 z+ b_2) (a_1 b_2 z+ a_2b_2-
a_2 b_1 z)} \cr & \qquad \qquad - {b_2( b_2 c_1-b_1 c_2)\over (2 a_2 b_1-a_1
b_2)}{1\over (b_2+ z b_1)^2} \bigg\} \cr & = - \int_0^1 dz {b_2( b_2 c_1-b_1
c_2)\over (2 a_2 b_1-a_1 b_2)}{1\over (b_2+ z b_1)^2}  + \int_0^1 dz
{  (a_2 b_2 c_1+a_2 b_1 c_2-a_1 b_2 c_2)\over b_2(2 a_2 b_1-a_1
b_2)}{1\over a_0 z^2+ a_1 z+a_2}. }} Of these two terms, the first
one gives a rational function while the second one gives a
logarithmic function. However, we can see that
$$
 b_2 c_1-c_2 b_1  =
-\vev{\ell~\eta}[\ell|P_{cut} Q|\ell],
$$
 so that, at our pole,
\eqn\lachesis{
 b_2 c_1-c_2 b_1  = 0,
} and hence the first term vanishes. 
We can solve \lachesis\ for
$c_1$ and simplify the coefficient: 
\eqn\rtwo{ R_2[\ell,\eta,
P_{cut}]={ (a_2 b_2 c_1+a_2 b_1 c_2-a_1 b_2 c_2)\over b_2(2 a_2
b_1-a_1 b_2)}= {c_2\over b_2}={\gb{\eta|P_{cut}|\ell}\over
\gb{\ell|P_{cut}|\ell}}. } 
Unlike the
function $R_1$ defined in \rone\ in which  $Q$ appears explicitly, $R_2$ does
not involve $Q$. However, in the formula for the coefficient, $R_2$ will depend on $Q$ through $\ell$
when evaluated at $\eta_{1,2}$, the solutions of $\vev{\ell|P_{cut}
Q|\ell}=0$.

{\bf The case of $n=3$:}

\bigskip

 For the case $n=3$ we have a double pole. Using our residue formula
\twopole\  we
 find that the new  $z$-integral
may be expressed as
 $$
 Z_3 = \int_0^1 dz { ( z c_1+c_2)(z d_1+d_2)\over (a_0 z^2+ a_1 z+ a_2)
 (z b_1+b_2)^2},
$$
where $a_i,b_i,c_i$ are the same as in \nocc, and \eqn\noccx{
 d_1  =
\gb{\xi|(Q-P_{cut})|\ell},~~~~d_2=\gb{\xi|P_{cut}|\ell}. } By
calculations similar to those in the case of $n=2$, we reach
\eqn\vvbenv{\eqalign{ Z_3&=
 \int_0^1 d z{ b_2^2 ( z c_1+c_2)(z d_1+d_2)\over (a_1 b_2
z- a_2 b_1 z+ a_2 b_2) (z b_1+b_2)^3}\cr & =  {c_1(-b_2  d_1+ b_1
d_2)\over b_1 (b_1+b_2) (2 a_2 b_1-a_1 b_2)} \cr &+ { c_1(a_2 b_2 d_1+a_2
b_1 d_2-a_1 b_2 d_2)\over b_2 b_1 (2 a_2 b_1-a_1 b_2)} \int_0^1
{b_2^2 \over (b_1 z+ b_2) (a_1 b_2 z+ a_2b_2- a_2 b_1 z)}, }} where
we have used the condition \lachesis\ to simplify the result. We see
that
$$
 -b_2 d_1+ b_1 d_2  =
 \vev{\ell~\xi}[\ell|P Q|\ell],
$$
 so that at our pole
\eqn\atropos{ -b_2 d_1+ b_1 d_2  = 0, }
 so the
rational contribution is zero. The relevant nonzero coefficient is
given by
\eqn\name{\eqalign{
 R_3[\ell,\eta,\xi,P_{cut}] = &{ c_1(a_2 b_2 d_1+a_2 b_1
d_2-a_1 b_2 d_2)\over b_2 b_1 (2 a_2 b_1-a_1 b_2)} \cr =&
-{\gb{\eta|(Q-P_{cut})|\ell}
\gb{\xi|P_{cut}(P_{cut}Q+QP_{cut})|\ell} \over
\gb{\ell|(Q-P_{cut})|\ell} (2 P_{cut}^2\gb{\ell|Q|\ell}-2
P_{cut}\cdot Q \gb{\ell|P_{cut}|\ell})}.
}}
Using the results \lachesis\ and \atropos, this can be simplified to
the formula \eqn\rthree{
 R_3[\ell,\eta,\xi,P_{cut}]= {c_2 d_2 \over
b_2^2}= {\gb{\eta|P_{cut}|\ell} \gb{\xi|P_{cut}|\ell}\over
\gb{\ell|P_{cut}|\ell}^2}. }

{\bf The case of $n=4$:}

For $n=4$ we have a triple pole; the new $z$-integration pattern is
given by
$$
Z_4= \int_0^1 dz { ( z c_1+c_2)(z d_1+d_2)^2\over (a_0 z^2+ a_1 z+
a_2)
 (z b_1+b_2)^3},
$$
where $a_i,b_i,c_i,d_i$ are given above in \nocc\ and \noccx.

Now we perform similar manipulations as in the case $n=3$.  Using the
three zero-conditions \clotho, \lachesis, and \atropos,
 we find once again that
the rational contribution is zero and we are left with
$$ Z_4  =
R_4 \int_0^1 dz {1\over a_0 z^2+a_1 z+ a_2},
$$
 where
\eqn\rfour{ R_4[\ell,\eta,\xi,P_{cut}] = {c_2 d_2^2\over b_2^3}=
{\gb{\eta|P_{cut}|\ell} \gb{\xi|P_{cut}|\ell}^2\over
\gb{\ell|P_{cut}|\ell}^3}. }

For general situations, it can be done by similar way. Observing
above patterns for $R_2, R_3, R_4$ we conjecture that for the
pattern
\eqn\Rpatt{ Z_{n+1} \equiv \int_0^1 dz { \prod_{i=1}^{n}
\gb{\xi_i|R|\ell}\over R^2 \gb{\ell|R|\ell}^n},}
the contribution of poles of the solution $\vev{\ell|P_{cut}
Q|\ell}=0$ to three-mass triangles or four-mass boxes will be given
by
\eqn\Rpattcon{ R_{n+1}[\ell,\{\xi_1,...,\xi_n\},P_{cut}] \equiv {
\prod_{i=1}^{n} \gb{\xi_i|P_{cut}|\ell}\over
\gb{\ell|P_{cut}|\ell}^n}.}
%

\appendix{C}{Coefficients of Three-Mass Triangles}

In this appendix we discuss the presentation of coefficients of
three-mass triangles. After the $t$-integration and splitting we end
up with an integral of the form \eqn\threemgenform{ S_3=\int
\vev{\ell~d\ell}[\ell~d\ell] {1\over
\gb{\ell|P|\ell}\gb{\ell|Q|\ell}} {\prod_{i=1}^N \vev{\ell~a_i}
\over \vev{\ell|P Q|\ell}^{(N-m)/2}\prod_{j=1}^m \vev{\ell~b_j}}, }
where the factor $\vev{\ell|P Q|\ell}$ is special to this case (here
$P$ is the $P_{cut}$ in Appendix B, but for simplicity we write $P$
in this part).

The first thing is to write it using Feynman parametrization as
\eqn\threemgenformz{\eqalign{ S_3 & =  \int_0^1 dz \int
\vev{\ell~d\ell}[\ell~d\ell] {1\over \gb{\ell|R|\ell}^2}
{\prod_{i=1}^N \vev{\ell~a_i} \over \vev{\ell|P
Q|\ell}^{(N-m)/2}\prod_{j=1}^m \vev{\ell~b_j}},~~R=zQ+(1-z)P, \cr &
=  \int_0^1 dz \int \vev{\ell~d\ell}[d\ell~\partial_\ell] {[\W
\eta~\ell]\over \gb{\ell|R|\ell}\gb{\ell|R|\W\eta}} {\prod_{i=1}^N
\vev{\ell~a_i} \over \vev{\ell|P Q|\ell}^{(N-m)/2}\prod_{j=1}^m
\vev{\ell~b_j}}, }} where $\W\eta$ is an arbitrary auxiliary spinor
of negative chirality. In this form we have a pole at
$|R|\W\eta\rangle$, which depends on $z$. To simplify calculations,
we take $|\W\eta]=|R|\eta\rangle$ so that \eqn\thrmgenformtwo{ S_3 =
\int_0^1 dz \int \vev{\ell~d\ell}[d\ell~\partial_\ell]
{\gb{\eta|R|\ell}\over \gb{\ell|R|\ell} R^2 \vev{\ell~\eta}}
{\prod_{i=1}^N \vev{\ell~a_i} \over \vev{\ell|P
Q|\ell}^{(N-m)/2}\prod_{j=1}^m \vev{\ell~b_j}}. } The reason for
this choice is that now all poles are independent of $z$ (notice the
extra pole $\vev{\ell~\eta}$); all the $z$-dependence is in the
first factor.

First, we deal with the single poles at $b_i,\eta$. For a single
pole, the $z$-integration pattern is given by $Z_2$ and $R_1$
defined in Appendix B. Thus we get the following result:
\eqn\threemsing{ C_3^{(sing)}  =  -\sum_{p=1}^{m+1} {\prod_{i=1}^N
\vev{b_p~a_i} \over \vev{b_p|P
Q|b_p}^{(N-m)/2}\prod_{j=1}^{\prime~m+1} \vev{b_p~b_j}}
R_1[b_p,\eta,P,Q] } There are several remarks about these results:
(1) the minus sign appears because we need to take the negative
residues of poles; (2) we have set $b_{m+1}=\eta$; (3) in the
denominator, the prime symbol $\prime$ means that we need to omit
the factor with $j=p$ in the product; (4) the factor
$R_1[b_p,\eta,P,Q]$ is the contribution of the $z$-integration
defined by \rone.

Now we need to deal with the pole from $\vev{\ell|P
Q|\ell}^{(N-m)/2}$. We proceed case by case. If $(N-m)/2=0$ there is
no such pole. The next case is that $(N-m)/2=1$, i.e., two single
poles. Then we have the following result: \eqn\threemmone{ S_{3;1} =
-\sum_{p=1,2}\lim_{\ell\to \eta_p}{\vev{\ell~\eta_p}\over
\vev{\ell|PQ|\ell}}{\prod_{i=1}^N \vev{\ell~a_i} \over
\vev{\ell~\eta}\prod_{j=1}^{N-2} \vev{\ell~b_j}} R_2[\ell,\eta,P] ,}
where $\eta_{1,2}$ are solutions of $\vev{\ell|PQ|\ell}=0$ and $R_2$
is defined by \rtwo. We will stick to the notation $\eta_{1,2}$.

Now we proceed to the case $(N-m)/2=2$, i.e., double poles. Using
$$
\vev{\ell|PQ|\ell}= \vev{\ell~\eta_1}\vev{\ell~\eta_2} {\vev{b|PQ|b}
\over \vev{a~b}^2}
$$
where $a,b$ are two spinors used to solve $\eta_1,\eta_2$ (see also
equations \newexp\ ), we get the following expression:
$$
S_3  =  \int_0^1 dz \int \vev{\ell~d\ell}[d\ell~\partial_\ell]
{\gb{\eta|R|\ell}\over \gb{\ell|R|\ell} R^2 \vev{\ell~\eta}}
{\prod_{i=1}^N \vev{\ell~a_i} \over
\vev{\ell~\eta_1}^2\vev{\ell~\eta_2}^2 {\vev{b|PQ|b}^2 \over
\vev{a~b}^4}\prod_{j=1}^{N-4} \vev{\ell~b_j}}
$$
Now using the formula \twopole\ for the residue of double poles, we
find (here the limit is more like replacement, but we use this
notation to simplify our discussion)
\eqn\missy{\eqalign{ & -\lim_{\ell\to
\eta_1}{\gb{\eta|R|\ell}\over \gb{\ell|R|\ell} R^2 \vev{\ell~\eta}}
{\prod_{i=1}^N \vev{\ell~a_i} \over \vev{\ell~\eta_2}^2
{\vev{b|PQ|b}^2 \over \vev{a~b}^4}\prod_{j=1}^{N-4}
\vev{\ell~b_j}}\left(\sum_{i=1}^{N-1} {\vev{L_{1i}~L_{2i}}\over
\vev{\eta_1~L_{1i}}\vev{\eta_1~L_{2i}}}+ {\gb{a_N|R|\eta_1}\over
\gb{\eta_1|R|\eta_1}\vev{\eta_1~a_N}} \right) \cr & -\{\eta_1\to
\eta_2, L_2\to L_3 \}, }} with \eqn\poppy{\eqalign{ L_1 & =
 \{ a_1,a_2,...,a_N\}, \cr
L_2 & =  \{ b_1, b_2,..., b_{N-4}, \eta, \eta_2,\eta_2, |R|\eta_1]
\}, \cr L_3 & =  \{ b_1, b_2,..., b_{N-4}, \eta, \eta_1,\eta_1,
|R|\eta_2] \}. }}
 There are two terms. The first term is the
contribution of double pole $\eta_1$ and the second term is that of
$\eta_2$. We got the latter from the former by replacing $\eta_1\to
\eta_2$ and $L_2\to L_3$. In the first line,  the most significant
manipulation is that in the brackets we have separated $N$ terms
into two parts: the first $N-1$ terms are independent of $R$, while
the last term depends on $R$ and thus on $z$. This will make the
$z$-integration different as we will see shortly.

Having residues given by \missy\  we need to do the Feynman
parameter integration and read out contribution to three-mass
triangles. This has essentially been done in Appendix B. Using
\rtwo\ and \rthree, we get
\eqn\manpan{\eqalign{ & -\lim_{\ell\to \eta_1}
 {\prod_{i=1}^N \vev{\ell~a_i} \over \vev{\ell~\eta}
\vev{\ell~\eta_2}^2 {\vev{b|PQ|b}^2 \over
\vev{a~b}^4}\prod_{j=1}^{N-4} \vev{\ell~b_j}}\left(\sum_{i=1}^{N-1}
{\vev{L_{1i}~L_{2i}}\over \vev{\ell~L_{1i}}\vev{\ell~L_{2i}}}
R_2[\ell,\eta,P]\right. \cr & \left. + {1\over \vev{\ell~a_N}}
R_3[\ell,\eta,a_N,P]\right)  -\{\eta_1\to \eta_2, L_2\to L_3 \} }}

Now we discuss the last case needed for our calculation, namely
$(N-m)/2=3$. We have \eqn\kitny{\eqalign{ S_3 & =  \int_0^1 dz \int
\vev{\ell~d\ell}[d\ell~\partial_\ell] {\gb{\eta|R|\ell}\over
\gb{\ell|R|\ell} R^2 \vev{\ell~\eta}} {\prod_{i=1}^N \vev{\ell~a_i}
\over \vev{\ell~\eta_1}^3\vev{\ell~\eta_2}^3 {\vev{b|PQ|b}^3 \over
\vev{a~b}^6}\prod_{j=1}^{N-6} \vev{\ell~b_j}}. }} Here we have two
triple poles. Again we need to read out the negative residue first
and then use previous results to get the coefficients. For a triple
pole $\eta_1$ there are several parts in the denominator: (1)
factors with $b_i,~i=1,...,N-6$, (2) one factor with $\eta$, (3)
three factors with $\eta_2$, and (4) one factor with $|R|\eta_1]$.
We group the first three together into the list $L_{b,1}^{IV}$
defined below. The reason doing that is that only the last one has
$z$-dependence. Using this notation we find that the residue of
$\eta_1$ is given by \eqn\tappan{\eqalign{ &
-{\gb{\eta_1|R|\eta_1}\over \gb{\eta_1|R|\eta_1} R^2
\vev{\eta_1~L_{b,1,N-5}^{IV}}} {\prod_{i=1}^N \vev{\eta_1~L_{a,i}}
\over \vev{\eta_1~\eta_2}^3 {\vev{b|PQ|b}^3 \over
\vev{a~b}^6}\prod_{j=1}^{N-6} \vev{\eta_1~L_{b,1,j}^{IV}}} \cr &
\left[ \sum_{i=1}^{N-2} {\vev{L_{a,i}~L_{b,1,i}^{IV}}\over
\vev{\eta_1~L_{a,i}}\vev{\eta_1~L_{b,1,i}^{IV}}}
{\vev{L_{a,N}~L_{a,i}}\over
\vev{\eta_1~L_{a,N}}\vev{\eta_1~L_{a,i}}}\right. \cr  &  +
{\gb{a_{N-1}|R|\eta_1}\over
\gb{\eta_1|R|\eta_1}\vev{\eta_1~a_{N-1}}} {\vev{a_N~a_{N-1}}\over
\vev{\eta_1~a_N}\vev{\eta_1~a_{N-1}}}+\sum_{1\leq i\leq j\leq N-2}
{\vev{L_{a,i}~L_{b,1,i}^{IV}}\over
\vev{\eta_1~L_{a,i}}\vev{\eta_1~L_{b,1,i}^{IV}}}
{\vev{L_{a,j}~L_{b,1,j}^{IV}}\over
\vev{\eta_1~L_{a,j}}\vev{\eta_1~L_{b,1,j}^{IV}}} \cr  & \left. +
\sum_{1\leq i\leq N-2}{\vev{L_{a,i}~L_{b,1,i}^{IV}}\over
\vev{\eta_1~L_{a,i}}\vev{\eta_1~L_{b,1,i}^{IV}}}{\gb{L_{a,N-1}|R|\eta_1}\over
\vev{\eta_1~L_{a,N-1}}\gb{\eta_1|R|\eta_1}}+\left({\gb{L_{a,N-1}|R|\eta_1}\over
\vev{\eta_1~L_{a,N-1}}\gb{\eta_1|R|\eta_1}}\right)^2 \right] }}
 In
this result, we have separated terms having different $R$ factors.
From this we can read the coefficient of the three-mass triangle as
\eqn\noodba{\eqalign{ & \left.{-\prod_{i=1}^N \vev{\eta_1~L_{a,i}}
\over \vev{\eta_1~L_{b,1,N-5}^{IV}}\vev{\eta_1~\eta_2}^3
{\vev{b|PQ|b}^3 \over \vev{a~b}^6}\prod_{j=1}^{N-6}
\vev{\eta_1~L_{b,1,j}^{IV}}}\right[ R_2[\eta_1,L_{b,1,N-5}^{IV},P]
\cr & \left( \sum_{i=1}^{N-2} {\vev{L_{a,i}~L_{b,1,i}^{IV}}\over
\vev{\eta_1~L_{a,i}}\vev{\eta_1~L_{b,1,i}^{IV}}}
{\vev{L_{a,N}~L_{a,i}}\over
\vev{\eta_1~L_{a,N}}\vev{\eta_1~L_{a,i}}}
\right. \cr &
+ \left. \sum_{1\leq i\leq j\leq
N-2} {\vev{L_{a,i}~L_{b,1,i}^{IV}}\over
\vev{\eta_1~L_{a,i}}\vev{\eta_1~L_{b,1,i}^{IV}}}
{\vev{L_{a,j}~L_{b,1,j}^{IV}}\over
\vev{\eta_1~L_{a,j}}\vev{\eta_1~L_{b,1,j}^{IV}}}\right) \cr & +
R_3[\eta_1,L_{b,N-5}^{IV},L_{a,N-1},P] \cr & \left(
{\vev{L_{a,N}~L_{a,{N-1}}}\over\vev{\eta_1~L_{a,{N-1}}}
\vev{\eta_1~L_{a,N}}\vev{\eta_1~L_{a,{N-1}}}} + \sum_{1\leq i\leq
N-2}{\vev{L_{a,i}~L_{b,1,i}^{IV}}\over\vev{\eta_1~L_{a,N-1}}
\vev{\eta_1~L_{a,i}}\vev{\eta_1~L_{b,1,i}^{IV}}}\right) \cr & \left.
+{1\over \vev{\eta_1~L_{a,N-1}}^2}
R_4[\eta_1,L_{b,N-5}^{IV},L_{a,N-1},P] \right]. }}

{\bf Summary:}

The reader who wishes to skip the derivations may simply use the
formulas given below.  Where these functions appear in the bulk of the paper with a tilde, this means to take the complex conjugate.

We define the following lists. \eqn\lucy{\eqalign{ L_a & =  \{ a_1,
a_2,..., a_N\}, \cr L_b^{I} & =  \{ b_1, b_2,..., b_N,\eta \}, \cr
L_b^{II} & =  \{b_1, b_2,...,b_{N-2},\eta \}, \cr L_{b,1}^{III} & =
\{b_1, b_2,...,b_{N-4},\eta, \eta_2,\eta_2 \}, \cr L_{b,2}^{III} & =
\{b_1, b_2,...,b_{N-4},\eta, \eta_1,\eta_1 \}, \cr L_{b,1}^{IV} & =
\{b_1, b_2,...,b_{N-6},\eta, \eta_2, \eta_2, \eta_2 \}, \cr
L_{b,2}^{IV} & =  \{b_1, b_2,...,b_{N-6},\eta, \eta_1, \eta_1,
\eta_1 \}, }}
 where $\eta$ is an arbitrary auxiliary spinor.

(1) Case one: for the integral
$$
S_3^{I}=\int \vev{\ell~d\ell}[\ell~d\ell] {1\over
\gb{\ell|P|\ell}\gb{\ell|Q|\ell}} {\prod_{i=1}^N \vev{\ell~a_i}
\over \prod_{j=1}^N \vev{\ell~b_j}},
$$
we have the coefficient \eqn\ethel{
 C_3^{I}
[L_a,L_b^{I},P,Q] =  \sum_{p=1}^{N+1} {-\prod_{i=1}^N
\vev{L_{b,p}^{I}~L_{a,i}} \over \prod_{j=1}^{\prime~N+1}
\vev{L_{b,p}^{I}~L_{b,j}^{I}}} R_1[L_{b,p}^{I},L_{b,N+1}^{I},P,Q]. }
Although in this paper we have not encountered this situation, we include it
for completeness.

(2) Case two: for the integral
$$
S_3=\int \vev{\ell~d\ell}[\ell~d\ell] {1\over
\gb{\ell|P|\ell}\gb{\ell|Q|\ell}} {\prod_{i=1}^N \vev{\ell~a_i}
\over \vev{\ell|P Q|\ell}\prod_{j=1}^{N-2} \vev{\ell~b_j}},
$$
we have the coefficient \eqn\milly{\eqalign{
C_3^{II}[L_a,L_b^{II},P,Q]  = & \sum_{p=1}^{N-1} {-\prod_{i=1}^N
\vev{L_{b,p}^{II}~L_{a,i}} \over \vev{L_{b,p}^{II}|P
Q|L_{b,p}^{II}}\prod_{j=1}^{\prime~N-1}
\vev{L_{b,p}^{II}~L_{b,j}^{II}}}
R_1[L_{b,p}^{II},L_{b,N-1}^{II},P,Q] \cr &
-\sum_{p=1,2}\lim_{\ell\to \eta_p}{\vev{\ell~\eta_p}\over
\vev{\ell|PQ|\ell}}{\prod_{i=1}^N \vev{\ell~L_{a,i}} \over
\vev{\ell~L_{b,N-1}^{II}}\prod_{j=1}^{N-2} \vev{\ell~L_{b,j}^{II}}}
R_2[\ell,L_{b,N-1}^{II},P]. }}

(3) Case three: for the integral
$$
S_3=\int \vev{\ell~d\ell}[\ell~d\ell] {1\over
\gb{\ell|P|\ell}\gb{\ell|Q|\ell}} {\prod_{i=1}^N \vev{\ell~a_i}
\over \vev{\ell|P Q|\ell}^2\prod_{j=1}^{N-4} \vev{\ell~b_j}},
$$
we have the coefficient \eqn\moutai{\eqalign{
 & C_3^{III}[L_a,L_{b,1}^{III},L_{b,2}^{III},P,Q]
\cr & =  \sum_{p=1}^{N-3} {-\prod_{i=1}^N
\vev{L_{b,p}^{III}~L_{a,i}} \over \vev{L_{b,p}^{III}|P
Q|L_{b,p}^{III}}^2\prod_{j=1}^{\prime~N-3}
\vev{L_{b,p}^{III}~L_{b,j}^{III}}}
R_1[L_{b,1,p}^{III},P,Q,L_{b,1,N-3}^{III}] \cr &
-\left[\lim_{\ell\to \eta_1}
 {\prod_{i=1}^N \vev{\ell~L_{a,i}} \over \vev{\ell~L_{b,1,N-3}^{III}}
\vev{\ell~\eta_2}^2 {\vev{b|PQ|b}^2 \over
\vev{a~b}^4}\prod_{j=1}^{N-4} \vev{\ell~L_{b,1,j}^{III}}}\right. \cr
& \left( R_2[\ell,L_{b,1,N-3}^{III},P]\sum_{i=1}^{N-1}
{\vev{L_{a,i}~L_{b,1,i}^{III}}\over
\vev{\eta_1~L_{a,i}}\vev{\eta_1~L_{b,1,i}^{III}}}  + {1\over
\vev{\eta_1~L_{a,N}}} R_3[\eta_1,L_{b,1,
N-3}^{III},L_{a,N},P]\right)
 \cr &  \left. +\{\eta_1\to \eta_2,L_{b,1}^{III}\to
L_{b,2}^{III}  \}\right]. }}

(4) Case four: for the integral
$$
S_3=\int \vev{\ell~d\ell}[\ell~d\ell] {1\over
\gb{\ell|P|\ell}\gb{\ell|Q|\ell}} {\prod_{i=1}^N \vev{\ell~a_i}
\over \vev{\ell|P Q|\ell}^{3}\prod_{j=1}^{N-6} \vev{\ell~b_j}},
$$
we have the coefficient \eqn\biimpp{\eqalign{ &
C_3^{IV}[L_a,L_{b,1}^{IV},L_{b,2}^{IV},P,Q] \cr & =
\sum_{p=1}^{N-5} {-\prod_{i=1}^N \vev{L_{b,p}^{IV}~L_{a,i}} \over
\vev{L_{b,p}^{IV}|P Q|L_{b,p}^{IV}}^3\prod_{j=1}^{\prime~N-5}
\vev{L_{b,p}^{IV}~L_{b,j}^{IV}}}
R_1[L_{b,1,p}^{IV},P,Q,L_{b,1,N-5}^{IV}] \cr & -\left[
\left.{\prod_{i=1}^N \vev{\eta_1~L_{a,i}} \over
\vev{\eta_1~L_{b,1,N-5}^{IV}}\vev{\eta_1~\eta_2}^3 {\vev{b|PQ|b}^3
\over \vev{a~b}^6}\prod_{j=1}^{N-6}
\vev{\eta_1~L_{b,1,j}^{IV}}}\right(
R_2[\eta_1,L_{b,1,N-5}^{IV},P]\right. \cr &  \left( \sum_{i=1}^{N-2}
{\vev{L_{a,i}~L_{b,1,i}^{IV}}\over
\vev{\eta_1~L_{a,i}}\vev{\eta_1~L_{b,1,i}^{IV}}}
{\vev{L_{a,N}~L_{a,i}}\over
\vev{\eta_1~L_{a,N}}\vev{\eta_1~L_{a,i}}}
\right. \cr &
+ \left.\sum_{1\leq i\leq j\leq
N-2} {\vev{L_{a,i}~L_{b,1,i}^{IV}}\over
\vev{\eta_1~L_{a,i}}\vev{\eta_1~L_{b,1,i}^{IV}}}
{\vev{L_{a,j}~L_{b,1,j}^{IV}}\over
\vev{\eta_1~L_{a,j}}\vev{\eta_1~L_{b,1,j}^{IV}}}\right) \cr &
 + \left(
{\vev{L_{a,N}~L_{a,{N-1}}}\over\vev{\eta_1~L_{a,{N-1}}}
\vev{\eta_1~L_{a,N}}\vev{\eta_1~L_{a,{N-1}}}} + \sum_{1\leq i\leq
N-2}{\vev{L_{a,i}~L_{b,1,i}^{IV}}\over\vev{\eta_1~L_{a,N-1}}
\vev{\eta_1~L_{a,i}}\vev{\eta_1~L_{b,1,i}^{IV}}}\right) \cr &
 \left.\left.R_3[\eta_1,L_{b,N-5}^{IV},L_{a,N-1},P] +{1\over
\vev{\eta_1~L_{a,N-1}}^2} R_4[\eta_1,L_{b,N-5}^{IV},L_{a,N-1},P]
\right) 
\right. \cr &
+ \left. \{\eta_1\to \eta_2, L_{b,1}^{IV} \to L_{b,2}^{IV} \}\right].
}}


\listrefs

\end